\documentclass[aps,prl,superscriptaddress,showpacs,twocolumn,nourl]{revtex4-2}
\usepackage[left=1.6cm,right=1.6cm,bottom=1.7cm,top=1.7cm,ignoreall]{geometry}
\usepackage[utf8]{inputenc}
\usepackage{graphicx}
\usepackage{amsmath}
\usepackage{amssymb}
\usepackage{xspace}
\usepackage{bm}
\usepackage{color}
\usepackage[squaren,Gray]{SIunits}
\usepackage[hyperindex=true]{hyperref}
\hypersetup{linktocpage,colorlinks=true,citecolor=blue,linkcolor=blue}
\usepackage{empheq}
\usepackage{braket}
\usepackage{physics}
\usepackage{multirow}
\usepackage{array}
\usepackage{makecell}
\usepackage{enumitem}
\usepackage{placeins}

\definecolor{Green}{RGB}{34,139,34}

\begin{document}

\title{Nonlinear integrated quantum photonics with AlGaAs}

\author{F.~Baboux}
\email{Corr. author: florent.baboux@u-paris.fr}
\affiliation{Université Paris Cité, CNRS, Laboratoire Matériaux et Phénomènes Quantiques, 75013 Paris, France}

\author{G.~Moody}
\email{moody@ucsb.edu}
\affiliation{Electrical and Computer Engineering Department, University of California, Santa Barbara, CA 93106, USA}

\author{S.~Ducci}
\email{sara.ducci@u-paris.fr}
\affiliation{Université Paris Cité, CNRS, Laboratoire Matériaux et Phénomènes Quantiques, 75013 Paris, France}

\begin{abstract}

Integrated photonics provides a powerful approach for developing compact, stable and scalable architectures for the generation, manipulation and detection of quantum states of light.  To this end, several material platforms are being developed in parallel, each providing its specific assets, and hybridization techniques to combine their strengths are now possible. This review focuses on AlGaAs, a III-V semiconductor platform combining a mature fabrication technology, direct band-gap compliant with electrical injection, low-loss operation, large electro-optic effect, and compatibility with superconducting detectors for on-chip detection. We detail recent implementations of room-temperature sources of quantum light based on the high second- and third-order optical nonlinearities of the material, as well as photonic circuits embedding various functionalities ranging from polarizing beamsplitters to Mach-Zehnder interferometers, modulators and tunable filters. We then present several realizations of quantum state engineering enabled by these recent advances and discuss open perspectives and remaining challenges in the field of integrated quantum photonics with AlGaAs.

\end{abstract}

\maketitle

\section{Introduction}

Integrated photonics is playing a major role in the advancement of quantum technologies, with applications including communication, computing, simulation, and metrology \cite{Flamini2018,Slussarenko2019,Polino2020,moody20222022} that are expected to radically change the way we transmit, process, measure, and store information. Indeed, photons are ideal low-noise carriers of information, and they can be used by directly exploiting their properties or to probe/interact with the state of other quantum systems; moreover, they maintain a high degree of coherence without the need of vacuum or cooling systems, thus offering significant advantages for practical implementations. Over the years, important milestones have been reached, culminating in the recent demonstrations of space-to-ground quantum communications \cite{Liao2017} and quantum computational advantage \cite{Zhong2020,Madsen2022}. 

\begin{figure*}[t]
	\centering
	\includegraphics[width=0.8\textwidth]{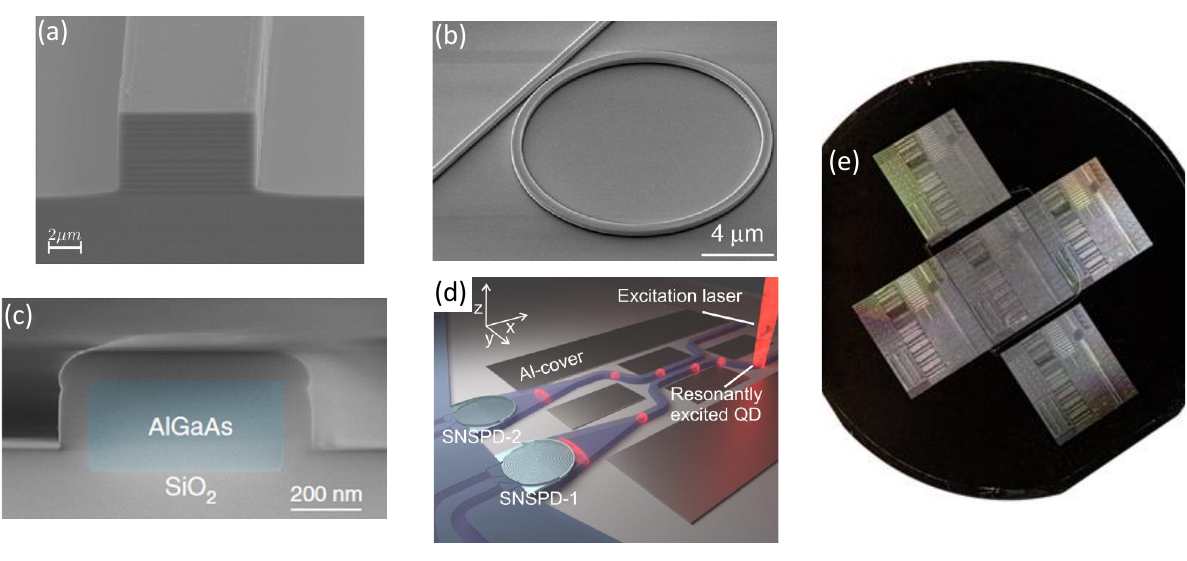}
	\caption{
		Examples of AlGaAs chips for quantum photonics. (a) AlGaAs Bragg reflection waveguide \cite{Autebert16}. (b) AlGaAsOI microring resonator  \cite{Chang2020}. (c) Cross-section of an AlGaAsOI waveguide \cite{Chang2020}. 
		(d) GaAs chip combining quantum dot, beamsplitter and superconducting nanowire single photon detector \cite{Schwartz2018}.
		(e) Fabrication of AlGaAsOI quantum photonic circuits on a 4" wafer \cite{Steiner2021}.
	}
	\label{Fig_Intro}
\end{figure*}

\begin{figure}[!t]
	\centering
	\includegraphics[width=0.9\columnwidth]{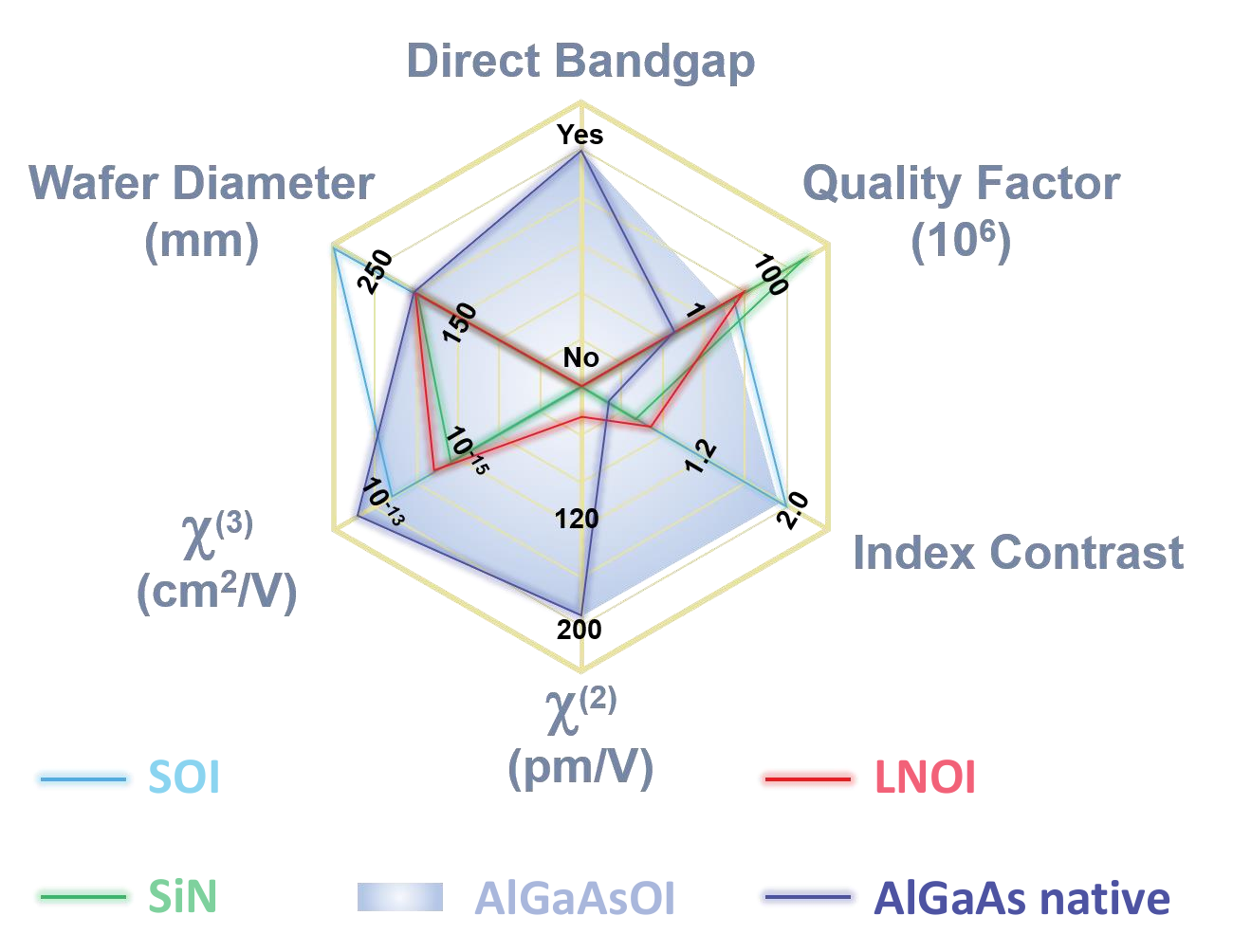}
	\caption{Comparison of the relevant metrics for quantum photonics for different nonlinear materials: silicon on insulator (SOI), silicon nitride (SiN), lithium niobate on insulator (LNOI), AlGaAs and AlGaAs on insulator (AlGaAsOI).}
	\label{Fig_Comparison}
\end{figure}

\begin{figure*}[!t]
	\centering
	\includegraphics[width=0.7\textwidth]{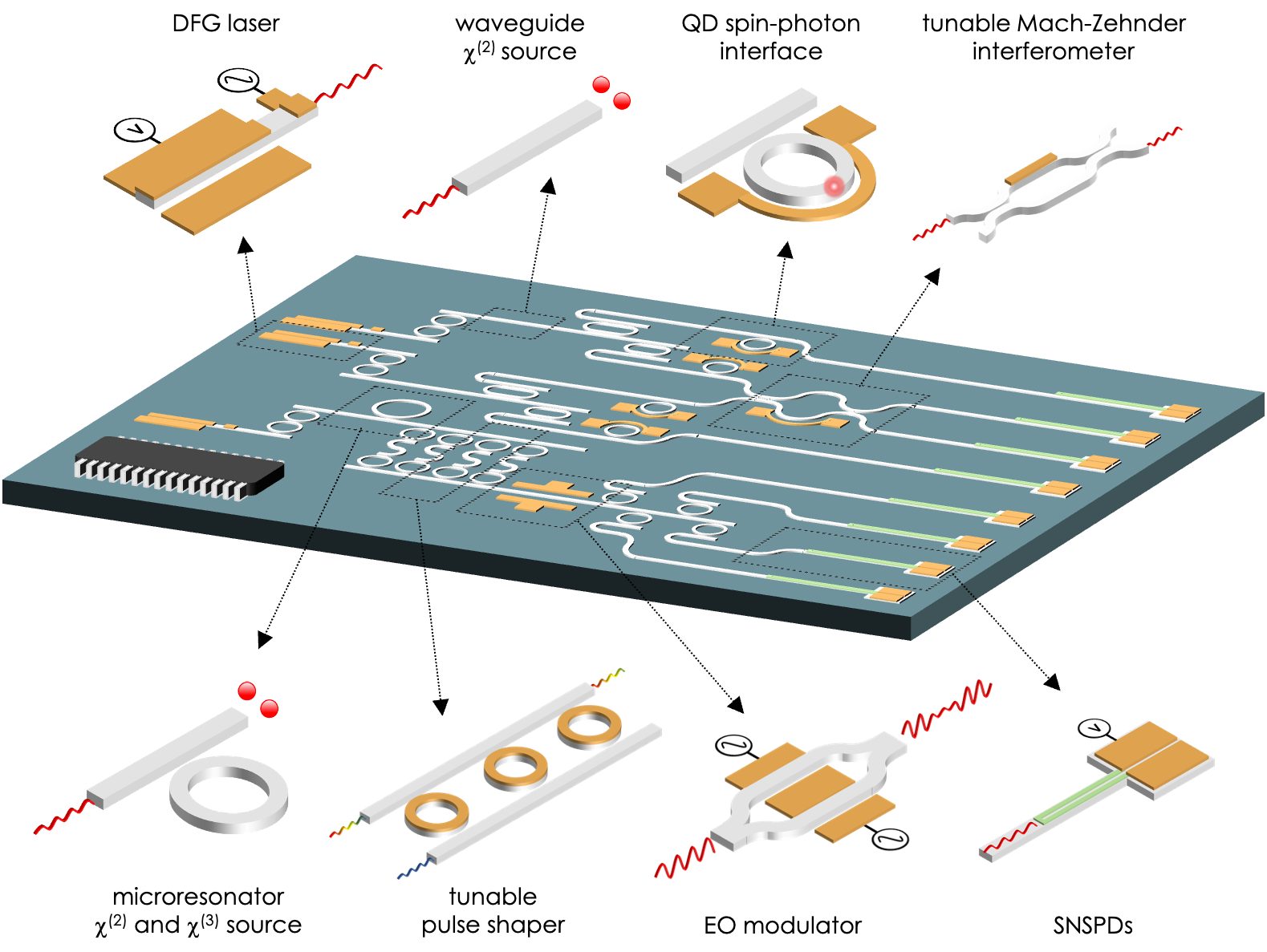}
	\caption{AlGaAs-based prospective integrated quantum photonic circuit, including a variety of components needed for specific functionalities as classical and quantum light sources, single photon detectors, modulators, interferometers, filters and light-matter interfaces.
	}
	\label{Fig_QPICs_illustration}
\end{figure*}

\begin{figure*}[!t]
	\centering
	\includegraphics[width=0.9\textwidth]{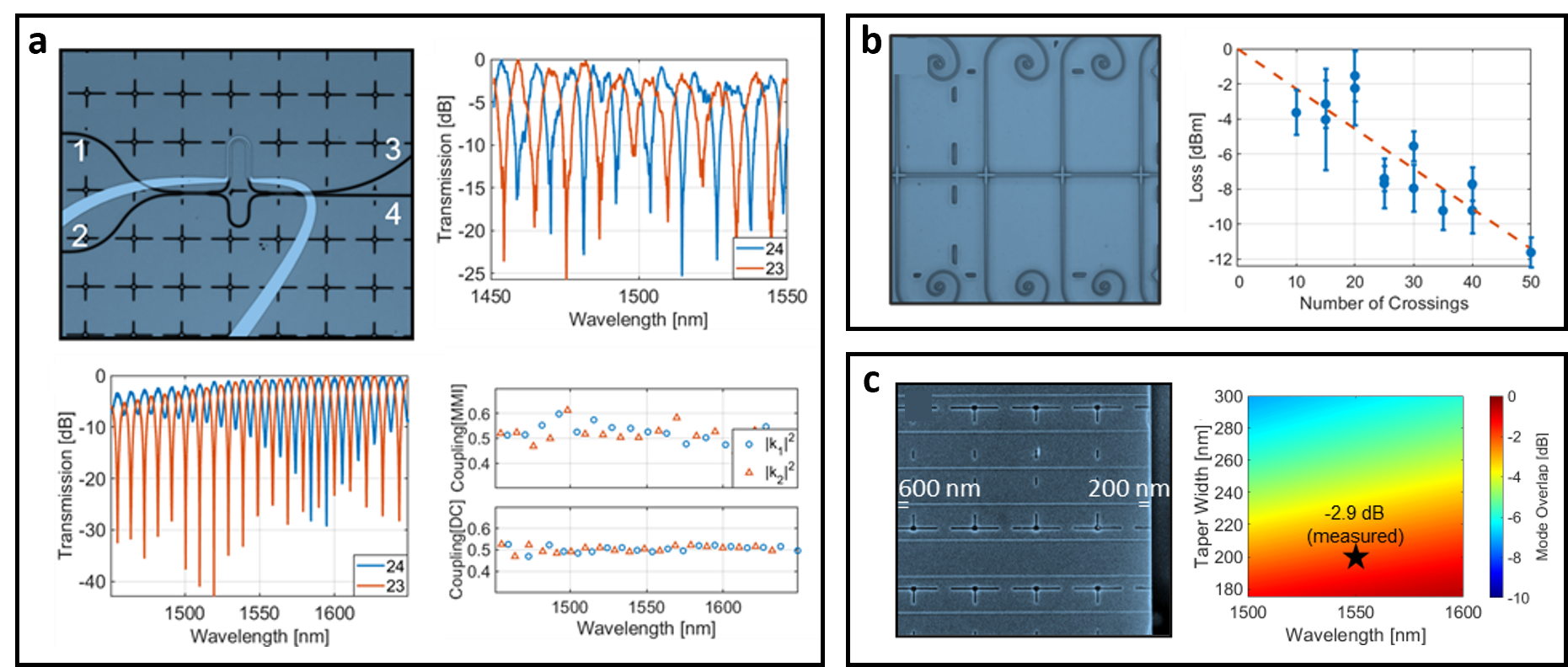}
	\caption{(a) Tunable unbalanced Mach Zehnder Interferometer (MZI) exhibiting $> $ 40 dB ($> $23 dB) extinction for MZIs with directional coupler (MMI) 3-dB splitters. The top-right (bottom-left) panels show the transmission spectra for MZIs with MMIs (directional couplers). The bottom-right panel shows the coupling coefficients measured from 1450 nm to 1650 nm. (b) Waveguide crossers with $< $ 0.2 dB/crosser loss and $> $ 40 dB extinction between cross ports. (c) Inverse tapers for chip-to-fiber coupling with $< $ 3 dB loss. Data shown is modified with permission from \cite{CastroAlGaAsOI}.
	}
	\label{Fig_QPICs}
\end{figure*}

\begin{figure*}[!t]
	\centering
	\includegraphics[width=0.9\textwidth]{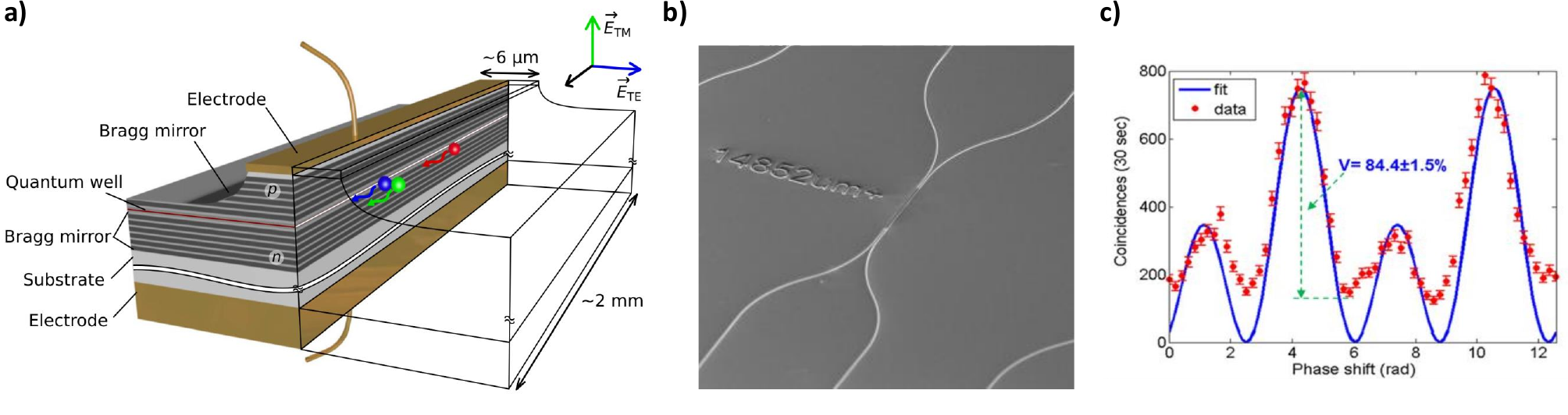}
	\caption{
		(a) Sketch of an AlGaAs electrically pumped source of photon pairs \cite{Boitier14}.
		(b) SEM image of an AlGaAs parametric source integrated with a 50/50 beamsplitter \cite{Belhassen18}.
		(c) Quantum interference pattern of a two-photon state measured in an integrated GaAs-based Mach-Zehnder interferometer \cite{Wang14}.
	}
	\label{Fig_Circuits}
\end{figure*}

\begin{figure*}[!th]
	\centering
	\includegraphics[width=0.7\textwidth]{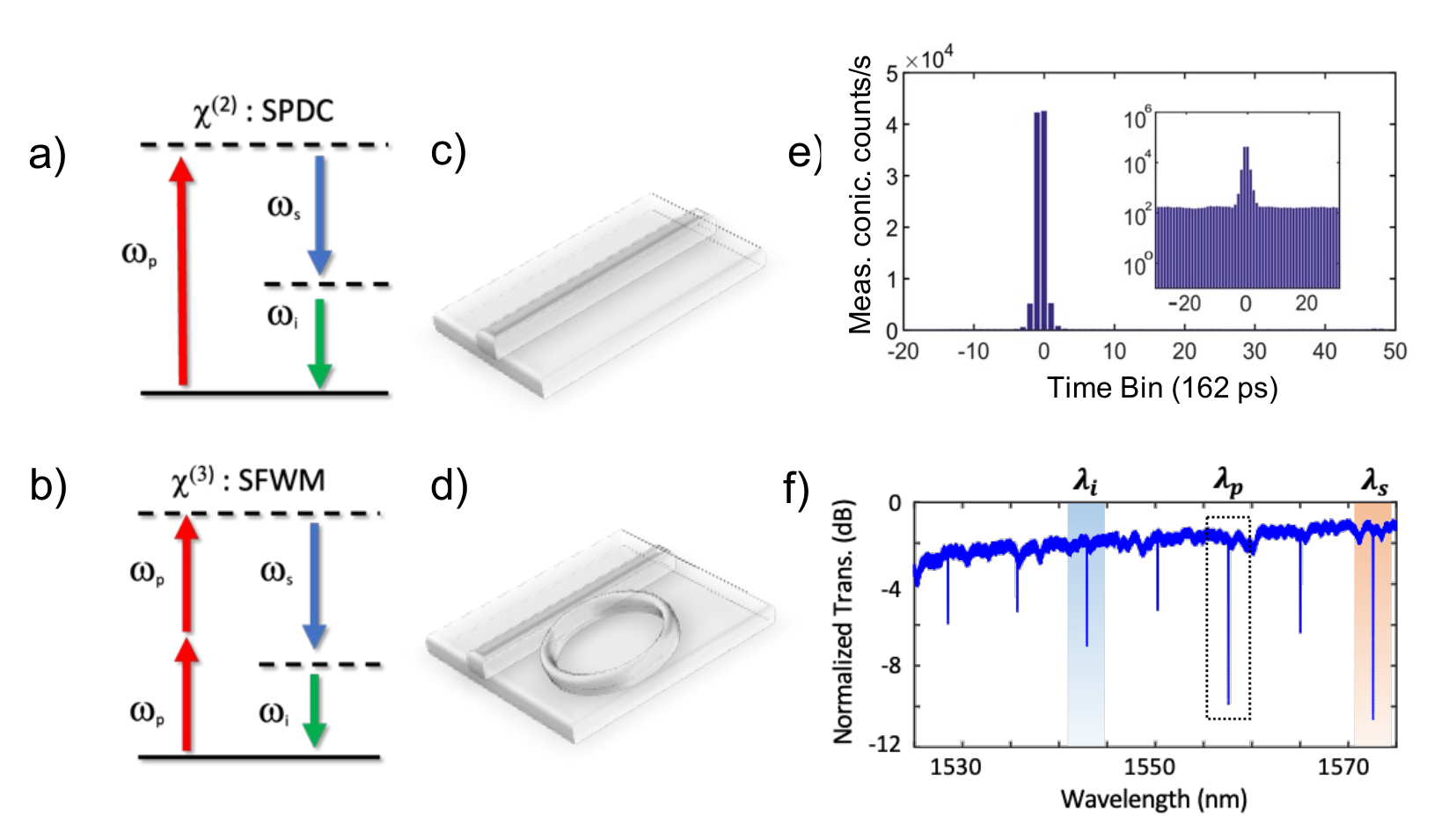}
	\caption{Quantum light generation with integrated nonlinear photonics relies on (a) $\chi^2$(SPDC) or (b) $\chi^3$ (SFWM) processes in which one or two pump photons $\omega_p$ are converted into correlated signal $\omega_s$ and idler $\omega_i$ photons, respectively. Two main device geometries implemented to generate photon pairs in AlGaAs: (c) linear waveguides and (d) microring resonators.
		(e) Experimental time correlation histogram of photon pairs generated by SPDC in an AlGaAs waveguide in linear and logarithmic (inset) scale for an internal pump power of 2 mW demonstrating a PGR of $3.4\times 10^6$ /s/mW at the chip output \cite{Appas2022}. (f) Experimental transmission spectrum of an AlGaAsOI ring-resonator with a $30\mu m$ radius designed to generate photon pairs via SFWM. The signal (1572 nm) and idler (1542 nm) wavelengths are two free-spectral ranges away from the pump (1557 nm) resonance \cite{Steiner2021}.
	}
	\label{Fig_SPDC_SFWM}
\end{figure*}

\begin{figure*}[!t]
	\centering
	\includegraphics[width=0.65\textwidth]{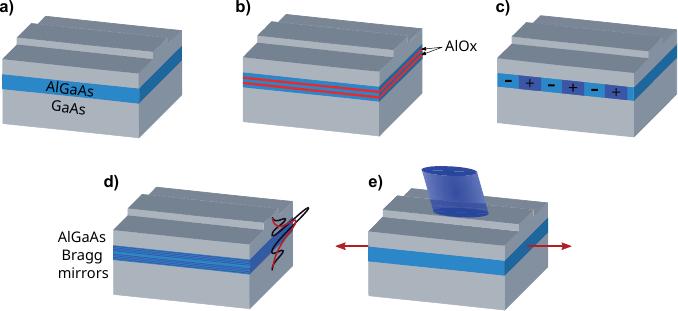}
	\caption{Various types of phase-matching schemes implemented in AlGaAs microstructures: (a) modal phase-matching in simple GaAs/AlGaAs waveguides based on total internal reflection, (b) form birefringence phase-matching, (c) quasi-phase-matching, (d) modal phase-matching in Bragg reflection waveguides, (e) counter-propagating phase-matching.
	}
	\label{Fig_PM}
\end{figure*}

As with classical photonics, it has become crucial to integrate components for quantum photonics at the chip scale in order to advance from laboratory experiments to large-scale implementations and real-world technologies. Unlike the case of electronics, where the basic device is the transistor and the dominant material is silicon, the variety of requirements needed for each intended application prevents the choice of a single material platform and device geometry for the realization of integrated quantum photonic devices; for this reason, a range of solutions has been developed in the last two decades including quantum emitters (such as semiconductor quantum dots, color centers in diamond, and molecules), nonlinear optical platforms (such as lithium niobate, silicon-based ones, and III-V's) and several others \cite{Flamini18,Bogdanov2017}. One of the major challenges for years to come is the implementation of fully on-chip photonic quantum information processing \cite{Pelucchi2022,Wang20}; however, competing functionalities impede the realization of integrated quantum photonic circuits including the generation, manipulation, and detection of quantum states of light from a single material platform (monolithic integration). For this reason, important efforts are currently dedicated to the development of hybrid (the integration of post-processed components onto a specific chip platform) and heterogeneous (the direct deposition of various active materials onto the same wafer, and different from the native wafer composition) manufacturing processes. A review on this topic can be found in \cite{Kim20}.

In this paper, we review the field of nonlinear integrated quantum photonics in AlGaAs. This direct bandgap-semiconductor platform offers a wide range of functionalities, including quantum state generation, low-loss routing, electro-optic modulation, and on-chip single-photon detection through hybrid integration of superconducting detectors \cite{Dietrich16,Orieux2017,Schwartz2018,mcdonald2019iii}. The generation of non-classical states of light in this platform can be achieved either via optical excitation of embedded quantum dots (QDs) \cite{Schimpf2021,Somaschi2016}, or by exploiting its strong nonlinear optical effects \cite{moody2020}. The first approach combines the potential of efficient deterministic generation of pure quantum states with the possibility of exploiting the electron or hole spin as a matter qubit; a review on this topic can be found in \cite{michler2017quantum}. In this article, we focus on the second approach, i.e. nonlinear photonics, that despite the nondeterministic nature of the generation process leading to a compromise between brightness and multi-photon emission suppression, presents the advantages of room temperature operation, an extremely high quality of the produced quantum state combined with device homogeneity on an entire wafer, and the experimental simplicity of operation and integration with other photonic components and devices. An exciting prospect for the realization of complex photonic circuits is to combine the advances described in this review with those achieved with QDs, thus merging the two approaches to integrate a variety of functionalities on the same chip. After having introduced the assets of AlGaAs in the landscape of material platforms for quantum photonics, we discuss the implementation of quantum photonic circuits. We then describe various device geometries implemented up to now to generate non-classical states of light exploiting either second- or third-order nonlinearities, and we highlight the criteria used to quantify the performances of photon pair sources. Finally, we present the possibilities of quantum state engineering offered by this platform, and we conclude by drawing up perspectives on future developments from the point of view of both fundamental and technological research enabled by advances in these last few years.

\section{Assets of AlGaAs}

Silicon-on-insulator (SOI), silicon nitride (SiN), and silica are the central platforms upon which many essential photonic components are built \cite{margalit2021perspective,moody2022materials}. Besides their ubiquitous impact on classical integrated photonics, silicon-based platforms have also driven the field of quantum photonics from few-component prototypes to multi-qubit optical processors \cite{Adcock2020,Wang2018}, at least partially due to the existing silicon foundry infrastructure. Processing Photonic Integrated Circuits (PICs) in a foundry provides an advantage for scalability; however, silicon-based photonics have several fundamental limitations for nonlinear quantum photonics. Silicon in particular suffers from large nonlinear losses due to two-photon absorption (TPA) and induced free carrier absorption (FCA) at 1550 nm. These place limits on the power density inside waveguides, which in turn restrict the on-chip generation rates of quantum light. Additionally, they do not exhibit an electro-optic effect due to a centrosymmetric crystalline structure, which leads to challenges in efficient tunable elements and high-speed manipulation. The majority of tunable elements thus rely on thermal tuning at room temperature, which precludes the integration of superconducting detectors. Considering each of these aspects, an ideal Quantum PIC (QPIC) platform would exhibit high $\chi^{\left(2\right)}$ and $\chi^{\left(3\right)}$ optical nonlinearities, a wide and tunable direct bandgap, high index contrast for tight modal confinement, native integration of solid-state single-photon emitters and gain for on-chip amplification and lasing, as well as wafer-scale fabrication for scalable manufacturing.

Advances in the growth and fabrication of a variety of emerging photonic materials over the last decade have opened many new and exciting opportunities for nonlinear integrated quantum photonics \cite{Bogdanov2017}. AlGaAs ternary alloys stand out among a growing family of highly nonlinear photonic materials that include dielectrics such as lithium niobate or lithium niobate on insulator (LNOI), and other III-V materials like AlN (Fig. \ref{Fig_Comparison} and Table \ref{tab:PICcomparison}). Using high-quality molecular beam epitaxy (MBE) with atomically smooth interfaces, AlGaAs/GaAs heterostructures can be grown with tunable composition and thickness (Fig. \ref{Fig_Intro}a), enabling complex structures such as distributed Bragg reflectors, \textit{p-i-n} diodes, and lasers. The lattice constants of GaAs and AlAs are within 1 pm, and lattice matching is maintained for all AlGaAs compositions. 
This enables the defect-free growth of complex, multi-layered heterostructures \cite{Mobini2022} thanks to advanced molecular beam epitaxy (MBE) techniques, allowing for precisions on compound concentrations and layer heights well below 1\%. The patterning of these high-quality AlGaAs-based epitaxial structures into waveguides, resonators, and complex PICs can be performed with high precision, high directionality and low roughness using e.g. inductively coupled plasma (ICP) etching with HSQ resist masks \cite{Liao17b}.
Additionally, advances in chip- and wafer-scale bonding have enabled high-quality interfaces between AlGaAs and SiO$_2$, which is not native to AlGaAs growth. Using a combination of surface plasma activation, thin passivation layers such as alumina or aluminum nitride, and thermal annealing under high-pressure to enhance direct bond strengths, AlGaAs can be directly contacted with SiO$_2$ for high-quality AlGaAsOI designs. Using atomic-layer deposition or plasma enhanced chemical vapor deposition, SiO$_2$ thin films can also be directly deposited on AlGaAs, which can serve as a hard mask for optical lithography and the top cladding layer to fully encompass the AlGaAs waveguide. Further details of the AlGaAsOI fabrication process can be found in Refs. \cite{Semenova2016,Chang2020,Steiner2021}.

AlGaAs exhibits a wide transparency window that can be tuned by varying the aluminum concentration \textit{x}. A direct bandgap is maintained for $x < 0.45$, and the bandgap varies as $E_g\left(x\right) = 1.422 + 1.2475x$ eV. Waveguides with $x < 0.45$ thus have a wide transparency window spanning 0.62-17 $\mu$m, and this leads to low intrinsic optical losses due to TPA- and FCA-free operation throughout the telecommunications bands. Alloy engineering also provides a control knob for tuning the refractive index from ~2.9 (AlAs) to ~3.4 (GaAs), which allows for modal and dispersion engineering between normal and anomalous regimes using a variety of photonic waveguide geometries. The combination of negligible TPA and FCA, defect-free growth, and improved waveguide fabrication in recent years has resulted in high-quality waveguides with as low as 0.2 dB/cm propagation loss, which now rivals SOI and SiN PICs that are designed for integrated quantum photonics \cite{ChangCSOI}. While several material platforms, such as LNOI \cite{gao2021broadband,gao2022lithium} and SiN \cite{liu2022ultralow}, have achieved lower propagation loss better than ~$\sim$0.3 dB/m (corresponding to resonator quality factors $>10^8$ shown in Fig. \ref{Fig_Comparison}), these waveguides are designed with either high aspect ratio, where the optical mode is mostly in the surrounding cladding, or large cross-sectional area to avoid overlap of the mode with the core-cladding interfaces. Such waveguide geometries are typically not used for integrated quantum photonics, since the footprint of PIC components and circuitry can be significantly larger and the weak optical confinement reduces the nonlinear conversion efficiency for on-chip quantum light generation.

AlGaAs also offers various tuning mechanisms. The thermo-optic tuning efficiency is similar to silicon ($\simeq 20$ mW/$\pi$ phase tuning); however, this mechanism is not compatible with operation at cryogenic temperatures (which would be required in a chip including QDs and/or superconducting detectors), nor with fast quantum state modulation. From this point of view, AlGaAs, unlike silicon, presents two attractive properties: piezoelectricity, which has been exploited, for example, to run single-photon qubits created from the light emitted by InAs QDs at 4 K \cite{buhler2022chip}, and a high electro-optical effect allowing a fast modulation (GHz) of its refractive index \cite{Spickermann96}.
The ability to grow \textit{p}-type and \textit{n}-type doped heterostructures in direct-gap AlGaAs enables the direct integration of a variety of classical and quantum light sources for on- and off-chip emission. Surface-emitting and edge-emitting laser diodes based on III-V heterostructures find applications in precision timekeeping, laser cooling, metrology, LIDAR, and optical communications. Many configurations and geometries exist with single-mode operation, single polarization, low threshold current and power consumption, and narrow linewidths with excellent side-mode suppression \cite{norman2018perspective}. Using etched gratings for distributed feedback, on-chip emission into waveguides could enable on-chip pumping of quantum light sources, such as InAs QDs or nonlinear resonators for spontaneous parametric down-conversion (SPDC) and spontaneous four-wave-mixing (SFWM). AlGaAs-based materials exhibit some of the highest $\chi^{\left(2\right)}$ (180 pm/V) and $\chi^{\left(3\right)}$ (2.6 $\times 10^{-13}$ cm$^2$/V) nonlinearities of any photonic material platform, which have been exploited to realize efficient chip-integrated sources of quantum states of light, as detailed in section \ref{sect 4}. By embedding In(Ga)P QDs into the gain region, telecommunications wavelength laser diodes with $> $ 100 nm tunability and sub-10 kHz linewidth can be fabricated and directly integrated with III-V and silicon photonics \cite{liang2021recent}. At low QD density, individual QDs are isolated and can be optically and electrically pumped to deterministically generate single or entangled-photon pairs--currently considered the state-of-the-art for on-demand single-photon generation \cite{Senellart17}.

\section{Quantum photonic circuits}

Besides the generation of quantum states of light, which will be treated in detail in the following sections, improvements in waveguide fabrication resulting in as low as 0.2 dB/cm propagation loss have shifted the focus towards scalable QPIC technologies \cite{Chang2020}. The efficient coupling of light between photonic chips and optical fibers requires a matching of both the effective refractive index and the spatial mode. This can be achieved through several means, including grating couplers \cite{Taillaert02}, inverse tapers \cite{Almeida03}, spot size converters \cite{Liao17}, or evanescent coupling to tapered fibers \cite{Groblacher13}.
Besides in- and out-couplers and light sources, a functional on-chip platform for both QPICs and classical PICs generally relies on a set of components that include waveguide crossers, 3-dB splitters and polarization splitters, tunable interferometers, resonant and non-resonant filters, modulators, and detectors as illustrated in Fig. \ref{Fig_QPICs_illustration}. In \cite{Wang14}, GaAs-based Mach-Zehnder interferometers (MZIs) formed by two directional couplers and two electro-optical phase shifters (Fig. \ref{Fig_Circuits}c) have been demonstrated for the on-chip manipulation of single-photon and two-photon states produced externally. Subsequently, higher degrees of compactness have been achieved with the monolithic integration of lasing action and photon pair production under electrical injection at room temperature (see Fig. \ref{Fig_Circuits}a) and the integration of photon-pair sources with elementary optical components such as spatial beamsplitters \cite{Belhassen18} (see Fig. \ref{Fig_Circuits}b) and polarization beamsplitters  \cite{Appas22}, allowing in particular to implement an on-chip Hanbury Brown and Twiss experiment \cite{Belhassen18}.

As illustrated in Fig. \ref{Fig_QPICs}, many chip-integrated optical components have also been developed in an AlGaAsOI platform optimized for efficient entangled-photon pair generation \cite{CastroAlGaAsOI}, which are summarized in Table \ref{tab:PICcomparison} and compared to the state-of-the-art in other photonic platforms.
 Chip-to-fiber edge coupling with inverse tapers from 600 nm to 200 nm wide waveguides (mode matched to lensed fiber arrays) results in $<$ 3 dB coupling loss. From simulations, this can be $< $ 1 dB, which is possible by using smaller tapers and an anti-reflection coating on the facet. Low-loss waveguide crossings with $< $ 0.2 dB/cross loss and better than -40 dB of cross-port extinction was reported with a simple linear adiabatic taper design. These allow for planar single-mode PIC structures for non-nearest neighbor couplings between waveguides, enabling complex circuit design where, for example, interactions between non-next nearest-neighbor optical qubits are necessary. 

\begin{table*}[]
\footnotesize
    \caption{Table comparing passive components in the AlGaAsOI platform with SOI, SiN, LNOI, and AlN. The values reported are for PICs designed specifically for integrated quantum photonics applications.}
    \label{tab:PICcomparison}
    \centering

    \begin{tabular}{|c|c|c|c|c|c|}
           \hline
           & AlGaAsOI  & SOI & SiN & LNOI & AlN\\
         \hline
         \rule{0pt}{4ex}  Inverse Taper coupling Loss  & \textcolor{black}{$2.9$ dB} \cite{CastroAlGaAsOI} & $<3$ dB \cite{Mu2020} & $2-3$ dB \cite{Ramelow2015} & $<2$ dB \cite{chen2022low} & $<2$dB \cite{zhao2020high}\\ \hline
         \rule{0pt}{4ex}  Waveguide Crossing loss & $0.23$ dB \cite{CastroAlGaAsOI} & $0.2$ dB \cite{Sanchis2009} & $0.3$ dB \cite{Yang2019} & 0.18 dB \cite{zhang2022power} & NA\\ \hline
        
         \rule{0pt}{4ex}  MZI Extinction ratio& $> 30$ dB \cite{CastroAlGaAsOI} & $> 30$ dB \cite{Wang2016} & $> 40$ dB \cite{Rao2021} & $>40$dB \cite{he2019high} & 25 dB \cite{zhu2016aluminum}\\ \hline
         \rule{0pt}{4ex}  MZI Optical bandwidth ($>10$ dB ER) & $200$ nm \cite{CastroAlGaAsOI} & $> 40$ nm \cite{Lee2019} &$180$ nm \cite{Rao2021} & $>40$ nm \cite{he2019high} & $>20$ nm \cite{zhu2016aluminum} \\ \hline
         \rule{0pt}{4ex} MZI Heater efficiency & 20 mW/$\pi$ \cite{CastroAlGaAsOI} & $12$ mW/$\pi$ \cite{Lee2017} & $200$ mW/$\pi$  \cite{Lee2022} & 7.5 mW/$\pi$ \cite{liu2020highly} & 17 mW/$\pi$ \cite{shin2021demonstration} \\ 
          & (10.2 nm FSR) & (5.8 nm FSR) & (NA) & (0.8 nm FSR) & ($\sim0.1$ nm FSR) \\ \hline
    \end{tabular}

\end{table*}

Tunable interferometers serve many purposes including power distribution across PICs, optical filtering, demultiplexing, single-qubit operations, and linear optical classical and quantum programming \cite{carolan2015universal,o2007optical}. Thermally tunable MZIs were demonstrated recently in AlGaAsOI based on both 3-dB directional couplers (DCs) and 3-dB multi-mode interferometer (MMI) splitters. The thermal tuning power required for a $\pi$ phase shift, and the corresponding MZI free-spectral range (FSR), are shown for each platform in Table 1. Near a centered wavelength of 1550 nm, an extinction ratio $> $ 30 dB and a bandwidth of 200 nm was reported for DC-based MZIs, which is similar to the performance of SOI and Si$_3$N$_4$ \cite{CastroAlGaAsOI}. By designing the MZI with a path length imbalance, the free-spectral range (FSR) can be tuned to match the mode spacing from resonator-based classical and quantum light sources. This enables frequency qubit demultiplexing, which was demonstrated with $>$ 23 dB extinction using a monolithically integrated AlGaAsOI microring resonator and qubit demultiplexer \cite{CastroAlGaAsOI}. This experiment points to the possibility of creating scalable QPIC circuitry in a monolithic AlGaAs platform. Combined with recent progress in 3" wafer-scale nanofabrication \cite{ChangCSOI,Stanton2020}, a route exists for AlGaAs QPIC manufacturing.

The availability of compact chips for the generation and manipulation of quantum states of light that are robust to manufacturing and that operate at room temperature is of interest in itself, however certain applications require the on-chip integration of single photon detectors. The two leading technologies for quantum photonics is silicon or InGaAs single-photon avalanche photodiodes (SPADs) and superconducting nanowire single-photon detectors (SNSPDs). SPADs are advantageous in that they operate at room temperature and are inexpensive; however, SNSPDs have superior performance in nearly all metrics, including detection efficiency, timing jitter, detector reset time, and dark count rates \cite{bienfang2022materials}. SNSPDs rely on absorption of a photon in a thin film of superconducting material, which traditionally is embedded in a normal-incidence dielectric cavity and fiber-coupled package. SNSPDs have also been integrated with III-V GaAs/AlGaAs Bragg-reflection waveguides  \cite{mcdonald2019iii}. They comprise a WSi or NbN thin film deposited on top of the waveguide. Nanowires are patterned with electron-beam lithography, followed by dry etching and metal evaporation for the wire-bond pads. A photon propagating in the waveguide can be evanescently absorbed by the SNSPD with near-unity quantum efficiency with long nanowires, sub-millihertz dark count rates, and better than -40 dB nearest-neighbor crosstalk between detectors \cite{mcdonald2019iii}. This opens the possibility to realize AlGaAs-based photonic circuits integrating all three steps of generation, manipulation and detection of quantum states of light \cite{Schwartz2018} (Fig.~\ref{Fig_Intro}d).

A challenge with any photonic-integrated SNSPD  technology is cryogenic operation, which prevents their integration with other tunable elements that rely on thermo-optic tuning. One approach around this issue is to rely on modular architectures in which room-temperature, tunable components perform complex, programmable operations, which are connected to SNSPD arrays through chip-to-fiber interconnects; however, even the best couplers typically exhibit 1 dB of loss, compared to the nearly lossless detection with on-chip SNSPDs \cite{khan2020low}. Thus, much research and development is being devoted to cryo-compatible photonic elements that rely on alternative tuning mechanisms, including electro-optic \cite{shin2013ultralow}, magneto-optic \cite{pintus2022integrated} and piezo-electric \cite{khurana2022piezo}. These are active areas of research for many material platforms, including AlGaAsOI and AlGaAs Bragg-reflection waveguides. In addition to their cryo-compatibility, these approaches provide new opportunities for high-speed communications and computing. For example, Mach-Zehnder modulators based on travel-wave electrode designs with InAlGaAs waveguides can operate with 25 GHz 3-dB bandwidth and $<0.4$ V.cm $V_{\pi}$ half-wave voltage-length product \cite{dogru2013electrodes}. Likewise, electro-optic tuning in AlGaAsOI is expected to reach 30 GHz bandwidth with better than $< 5$ V.cm due to the large electro-optic coefficient, large index contrast, and small electrode spacing enabled by tight modal confinement, which would be competitive with the best photonic modulators based on lithium niobate \cite{wang2018integrated}.

\section{Generation of quantum states of light in nonlinear AlGaAs chips}\label{sect 4}

The large second and third-order susceptibilities exhibited by  the AlGaAs platform provide an efficient way for the generation of quantum states of light through SPDC and SFWM, respectively. In the first process, single photons of a pump beam of frequency $\omega_p$ get converted into pairs of photons called signal and idler of frequencies $\omega_s$ and $\omega_i$ (Fig. \ref{Fig_SPDC_SFWM}a); in the second process, each photon pair is generated through the annihilation of two pump photons (Fig. \ref{Fig_SPDC_SFWM}b).

The two processes must conserve energy, as well as wavevector, so that the involved fields (pump, signal and idler) remain in phase during their interaction (phase-matching condition) (Fig.~\ref{Fig_SPDC_SFWM}). If we neglect optical losses, and assuming perfect phase matching, the pair generation rate (PGR) can be shown to have the following expressions \cite{Wang2021}:
\begin{equation}\label{Eq_PGR_SPDC}
PGR_{\rm SPDC}\propto \frac{\chi^{\left(2\right)}_{\rm eff} P_p L^2} {\epsilon_0^2 c^2 A_{\rm eff}}
\end{equation}
\begin{equation}\label{Eq_PGR_SFWM}
PGR_{\rm SFWM}\propto \left|\frac{\omega_p\chi^{\left(3\right)}} {n_0^2\epsilon_0 c^2 A_{\rm eff}} P_p L \right|^2
\end{equation}
where $P_p$ is the pump power, $\chi^{\left(2\right)}_{\rm eff}$ is the effective value of the second order nonlinearity tensor, $A_{\rm eff}$ is the mode interaction overlap area, $L$ is the device length, $c$ the speed of light, $n_0$ the refractive index at the pump wavelength. In both cases, the conversion efficiency depends quadratically on the device length and the involved nonlinearity; while SPDC displays a linear (inverse linear) dependence on $P_p$ ($A_{\rm eff}$), in SFWM those dependencies are quadratic.

The desired material properties to realize efficient parametric sources of quantum light are thus a high $\chi^2$ and/or $\chi^3$ nonlinear coefficient, high index contrast to obtain well-confined optical modes (and thus low $A_{\rm eff}$) and low linear and non-linear absorption losses; these figures of merit for AlGaAs are given in Fig.~\ref{Fig_Comparison} and Table \ref{tab:Sources}.
Two main device geometries have been implemented up to now to generate photon pairs in AlGaAs: linear (Fig.~\ref{Fig_SPDC_SFWM}c) or spiral waveguides (displaying typical lengths of a few millimeters), and microresonators (at the micrometer scale) (Fig.~\ref{Fig_SPDC_SFWM}d). 
Satisfying the phase-matching condition for parametric generation typically requires that the refractive index at the pump frequency matches the refractive index at the signal and idler frequency. In this respect, a key difference between second- and third-order parametric processes is that while SPDC tends to be more efficient thanks to the larger $\chi^{\left(2\right)}$ nonlinearity, SFWM allows for an easier achievement of the phase matching condition since in this case pump, signal, and idler frequencies are degenerate or quasi-degenerate, which relaxes the constraints for dispersion engineering requirements. The spectral difference between the pump photons and the pairs generated by SPDC, while on the one hand eases the spectral removal of the pump beam, on the other hand requires more sophisticated strategies to achieve phase matching.

In certain materials (such as barium borate), material birefringence can be exploited to achieve phase-matching for SPDC, by ensuring that the difference in refractive index between two orthogonal polarizations compensates for the chromatic dispersion \cite{Midwinter65}. However, this strategy cannot be employed in GaAs and AlGaAs, which are isotropic (zinc-blende structure) and thus not natively birefringent. The implementation of SPDC in few-layer GaAs/AlGaAs microstructures based on total internal reflection (Fig.~\ref{Fig_PM}a) is thus challenging \cite{Ducci04}. Alternative phase-matching techniques have been developed, relying on multilayer AlGaAs microstructures. In a waveguide, an artificial birefringence can be created by inserting thin layers of aluminium oxide, leading to a sub-wavelength modulation of the refractive index along the vertical direction, which breaks the original symmetry of the medium (form birefringence, Fig.~\ref{Fig_PM}b)~\cite{Fiore98}. Another approach consists in periodically inverting the sign of the nonlinearity so as to compensate for the dephasing of the three fields during their propagation (quasi phase-matching, Fig.~\ref{Fig_PM}c)~\cite{Armstrong62}, a technique commonly used in dielectric materials (periodically poled lithium niobate or KTP)~\cite{Tanzilli01}. In a waveguide, this requires to periodically modify the crystalline orientation of the material, which has been realized in AlGaAs but at the price of significant optical losses \cite{Yoo95,Skauli02}, while in a ring or disk resonator, this condition can be inherently achieved by the rotation of the field polarization along the circumference \cite{Dumeige06,Mariani14}. These various approaches allowed for second-harmonic generation and optical parametric oscillation \cite{Fu20} but not yet for the generation of biphoton states by SPDC.

Other solutions have been demonstrated over the last three decades to generate photon pairs via SPDC in AlGaAs (for a detailed review, see \cite{Orieux2017,Helmy2011}); among these, the most advanced results in quantum state generation and manipulation have been obtained in ridge Bragg-reflection waveguides (see Figs.~\ref{Fig_PM}d and \ref{Fig_Intro}a) based on a collinear modal phase-matching scheme in which the phase velocity mismatch is compensated by multimode waveguide dispersion \cite{Abolghasem10,Horn12,Zhukovsky2012,Boitier14}, and a counter-propagating phase-matching scheme based on a transverse pump configuration with the photon of each pair exiting from the opposite facets of the waveguide (Fig.~\ref{Fig_PM}e) \cite{Lanco06,Orieux13,Francesconi20}. While the implementation of both phase-matching geometries has led to the demonstration of bright photon pair sources displaying high two-photon indistinguishability \cite{Gunthner15,Francesconi20}, energy-time \cite{Autebert16, Chen18} and polarization entanglement \cite{Valles13,Horn13,Gunthner15,Kang16,Appas21}, Bragg-reflection waveguides have already enabled the monolithic integration with a diode laser leading to an electrically injected photon pair source working at room temperature (Fig. \ref{Fig_Circuits}a) \cite{Boitier14}, and the counter-propagating phase-matching scheme has demonstrated to be particularly versatile in the engineering of the two-photon wavefunction \cite{Francesconi20,Francesconi21,Francesconi22} (see section 6). Another  approach having led to the generation of entangled photon pairs relies on the implementation of a quasi-phase-matching technique based upon quantum-well intermixing in an AlGaAs superlattice waveguide \cite{Sarrafi13,Sarrafi14}.

Concerning the generation of quantum states of light via SFWM, more simple few-layer microstructures based on total internal reflection can be used, thanks to the more relaxed constraints of dispersion engineering.
SFWM has been successfully achieved with such simple waveguides in both native AlGaAs \cite{Kultavewuti2016} and AlGaAsOI \cite{Mahmudlu2021}, while AlGaAsOI microring resonators (Fig. \ref{Fig_Intro}b) with $Q >$ 1 million has led to an entangled-pair brightness $> $ 1,000 higher than SOI and nearly 500 higher than Si$_3$N$_4$ \cite{Steiner2021}. In addition to quantum state generation, we also remark that in both GaAs/AlGaAs BRWs and AlGaAsOI, the dispersion can be engineered to reach the anomalous regime across a broad bandwidth centered at telecommunications wavelengths. This has been leveraged for milliwatt \cite{pu2016efficient} and sub-30 $\mu$W threshold frequency comb generation in microrings \cite{Chang2020} and octave-spanning supercontinuum generation \cite{kuyken2020octave,zhang2022polarization,chiles2019multifunctional}, both utilizing the large $\chi^{\left(3\right)}$ nonlinearity of AlGaAs.

\bigskip

\section{Performance of photon pair sources}

To quantify the performance of AlGaAs photon-pair sources and situate them within the state-of-the-art, several parameters can be employed. As their definition can slightly vary according to the community, we start by defining hereafter the most used ones \cite{Wang2021}:

\begin{itemize}[itemsep=1pt,leftmargin=*]
    \item The coincidence ($C$) and accidental ($A$) count rates are the measured rates of temporal correlations and accidental temporal correlations (the latter due to "unwanted" correlation events related to dark counts of the detectors and various source of noise in the generation process). It can be useful to distinguish between raw and net coincidences rates values which are related by $C_{\rm net}=C_{\rm raw}-A$.
    \item The coincidence-to-accidental ratio (CAR) corresponds to the ratio between the net coincidence count rate and the accidental count rate: ${\rm CAR} = C_{\rm net}/A$.
    \item The pair generation rate (PGR) is the rate of pair generation events; in some cases, the authors refer to an on-chip value, in others to the pair events arriving at the first collection lens. It is usually inferred from the number of detected coincidences divided by the squared values of the signal collection and single photon detectors efficiencies.
    \item The brightness is a normalized PGR per unit pump power and unit spectral bandwidth.
\end{itemize}

An overview of these various figures of merit for the main types of SPDC and SFWM sources (see Section 3) implemented in AlGaAs chips is given in Table \ref{tab:Sources}. 
In addition, the characteristics of the produced quantum states are usually quantitatively evaluated via the following parameters:

\begin{itemize}[itemsep=1pt,leftmargin=*]
    \item The joint spectral amplitude (JSA) is a complex-valued function giving the probability amplitude to measure one photon of the pair at a given frequency and its twin at given other frequency (see section 6 for details); this function can be expanded according to Schmidt decomposition into a series based on frequency eigenmodes with a characteristic Schmidt number quantifying the dimension of the associated Hilbert space \cite{Parker2000}. A comparison among the different experimental techniques developed to reconstruct the joint spectrum can be found in Ref.~\cite{Zielnicki2018}.

     \item Two-photon interference effects, measured via a Hong-Ou-Mandel (HOM) interferometer \cite{Hong87} are widely used for the characterization of photons indistinguishability: when two indistinguishable independent photons enter a 50/50 beam splitter via two different input ports, they both exit from the same output port. In this case, the coincidence rate at the output of the interferometer as a function of the delay between the two photons imposed with a delay line displays a HOM dip with a visibility $V_{HOM}$ quantifying the indistinguishability and purity of the generated photons. In the case in which the joint spectrum of the two-photon state presents quantum correlations, they are imprinted in the two-photon interferogram which can be usefully employed to reveal and quantify such correlations (see section 6) \cite{Fedrizzi09, Douce13}.
     
     \item Entangled two-photon states are characterized either via a coincidence setup including a polarization stage analysis leading to the reconstruction of the density matrix \cite{James2001} (in the case of polarization entanglement) or via a Franson interferometer \cite{Franson1989} (in the case of time-bin or energy-time entanglement). Nonlocality can be quantified by measuring the violation of the Clauser-Horne-Shimony-Holt  (CHSH) inequality \cite{CHSH1969} as a generalization of the Bell’s test. 
\end{itemize}

That last two columns of Table \ref{tab:Sources} summarize the HOM indistinguishability values and various kinds of entanglement demonstrated in AlGaAs photon-pair sources in the recent years.

\begin{table*} \label{Table_sources}
\caption{Comparison of the main figures of merit for various SPDC and SFWM phase-matching schemes in AlGaAs photon-pair sources (given power values correspond to internal pump power, and losses are given at telecom wavelengths).}
\label{tab:Sources}
\footnotesize
\centering
\begin{tabular}{|c|c|c|c|c|c|c|c|}
	\hline
& Losses &  \begin{tabular}{@{}c@{}} PGR \\ (on-chip) \end{tabular} & CAR & Brightness & \begin{tabular}{@{}c@{}}Total \\ bandwidth \end{tabular} & \begin{tabular}{@{}c@{}} HOM \\ visibility \end{tabular}  & \begin{tabular}{@{}c@{}} Demonstrated \\ entanglement \end{tabular}\\
	\hline
\begin{tabular}{@{}c@{}} SPDC\\ Modal PM \\ \cite{Abolghasem10,Horn12,Boitier14} \end{tabular}	& \begin{tabular}{@{}c@{}}  $0.4$ dB/cm \\ \cite{Boitier14} \end{tabular}
& \begin{tabular}{@{}c@{}} $7 \times 10^6$ pairs.s$^{-1}$ \\ @$P=0.6$ mW \cite{Appas21} \end{tabular} & 80 \cite{Appas21} & $2\times 10^5$ s$^{-1}$mW$^{-1}$nm$^{-1}$ \cite{Appas21}  & 60 nm \cite{Appas21} & 95 \% \cite{Gunthner15} & \begin{tabular}{@{}c@{}} Energy-time \cite{Autebert16,Chen18}\\ Polarization \cite{Valles13,Horn13,Gunthner15,Kang16,Appas21}\end{tabular} \\
 	\hline
\begin{tabular}{@{}c@{}} SPDC\\ Quasi PM \\ \cite{Sarrafi13,Sarrafi14} \end{tabular} 	& \begin{tabular}{@{}c@{}} $3$ dB/cm \\ \cite{Hutchings10} \end{tabular}
& \begin{tabular}{@{}c@{}} $1.9\times 10^6$ pairs.s$^{-1}$ \\ @$P=8$ mW \cite{Sarrafi13} \end{tabular} & 115 \cite{Sarrafi13}& $2.6\times 10^5$ s$^{-1}$mW$^{-1}$nm$^{-1}$ \cite{Sarrafi13}  & 50 nm \cite{Sarrafi13}& --- & \begin{tabular}{@{}c@{}} Energy-time \cite{Sarrafi14} \\ Polarization \cite{Sarrafi13}  \end{tabular} \\
	\hline
\begin{tabular}{@{}c@{}} SPDC\\ Contra PM \\ \cite{Lanco06,Orieux13,Francesconi20} \end{tabular}	& \begin{tabular}{@{}c@{}} $0.4$ dB/cm \\ \cite{Caillet10} \end{tabular} 
& \begin{tabular}{@{}c@{}} $5\times 10^6$ pairs.s$^{-1}$ \\ @$P=10$ mW \cite{Francesconi20} \end{tabular} 
& $\sim$ 100 \cite{Francesconi20} & $4\times 10^5$ s$^{-1}$mW$^{-1}$nm$^{-1}$ \cite{Francesconi20}  & 0.5 nm \cite{Francesconi20} & 88 \% \cite{Francesconi20} & \begin{tabular}{@{}c@{}} Polarization \cite{Orieux13} \\ Frequency \cite{Francesconi20,Francesconi21}  \end{tabular} \\
	\hline
\begin{tabular}{@{}c@{}} SFWM\\ Nanowires \\ \cite{Kultavewuti2016,Mahmudlu2021} \end{tabular}	& \begin{tabular}{@{}c@{}} $2$ dB/cm \\ \cite{Mahmudlu2021}  \end{tabular}
& \begin{tabular}{@{}c@{}} $2.3\times 10^6$ pairs.s$^{-1}$ \\ @$P=32$ $\mu$W \cite{Kultavewuti2016} \end{tabular}
 & 177 \cite{Kultavewuti2016} & $3\times 10^7$ s$^{-1}$mW$^{-2}$nm$^{-1}$ \cite{Kultavewuti2016} &  80  nm \cite{Kultavewuti2016} & --- &--- \\
	\hline
\begin{tabular}{@{}c@{}} SFWM\\ Microrings \\ \cite{Steiner2021} \end{tabular} & \begin{tabular}{@{}c@{}} 0.3 dB/cm \\ \cite{Steiner2021} \end{tabular}
& \begin{tabular}{@{}c@{}} $12\times 10^6$ pairs.s$^{-1}$ \\ @$P=30$ $\mu$W \cite{Steiner2021} \end{tabular}
& 350 \cite{Steiner2021} & $2.5\times 10^{13}$ s$^{-1}$mW$^{-2}$nm$^{-1}$ \cite{Steiner2021} & $> $50 nm \cite{Steiner2021} & --- & Energy-time \cite{Steiner2021} \\
	\hline
\end{tabular}
\end{table*}

\section{Quantum state engineering}

Quantum information applications require, depending on their specificities, various types of quantum states, motivating the development of quantum state engineering techniques. Several of them have been implemented recently on the AlGaAs platform, exploiting various degrees of freedom of light.

\subsection{Polarization entanglement engineering}

Polarization is a paradigmatic two-dimensional photonic degree of freedom that has enabled pioneering experiments in quantum information, ranging from fundamental tests of the quantum theory \cite{Aspect81} to quantum computing \cite{Shor97} and communication applications \cite{Ekert91,Bouwmeester97}. Polarization Bell states, such as $\ket{\Phi}=\left (\ket{HH}+e^{i \phi}\ket{VV} \right) / \sqrt{2} $, where $H$ and $V$ stand for the horizontal and vertical polarizations of single photons, constitute a critical resource for many of these demonstrations.
While such states can be efficiently generated by optical nonlinear processes in bulk crystals combined with external components (such as walk-off compensators or Sagnac interferometers) \cite{Kwiat95,Shi04,Kim06}, there is a growing need for the development of compact, chip-integrated sources capable of directly emitting polarization-entangled states without resorting to external elements; in addition, a broad emission bandwidth is highly desirable in view of distributing these states to many users in quantum networks through wavelength demultiplexing \cite{Wengerowsky18}.

The AlGaAs platform provides several assets for developing such sources: its nonlinear tensor allows for a broad versatility of phase-matching processes, leading to various possible polarizations for the emitted photons, while its absence of bulk birefringence allows for a spectrally broadband emission and circumvents the usual need of compensating for the group delay between orthogonally polarized photons.

SPDC in AlGaAs waveguides has been demonstrated within three possible types of phase-matching processes \cite{Abolghasem10}, allowing the generation of both photons in TM polarization (type 0), both in TE polarization (type 1), or orthogonally polarized signal and idler photons (type 2). This versatility has been exploited to generate polarization Bell states of the form $\ket{\Phi}=\ket{HH}+e^{i \varphi}\ket{VV} $  (using simultaneously the type 0 and 1 processes) as well as $\ket{\Psi}=\ket{HV}+e^{i \varphi}\ket{VH} $ (using two concurrent type 2 processes) with the same chip-integrated source \cite{Kang15}.

In recent years, the direct generation of $\ket{\Psi}$ Bell states with AlGaAs chips \cite{Orieux13,Horn13,Valles13,Kang16,Schlager17,Appas21} has been demonstrated with increasing fidelity and bandwidth. In \cite{Kang16}, a total emission bandwidth of 95 nm was demonstrated, with a fidelity up to $99 \%$ for a 40 nm spectral separation between signal and idler photons without any post-manipulation; in \cite{Appas21}, a fidelity above $95\%$ has been demonstrated over a 50 nm bandwidth and with a high PGR of $1.2 \times 10^7$ s$^{-1}$mW$^{-1}$. These are valuable assets compared to alternative telecom-band chip-based sources of polarization-entangled photon pairs, either based on type 2 SPDC in PPLN waveguides (which leads to comparable fidelity and PGR, but two-order of magnitude narrower bandwidth, and requires off-chip walk-off compensation) \cite{Martin10} or SFWM in silicon waveguides (which displays a bandwidth comparable to AlGaAs sources, but a two order-of-magnitude smaller PGR) \cite{Matsuda12}. These performances are appealing in view of implementing communication tasks in quantum networks \cite{Appas21} with miniaturized and cost-efficient resources.

\subsection{Frequency entanglement engineering}

\begin{figure*}[!t]
\centering
\includegraphics[width=0.7\textwidth]{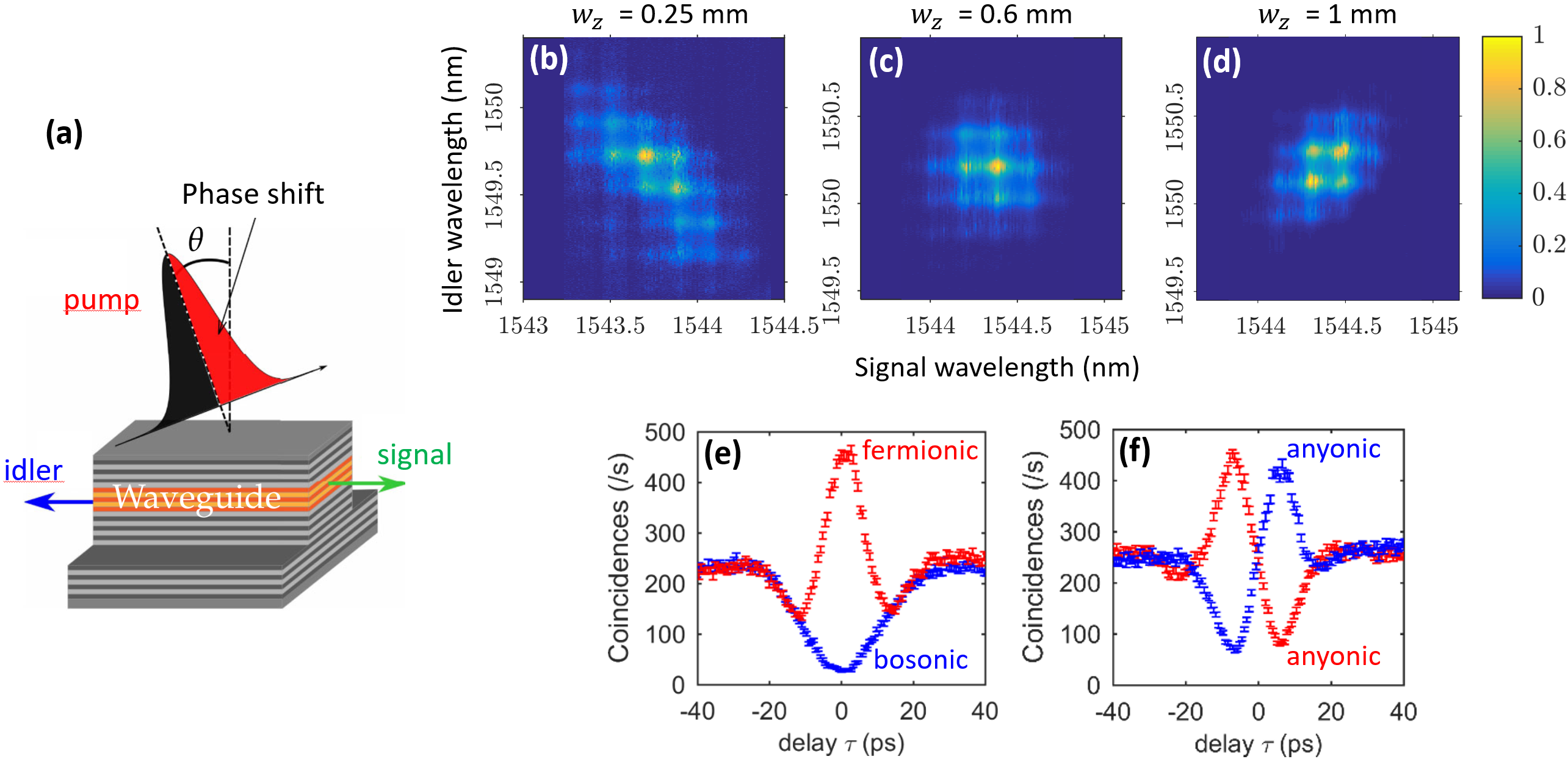}
\caption{
(a) Sketch of an AlGaAs source emitting counter-propagating photon pairs under transverse pumping, with controlled intensity and phase spatial profile. (b-d) Measured biphoton joint spectral intensity (JSI) for increasing values of the pump beam waist: (b) $0.25$ mm, (c) $0.6$ mm, (d) $1$ mm \cite{Francesconi20}.
(e-f) Hong-Ou-Mandel interferograms evidencing (e) bosonic (blue) and fermionic (red) behavior, and (f) anyonic behavior (blue and red traces correspond to two implementations of $\alpha=1/2$ anyons) \cite{Francesconi21}.
}
\label{Fig_frequency}
\end{figure*}

Besides two-dimensional degrees of freedom (DoF) such as polarization, great efforts have been focused recently onto the high-dimensional DoF of photons, such as orbital angular momentum, path, or frequency modes as a means to strengthen the violation of Bell inequalities \cite{Dada11}, increase the density and security of quantum communication \cite{Barreiro08} or enhance flexibility in quantum computing \cite{Lanyon09}. Among the various investigated DoF, frequency is particularly attractive due to its robustness to propagation in optical fibers and its capability to convey large-scale quantum information into a single spatial mode. Nonlinear parametric processes offer a high versatility for the generation of frequency-entangled
photon pairs \cite{Ansari18b,Kues19}, that are are described by the joint spectral amplitude (JSA, see section 5). 
Several techniques have been demonstrated recently to manipulate the JSA of photon pairs by post-manipulation, using e.g. pulse shapers and electro-optic modulators \cite{Kues19}, or directly at the generation stage by engineering the spectral \cite{Ansari18,Ansari18c} properties of the pump beam or by tailoring the material nonlinearity in domain-engineered crystals \cite{Graffitti18,Graffitti20} on the PPLN and PPKTP platforms.
Another approach, recently developed in AlGaAs waveguides, consists in tuning the spatial properties of the pump beam within a counter-propagating phase-matching scheme \cite{Francesconi20}, as sketched in Fig. \ref{Fig_frequency}a.

In this situation, the JSA of the generated biphoton state can be expressed as $\phi (\omega_s,\omega_i)=\phi_{\rm spectral}(\omega_s+\omega_i) \, \phi_{\rm PM}(\omega_s-\omega_i)$. The function $\phi_{\rm spectral}$, reflecting the condition of energy conservation, corresponds to the spectrum of the pump beam while $\phi_{\rm PM}$, reflecting the phase-matching condition, is governed by the spatial properties of the pump beam:
\begin{equation}\label{Eq_PM}
\phi_{\rm PM}(\omega_s-\omega_i)=\int_{-L/2}^{L/2} dz \,\mathcal{A}_p (z) e^{-i (k_{\rm deg}+(\omega_s-\omega_i)/v_{\rm g})z}
\end{equation}
where $\mathcal{A}_p (z)$ is the pump amplitude profile along the waveguide direction, $L$ is the waveguide length, $v_g$ is the harmonic mean of the group velocities of the twin photon modes and $k_{\rm deg}$ is governed by the modal birefringence of the device.
The JSA can thus be controlled by tailoring either the spectral or the spatial properties of the pump beam. 
Figures \ref{Fig_frequency}b-c-d shows the joint spectral intensity (JSI), i.e. the modulus squared of the JSA, measured at fixed pulse duration but increasing values of the pump waist. The generated quantum state goes successively from frequency-anticorrelated, to separable, to frequency correlated \cite{Francesconi20}.
These results demonstrate a flexible and reconfigurable control of frequency correlations, which could be exploited to adapt the source to different quantum information tasks such as clock synchronization \cite{Giovannetti01_PRL}, dispersion cancellation in the pulsed regime \cite{Erdmann00} or heralded single-photon generation \cite{Belhassen18}.

\subsection{Particle statistics control} ~

At a deeper level, it is desirable to manipulate the biphoton spectrum both in intensity and phase, and to control also the symmetry properties of the spectral wavefunction, i.e. how the the JSA is modified when exchanging signal and idler photons. This determines the effective particle statistics of the biphoton state, as probed e.g.~in a Hong-Ou-Mandel experiment \cite{Hong87}. Indeed, when two correlated photons are incident on a beamsplitter, they can either leave the beamsplitter through the same output port (bunching) or through opposite ports (antibunching). For a symmetric two-photon state, antibunching probability amplitudes cancel each other, leaving only bunching events and thus a dip in the HOM interferogram, typical of bosonic statistics.
For an antisymmetric two-photon state, the reverse scenario occurs, yielding a peak in the HOM interferogram, as would be the case for (independent) fermions \cite{Fedrizzi09}.

Using the previously considered counter-propagating AlGaAs photon pair source, and starting from a Gaussian pump profile yielding a symmetric quantum state, the application of a $\pi$ phase step of at the center of the pump spot (see Fig. \ref{Fig_frequency}a) renders the state antisymmetric.
Figure \ref{Fig_frequency}e shows the HOM interferogram measured with the biphoton state generated in these two configurations respectively (blue and red traces). A transition from a bosonic to an effectively fermionic behavior of the biphoton state is observed \cite{Francesconi20,Fabre22_HOM}.

Anyonic particles, displaying an exchange statistics intermediate between bosons and fermions and playing a key role in the fractional quantum Hall effect \cite{Halperin84,Bartolomei20} as well as spin lattice models \cite{Kitaev03}, can also be simulated through the Hong-Ou-Mandel effect of entangled photons. For this the JSA of the photon pairs needs to be engineered such that $\phi(\omega_s,\omega_i)=e^{\pm i \alpha \pi} \phi(\omega_i,\omega_s)$, with $\alpha$ a real number ranging between $0$ (bosons) and $1$ (fermions).
This relationship means that the spectral wavefunction acquires a phase $\pm \alpha \pi$ upon particle exchange, and the sign of this phase depends on the directionality of the exchange, which corresponds here to either increasing of decreasing the frequency difference $\omega_-=\omega_s-\omega_i$ between signal and idler photons \cite{Francesconi21,Fabre20}. 
By inverting Eq. \ref{Eq_PM}, a suitable pump phase profile can be found to satisfy this anyonic relationship, e.g. for $\alpha=1/2$. 
Pumping the counter-propagating phase-matched AlGaAs source with this pump profile leads to the HOM interferogram shown in Fig. \ref{Fig_frequency}f (blue). Quite differently from the one of bosons and fermions, it displays a coincidence dip at negative delay and a peak at positive delay, and it is point-symmetric with respect to the central point at $\tau=0$.
Another realization of $\alpha=1/2$ anyonic-like biphotons can be obtained by employing a pump beam profile symmetric to the previous one with respect to the waveguide center. The resulting HOM interferogram (red trace in Fig. \ref{Fig_frequency}f) is mirror-symmetric to the previous one, with a coincidence dip at negative delay and a peak at positive delay \cite{Francesconi21}. 

~
These results demonstrate the on-chip generation of biphoton states with fermionic or anyonic exchange statistics, in a reconfigurable manner, at room temperature and telecom wavelength.
An alternative method to control the symmetry properties of biphoton states was demonstrated in Refs.~\cite{Maltese20,Fabre20}, based on quantum frequency combs emitted by AlGaAs Bragg-reflection waveguides with modal phase-matching.
The tuning of the pump wavelength, combined with the introduction of a temporal shift between the two photons of each pair, allowed to control the symmetry of the frequency combs and to produce either bunching or antibunching behavior in a HOM experiment, thus opening complementary perspectives to the those developed in this section.
Overall, these results could be harnessed to study the effect of exchange statistics in various quantum simulation problems \cite{Crespi13,Matthews13,Crespi15} with a chip-integrated platform, and for communication and computation protocols making use of antisymmetric high-dimensional quantum states \cite{Jex03,Goyal14}.

\section{Perspectives}

In this review, we have summarized the main recent achievements in the field of nonlinear integrated quantum photonics based on the AlGaAs platform. The large second and third-order nonlinearities of this material, combined with its mature fabrication technologies have allowed the development of efficient chip-scale sources of quantum light based on parametric processes, and compliant with electrical pumping.
A variety of optical functionalities have been implemented in AlGaAs-based circuits, ranging from polarizing beamsplitters to waveguide crossers, Mach-Zehnder interferometers, filters or modulators. 
The production and manipulation of entanglement in different degrees of freedom has been demonstrated, allowing the engineering of useful quantum states in view of diverse applications in quantum information. We also notice that recently, nonlinear metasurfaces, i.e. arrays of nanoresonators, have been implemented in GaAs with the aim of relaxing the constraint of phase-matching and gaining in quantum state flexibility, at the expense of the pair production rate. This has led for example to the generation of complex frequency-multiplexed quantum states, in particular cluster states  \cite{Santiago2022}. From these results, various perspectives can be envisaged for the next years, both from a technological and from a fundamental point of view.

On the one hand, thanks to rapid progresses on the front of both sources and circuits, the road seems open now to the realization of more elaborated AlGaAs circuits combining monolithically the generation and manipulation of quantum states in view of complex operations, as illustrated in Fig. \ref{Fig_QPICs_illustration}. Thanks to the maturity and reproducibility of the III-V technological processes, such circuits could comprise a series of nominally identical nonlinear sources able to mutually interfere, similarly to recent realizations on the silica-on-silicon \cite{Spring17} and silicon-on-insulator \cite{Paesani19} platforms.
As an example of such multi-sources circuits, arrays of nonlinear AlGaAs waveguides appear as a promising candidate to investigate quantum simulation tasks \cite{Raymond24,Raymond25}. In such devices, photons can continuously tunnel from one waveguide to the other during their propagation, implementing quantum random walks \cite{Peruzzo10}. Thanks to the $\chi^{(2)}$ nonlinearity, photon pairs can be generated directly into the device, and in various waveguides simultaneously, allowing to realize compact and versatile sources of spatially entangled states \cite{Solntsev14}. By tuning statically or dynamically the parameters of the arrays, these devices can provide a workbench to simulate physical problems otherwise difficult to access in condensed matter systems, such as the Anderson localization of multi-particle states \cite{Crespi13} or the topological protection of quantum states \cite{BlancoRedondo18}.

Besides the realization of such monolithic quantum photonic circuits, it is also highly valuable to realize hybrid photonic circuits combining the assets of different materials, so as to reach enlarged capabilities \cite{Elshaari20,Kim20}.
Among the various possible combinations, merging the complementary assets of AlGaAs (e.g. electrical injection, electro-optic effect) with those of silicon-based materials (e.g CMOS production, wide library  of photonic components) holds great promise. 
In quantum photonics, lots of efforts have been devoted recently to the incorporation (by wafer bonding, transfer printing or pick-and-place techniques) \cite{Kim20} of GaAs-based QDs into silicon-based circuits, allowing to integrate the emission, routing and manipulation of single photons in hybrid circuits. 
Recent examples include the integration of QDs into silica-on-silicon microdisks \cite{Yue18}, silicon nitride waveguides \cite{Davanco17} or silicon-on-insulator waveguides \cite{Osada19}. By contrast, the integration of III-V parametric sources with silicon has not yet been demonstrated. In parallel with the single-emitter approach, such achievement would bring along unique advantages such as a high fabrication reproducibility allowing the realization of a large number of identical sources on the same chip, and access to a wider variety of quantum states, from heralded single photons to entangled photon pairs and squeezed states, that could be subsequently manipulated into high-quality, foundry-produced photonic circuits \cite{HybridConf2022}. Concerning the detection part, besides superconducting nanowires, the integration on AlGaAs of room-temperature single-photon detectors with high efficiency could also be envisioned in a near future using e.g. segmented detectors based on III-V single-photon avalanche-photodiodes \cite{Yu18} with potential photon-number-resolving capabilities \cite{Nehra20}.

So far --- and as reflected by the results presented in this review --- experimental demonstrations in integrated quantum photonics have mainly focused on the discrete variable (DV) approach, where quantum information is encoded in individual photons. The continuous-variable (CV) approach, where information is typically encoded in the quadratures of the electromagnetic field, has proven more challenging to transfer from table-top to chip-scale setups, since it typically requires very high-efficiency coupling and low-loss operation. But recent years have seen rapid progresses in this direction, by demonstrating the chip-integrated generation \cite{Dutt15,Mondain19,Kashiwazaki20}, manipulation \cite{Masada15,Mondain19} and detection \cite{Tasker21} of squeezed states of light. In particular, SFWM in silicon or silicon nitride microring resonators and SPDC in lithium niobate waveguides have been shown to generate up to $6$ dB continuous-wave squeezing  \cite{Kashiwazaki20}.
The strong second-order nonlinearity of AlGaAs, its compliance with electrical pumping and its high electro-optic effect make it an appealing platform to implement integrated CV quantum information tasks. Promising preliminary results in this direction have been obtained recently with AlGaAs Bragg-reflection waveguides, demonstrating phase-sensitive amplification with in-phase gain approaching 30 dB \cite{Yan22}, highlighting the potential of the AlGaAs platform for developing chip-integrated CV architectures \cite{Brodutch18}.

Interestingly, CV-based quantum information can also be investigated by exploiting high-dimensional degrees of freedom of single photons, such as frequency-time variables. Indeed, they display a perfect analogy with the continuous variables of a multiphoton mode of the electromagnetic field \cite{Abouraddy07,Fabre22}, opening the perspective to perform CV quantum information processing in the few-photon regime \cite{Tasca11,Fabre20}.
A promising example is provided by Gottesman-Kitaev-Preskill (GKP) states, which are powerful resources to implement quantum error correction schemes \cite{Gottesman01}. While their experimental realization is highly demanding in the quadrature representation, it has been shown that biphoton frequency combs generated by AlGaAs waveguides directly implement GKP states in the time-frequency degrees of freedom, making them an appealing testbed to investigate CV-like quantum information tasks \cite{Fabre20}.

Hybridizing various degrees of freedom of photons is also an emerging approach, allowing to increase the density and flexibility of information coding \cite{Kwiat97,Wang18_18qubit}. As detailed in section 6, AlGaAs waveguide sources can directly generate either polarization or frequency entangled photon pairs. 
But both degrees of freedom can also be combined, leading to the generation of hybrid polarization-frequency entangled photons without post-manipulation \cite{Francesconi22}. 
Such combination of DV and CV-like degrees of freedom could provide enlarged capabilities for quantum information tasks, allowing to switch from one degree of freedom to another and thus to adapt to different experimental conditions in a versatile manner.
Hyper-entangled polarization-frequency states, where polarization and frequency entanglement are fully independent, could also be produced with AlGaAs sources \cite{Francesconi22}, opening perspectives e.g. in the field quantum communication to improve bit rates and resilience to noise \cite{Steinlechner17,Vergyris19,Kim21}. 

\FloatBarrier

\section*{Acknowledgements}

G.M. acknowledges support from the NSF Quantum Foundry through Q-AMASE-i program Award No. DMR-1906325, AFOSR YIP Award No. FA9550-20-1-0150, and NSF Award CAREER-2045246.
S.D. and F.B. acknowledge  ANR (Agence Nationale de la Recherche) through Projects No. ANR-19-ASTR-0018-01, and through France 2030 Plan Projects No. ANR-22-PETQ-0006 and No. ANR-22-PETQ-0011; F.B. acknowledges Ville de Paris Emergence program (LATTICE project) and IdEx Université Paris Cité (ANR-18-IDEX-0001).

\section*{Disclosures}

The authors declare no conflicts of interest.

\section*{Data availability}

Data underlying the results presented in this paper are not publicly available at this time but may be obtained from the authors upon reasonable request.


\begin{thebibliography}{184}%
	\makeatletter
	\providecommand \@ifxundefined [1]{%
		\@ifx{#1\undefined}
	}%
	\providecommand \@ifnum [1]{%
		\ifnum #1\expandafter \@firstoftwo
		\else \expandafter \@secondoftwo
		\fi
	}%
	\providecommand \@ifx [1]{%
		\ifx #1\expandafter \@firstoftwo
		\else \expandafter \@secondoftwo
		\fi
	}%
	\providecommand \natexlab [1]{#1}%
	\providecommand \enquote  [1]{``#1''}%
	\providecommand \bibnamefont  [1]{#1}%
	\providecommand \bibfnamefont [1]{#1}%
	\providecommand \citenamefont [1]{#1}%
	\providecommand \href@noop [0]{\@secondoftwo}%
	\providecommand \href [0]{\begingroup \@sanitize@url \@href}%
	\providecommand \@href[1]{\@@startlink{#1}\@@href}%
	\providecommand \@@href[1]{\endgroup#1\@@endlink}%
	\providecommand \@sanitize@url [0]{\catcode `\\12\catcode `\$12\catcode
		`\&12\catcode `\#12\catcode `\^12\catcode `\_12\catcode `\%12\relax}%
	\providecommand \@@startlink[1]{}%
	\providecommand \@@endlink[0]{}%
	\providecommand \url  [0]{\begingroup\@sanitize@url \@url }%
	\providecommand \@url [1]{\endgroup\@href {#1}{\urlprefix }}%
	\providecommand \urlprefix  [0]{URL }%
	\providecommand \Eprint [0]{\href }%
	\providecommand \doibase [0]{https://doi.org/}%
	\providecommand \selectlanguage [0]{\@gobble}%
	\providecommand \bibinfo  [0]{\@secondoftwo}%
	\providecommand \bibfield  [0]{\@secondoftwo}%
	\providecommand \translation [1]{[#1]}%
	\providecommand \BibitemOpen [0]{}%
	\providecommand \bibitemStop [0]{}%
	\providecommand \bibitemNoStop [0]{.\EOS\space}%
	\providecommand \EOS [0]{\spacefactor3000\relax}%
	\providecommand \BibitemShut  [1]{\csname bibitem#1\endcsname}%
	\let\auto@bib@innerbib\@empty
	\bibitem [{\citenamefont {Flamini}\ \emph
		{et~al.}(2018{\natexlab{a}})\citenamefont {Flamini}, \citenamefont
		{Spagnolo},\ and\ \citenamefont {Sciarrino}}]{Flamini2018}%
	\BibitemOpen
	\bibfield  {author} {\bibinfo {author} {\bibfnamefont {F.}~\bibnamefont
			{Flamini}}, \bibinfo {author} {\bibfnamefont {N.}~\bibnamefont {Spagnolo}},\
		and\ \bibinfo {author} {\bibfnamefont {F.}~\bibnamefont {Sciarrino}},\
	}\bibfield  {title} {\emph {\bibinfo {title} {{\color{blue}Photonic quantum
					information processing: a review}}},\ }\href@noop {} {\bibfield  {journal}
		{\bibinfo  {journal} {Reports on Progress in Physics}\ }\textbf {\bibinfo
			{volume} {82}},\ \bibinfo {pages} {016001} (\bibinfo {year}
		{2018}{\natexlab{a}})}\BibitemShut {NoStop}%
	\bibitem [{\citenamefont {Slussarenko}\ and\ \citenamefont
		{Pryde}(2019)}]{Slussarenko2019}%
	\BibitemOpen
	\bibfield  {author} {\bibinfo {author} {\bibfnamefont {S.}~\bibnamefont
			{Slussarenko}}\ and\ \bibinfo {author} {\bibfnamefont {G.~J.}\ \bibnamefont
			{Pryde}},\ }\bibfield  {title} {\emph {\bibinfo {title}
			{{\color{blue}Photonic quantum information processing: A concise review}}},\
	}\href@noop {} {\bibfield  {journal} {\bibinfo  {journal} {Applied Physics
				Reviews}\ }\textbf {\bibinfo {volume} {6}},\ \bibinfo {pages} {041303}
		(\bibinfo {year} {2019})}\BibitemShut {NoStop}%
	\bibitem [{\citenamefont {Polino}\ \emph {et~al.}(2020)\citenamefont {Polino},
		\citenamefont {Valeri}, \citenamefont {Spagnolo},\ and\ \citenamefont
		{Sciarrino}}]{Polino2020}%
	\BibitemOpen
	\bibfield  {author} {\bibinfo {author} {\bibfnamefont {E.}~\bibnamefont
			{Polino}}, \bibinfo {author} {\bibfnamefont {M.}~\bibnamefont {Valeri}},
		\bibinfo {author} {\bibfnamefont {N.}~\bibnamefont {Spagnolo}},\ and\
		\bibinfo {author} {\bibfnamefont {F.}~\bibnamefont {Sciarrino}},\ }\bibfield
	{title} {\emph {\bibinfo {title} {{\color{blue}Photonic quantum
					metrology}}},\ }\href@noop {} {\bibfield  {journal} {\bibinfo  {journal} {AVS
				Quantum Science}\ }\textbf {\bibinfo {volume} {2}},\ \bibinfo {pages}
		{024703} (\bibinfo {year} {2020})}\BibitemShut {NoStop}%
	\bibitem [{\citenamefont {Moody}\ \emph {et~al.}(2022)\citenamefont {Moody},
		\citenamefont {Sorger}, \citenamefont {Blumenthal}, \citenamefont
		{Juodawlkis}, \citenamefont {Loh}, \citenamefont {Sorace-Agaskar},
		\citenamefont {Jones}, \citenamefont {Balram}, \citenamefont {Matthews},
		\citenamefont {Laing} \emph {et~al.}}]{moody20222022}%
	\BibitemOpen
	\bibfield  {author} {\bibinfo {author} {\bibfnamefont {G.}~\bibnamefont
			{Moody}}, \bibinfo {author} {\bibfnamefont {V.~J.}\ \bibnamefont {Sorger}},
		\bibinfo {author} {\bibfnamefont {D.~J.}\ \bibnamefont {Blumenthal}},
		\bibinfo {author} {\bibfnamefont {P.~W.}\ \bibnamefont {Juodawlkis}},
		\bibinfo {author} {\bibfnamefont {W.}~\bibnamefont {Loh}}, \bibinfo {author}
		{\bibfnamefont {C.}~\bibnamefont {Sorace-Agaskar}}, \bibinfo {author}
		{\bibfnamefont {A.~E.}\ \bibnamefont {Jones}}, \bibinfo {author}
		{\bibfnamefont {K.~C.}\ \bibnamefont {Balram}}, \bibinfo {author}
		{\bibfnamefont {J.~C.}\ \bibnamefont {Matthews}}, \bibinfo {author}
		{\bibfnamefont {A.}~\bibnamefont {Laing}}, \emph {et~al.},\ }\bibfield
	{title} {\emph {\bibinfo {title} {{\color{blue}2022 Roadmap on integrated
					quantum photonics}}},\ }\href@noop {} {\bibfield  {journal} {\bibinfo
			{journal} {Journal of Physics: Photonics}\ }\textbf {\bibinfo {volume} {4}},\
		\bibinfo {pages} {012501} (\bibinfo {year} {2022})}\BibitemShut {NoStop}%
	\bibitem [{\citenamefont {Liao}\ \emph
		{et~al.}(2017{\natexlab{a}})\citenamefont {Liao}, \citenamefont {Cai},
		\citenamefont {Liu}, \citenamefont {Zhang}, \citenamefont {Li}, \citenamefont
		{Ren}, \citenamefont {Yin}, \citenamefont {Shen}, \citenamefont {Cao},
		\citenamefont {Li}, \citenamefont {Li}, \citenamefont {Chen}, \citenamefont
		{Sun}, \citenamefont {Jia}, \citenamefont {Wu}, \citenamefont {Jiang},
		\citenamefont {Wang}, \citenamefont {Huang}, \citenamefont {Wang},
		\citenamefont {Zhou}, \citenamefont {Deng}, \citenamefont {Xi}, \citenamefont
		{Ma}, \citenamefont {Hu}, \citenamefont {Zhang}, \citenamefont {Chen},
		\citenamefont {Liu}, \citenamefont {Wang}, \citenamefont {Zhu}, \citenamefont
		{Lu}, \citenamefont {Shu}, \citenamefont {Peng}, \citenamefont {Wang},\ and\
		\citenamefont {Pan}}]{Liao2017}%
	\BibitemOpen
	\bibfield  {author} {\bibinfo {author} {\bibfnamefont {S.-K.}\ \bibnamefont
			{Liao}}, \bibinfo {author} {\bibfnamefont {W.-Q.}\ \bibnamefont {Cai}},
		\bibinfo {author} {\bibfnamefont {W.-Y.}\ \bibnamefont {Liu}}, \bibinfo
		{author} {\bibfnamefont {L.}~\bibnamefont {Zhang}}, \bibinfo {author}
		{\bibfnamefont {Y.}~\bibnamefont {Li}}, \bibinfo {author} {\bibfnamefont
			{J.-G.}\ \bibnamefont {Ren}}, \bibinfo {author} {\bibfnamefont
			{J.}~\bibnamefont {Yin}}, \bibinfo {author} {\bibfnamefont {Q.}~\bibnamefont
			{Shen}}, \bibinfo {author} {\bibfnamefont {Y.}~\bibnamefont {Cao}}, \bibinfo
		{author} {\bibfnamefont {Z.-P.}\ \bibnamefont {Li}}, \bibinfo {author}
		{\bibfnamefont {F.-Z.}\ \bibnamefont {Li}}, \bibinfo {author} {\bibfnamefont
			{X.-W.}\ \bibnamefont {Chen}}, \bibinfo {author} {\bibfnamefont {L.-H.}\
			\bibnamefont {Sun}}, \bibinfo {author} {\bibfnamefont {J.-J.}\ \bibnamefont
			{Jia}}, \bibinfo {author} {\bibfnamefont {J.-C.}\ \bibnamefont {Wu}},
		\bibinfo {author} {\bibfnamefont {X.-J.}\ \bibnamefont {Jiang}}, \bibinfo
		{author} {\bibfnamefont {J.-F.}\ \bibnamefont {Wang}}, \bibinfo {author}
		{\bibfnamefont {Y.-M.}\ \bibnamefont {Huang}}, \bibinfo {author}
		{\bibfnamefont {Q.}~\bibnamefont {Wang}}, \bibinfo {author} {\bibfnamefont
			{Y.-L.}\ \bibnamefont {Zhou}}, \bibinfo {author} {\bibfnamefont
			{L.}~\bibnamefont {Deng}}, \bibinfo {author} {\bibfnamefont {T.}~\bibnamefont
			{Xi}}, \bibinfo {author} {\bibfnamefont {L.}~\bibnamefont {Ma}}, \bibinfo
		{author} {\bibfnamefont {T.}~\bibnamefont {Hu}}, \bibinfo {author}
		{\bibfnamefont {Q.}~\bibnamefont {Zhang}}, \bibinfo {author} {\bibfnamefont
			{Y.-A.}\ \bibnamefont {Chen}}, \bibinfo {author} {\bibfnamefont {N.-L.}\
			\bibnamefont {Liu}}, \bibinfo {author} {\bibfnamefont {X.-B.}\ \bibnamefont
			{Wang}}, \bibinfo {author} {\bibfnamefont {Z.-C.}\ \bibnamefont {Zhu}},
		\bibinfo {author} {\bibfnamefont {C.-Y.}\ \bibnamefont {Lu}}, \bibinfo
		{author} {\bibfnamefont {R.}~\bibnamefont {Shu}}, \bibinfo {author}
		{\bibfnamefont {C.-Z.}\ \bibnamefont {Peng}}, \bibinfo {author}
		{\bibfnamefont {J.-Y.}\ \bibnamefont {Wang}},\ and\ \bibinfo {author}
		{\bibfnamefont {J.-W.}\ \bibnamefont {Pan}},\ }\bibfield  {title} {\emph
		{\bibinfo {title} {{\color{blue}Satellite-to-ground quantum key
					distribution}}},\ }\href@noop {} {\bibfield  {journal} {\bibinfo  {journal}
			{Nature}\ }\textbf {\bibinfo {volume} {549}},\ \bibinfo {pages} {43}
		(\bibinfo {year} {2017}{\natexlab{a}})}\BibitemShut {NoStop}%
	\bibitem [{\citenamefont {Zhong}\ \emph {et~al.}(2020)\citenamefont {Zhong},
		\citenamefont {Wang}, \citenamefont {Deng}, \citenamefont {Chen},
		\citenamefont {Peng}, \citenamefont {Luo}, \citenamefont {Qin}, \citenamefont
		{Wu}, \citenamefont {Ding}, \citenamefont {Hu}, \citenamefont {Hu},
		\citenamefont {Yang}, \citenamefont {Zhang}, \citenamefont {Li},
		\citenamefont {Li}, \citenamefont {Jiang}, \citenamefont {Gan}, \citenamefont
		{Yang}, \citenamefont {You}, \citenamefont {Wang}, \citenamefont {Li},
		\citenamefont {Liu}, \citenamefont {Lu},\ and\ \citenamefont
		{Pan}}]{Zhong2020}%
	\BibitemOpen
	\bibfield  {author} {\bibinfo {author} {\bibfnamefont {H.-S.}\ \bibnamefont
			{Zhong}}, \bibinfo {author} {\bibfnamefont {H.}~\bibnamefont {Wang}},
		\bibinfo {author} {\bibfnamefont {Y.-H.}\ \bibnamefont {Deng}}, \bibinfo
		{author} {\bibfnamefont {M.-C.}\ \bibnamefont {Chen}}, \bibinfo {author}
		{\bibfnamefont {L.-C.}\ \bibnamefont {Peng}}, \bibinfo {author}
		{\bibfnamefont {Y.-H.}\ \bibnamefont {Luo}}, \bibinfo {author} {\bibfnamefont
			{J.}~\bibnamefont {Qin}}, \bibinfo {author} {\bibfnamefont {D.}~\bibnamefont
			{Wu}}, \bibinfo {author} {\bibfnamefont {X.}~\bibnamefont {Ding}}, \bibinfo
		{author} {\bibfnamefont {Y.}~\bibnamefont {Hu}}, \bibinfo {author}
		{\bibfnamefont {P.}~\bibnamefont {Hu}}, \bibinfo {author} {\bibfnamefont
			{X.-Y.}\ \bibnamefont {Yang}}, \bibinfo {author} {\bibfnamefont {W.-J.}\
			\bibnamefont {Zhang}}, \bibinfo {author} {\bibfnamefont {H.}~\bibnamefont
			{Li}}, \bibinfo {author} {\bibfnamefont {Y.}~\bibnamefont {Li}}, \bibinfo
		{author} {\bibfnamefont {X.}~\bibnamefont {Jiang}}, \bibinfo {author}
		{\bibfnamefont {L.}~\bibnamefont {Gan}}, \bibinfo {author} {\bibfnamefont
			{G.}~\bibnamefont {Yang}}, \bibinfo {author} {\bibfnamefont {L.}~\bibnamefont
			{You}}, \bibinfo {author} {\bibfnamefont {Z.}~\bibnamefont {Wang}}, \bibinfo
		{author} {\bibfnamefont {L.}~\bibnamefont {Li}}, \bibinfo {author}
		{\bibfnamefont {N.-L.}\ \bibnamefont {Liu}}, \bibinfo {author} {\bibfnamefont
			{C.-Y.}\ \bibnamefont {Lu}},\ and\ \bibinfo {author} {\bibfnamefont {J.-W.}\
			\bibnamefont {Pan}},\ }\bibfield  {title} {\emph {\bibinfo {title}
			{{\color{blue}Quantum computational advantage using photons}}},\ }\href@noop
	{} {\bibfield  {journal} {\bibinfo  {journal} {Science}\ }\textbf {\bibinfo
			{volume} {370}},\ \bibinfo {pages} {1460} (\bibinfo {year}
		{2020})}\BibitemShut {NoStop}%
	\bibitem [{\citenamefont {Madsen}\ \emph {et~al.}(2022)\citenamefont {Madsen},
		\citenamefont {Laudenbach}, \citenamefont {Askarani}, \citenamefont
		{Rortais}, \citenamefont {Vincent}, \citenamefont {Bulmer}, \citenamefont
		{Miatto}, \citenamefont {Neuhaus}, \citenamefont {Helt}, \citenamefont
		{Collins}, \citenamefont {Lita}, \citenamefont {Gerrits}, \citenamefont
		{Nam}, \citenamefont {Vaidya}, \citenamefont {Menotti}, \citenamefont
		{Dhand}, \citenamefont {Vernon}, \citenamefont {Quesada},\ and\ \citenamefont
		{Lavoie}}]{Madsen2022}%
	\BibitemOpen
	\bibfield  {author} {\bibinfo {author} {\bibfnamefont {L.~S.}\ \bibnamefont
			{Madsen}}, \bibinfo {author} {\bibfnamefont {F.}~\bibnamefont {Laudenbach}},
		\bibinfo {author} {\bibfnamefont {M.~F.}\ \bibnamefont {Askarani}}, \bibinfo
		{author} {\bibfnamefont {F.}~\bibnamefont {Rortais}}, \bibinfo {author}
		{\bibfnamefont {T.}~\bibnamefont {Vincent}}, \bibinfo {author} {\bibfnamefont
			{J.~F.~F.}\ \bibnamefont {Bulmer}}, \bibinfo {author} {\bibfnamefont {F.~M.}\
			\bibnamefont {Miatto}}, \bibinfo {author} {\bibfnamefont {L.}~\bibnamefont
			{Neuhaus}}, \bibinfo {author} {\bibfnamefont {L.~G.}\ \bibnamefont {Helt}},
		\bibinfo {author} {\bibfnamefont {M.~J.}\ \bibnamefont {Collins}}, \bibinfo
		{author} {\bibfnamefont {A.~E.}\ \bibnamefont {Lita}}, \bibinfo {author}
		{\bibfnamefont {T.}~\bibnamefont {Gerrits}}, \bibinfo {author} {\bibfnamefont
			{S.~W.}\ \bibnamefont {Nam}}, \bibinfo {author} {\bibfnamefont {V.~D.}\
			\bibnamefont {Vaidya}}, \bibinfo {author} {\bibfnamefont {M.}~\bibnamefont
			{Menotti}}, \bibinfo {author} {\bibfnamefont {I.}~\bibnamefont {Dhand}},
		\bibinfo {author} {\bibfnamefont {Z.}~\bibnamefont {Vernon}}, \bibinfo
		{author} {\bibfnamefont {N.}~\bibnamefont {Quesada}},\ and\ \bibinfo {author}
		{\bibfnamefont {J.}~\bibnamefont {Lavoie}},\ }\bibfield  {title} {\emph
		{\bibinfo {title} {{\color{blue}Quantum computational advantage with a
					programmable photonic processor}}},\ }\href@noop {} {\bibfield  {journal}
		{\bibinfo  {journal} {Nature}\ }\textbf {\bibinfo {volume} {606}},\ \bibinfo
		{pages} {75} (\bibinfo {year} {2022})}\BibitemShut {NoStop}%
	\bibitem [{\citenamefont {Autebert}\ \emph {et~al.}(2016)\citenamefont
		{Autebert}, \citenamefont {Bruno}, \citenamefont {Martin}, \citenamefont
		{Lema{\^\i}tre}, \citenamefont {Carbonell}, \citenamefont {Favero},
		\citenamefont {Leo}, \citenamefont {Zbinden},\ and\ \citenamefont
		{Ducci}}]{Autebert16}%
	\BibitemOpen
	\bibfield  {author} {\bibinfo {author} {\bibfnamefont {C.}~\bibnamefont
			{Autebert}}, \bibinfo {author} {\bibfnamefont {N.}~\bibnamefont {Bruno}},
		\bibinfo {author} {\bibfnamefont {A.}~\bibnamefont {Martin}}, \bibinfo
		{author} {\bibfnamefont {A.}~\bibnamefont {Lema{\^\i}tre}}, \bibinfo {author}
		{\bibfnamefont {C.~G.}\ \bibnamefont {Carbonell}}, \bibinfo {author}
		{\bibfnamefont {I.}~\bibnamefont {Favero}}, \bibinfo {author} {\bibfnamefont
			{G.}~\bibnamefont {Leo}}, \bibinfo {author} {\bibfnamefont {H.}~\bibnamefont
			{Zbinden}},\ and\ \bibinfo {author} {\bibfnamefont {S.}~\bibnamefont
			{Ducci}},\ }\bibfield  {title} {\emph {\bibinfo {title}
			{{\color{blue}Integrated {AlGaAs} source of highly indistinguishable and
					energy-time entangled photons}}},\ }\href@noop {} {\bibfield  {journal}
		{\bibinfo  {journal} {Optica}\ }\textbf {\bibinfo {volume} {3}},\ \bibinfo
		{pages} {143} (\bibinfo {year} {2016})}\BibitemShut {NoStop}%
	\bibitem [{\citenamefont {Chang}\ \emph {et~al.}(2020)\citenamefont {Chang},
		\citenamefont {Xie}, \citenamefont {Shu}, \citenamefont {Yang}, \citenamefont
		{Shen}, \citenamefont {Boes}, \citenamefont {Peters}, \citenamefont {Jin},
		\citenamefont {Xiang}, \citenamefont {Liu}, \citenamefont {Moille},
		\citenamefont {Yu}, \citenamefont {Wang}, \citenamefont {Srinivasan},
		\citenamefont {Papp}, \citenamefont {Vahala},\ and\ \citenamefont
		{Bowers}}]{Chang2020}%
	\BibitemOpen
	\bibfield  {author} {\bibinfo {author} {\bibfnamefont {L.}~\bibnamefont
			{Chang}}, \bibinfo {author} {\bibfnamefont {W.}~\bibnamefont {Xie}}, \bibinfo
		{author} {\bibfnamefont {H.}~\bibnamefont {Shu}}, \bibinfo {author}
		{\bibfnamefont {Q.~F.}\ \bibnamefont {Yang}}, \bibinfo {author}
		{\bibfnamefont {B.}~\bibnamefont {Shen}}, \bibinfo {author} {\bibfnamefont
			{A.}~\bibnamefont {Boes}}, \bibinfo {author} {\bibfnamefont {J.~D.}\
			\bibnamefont {Peters}}, \bibinfo {author} {\bibfnamefont {W.}~\bibnamefont
			{Jin}}, \bibinfo {author} {\bibfnamefont {C.}~\bibnamefont {Xiang}}, \bibinfo
		{author} {\bibfnamefont {S.}~\bibnamefont {Liu}}, \bibinfo {author}
		{\bibfnamefont {G.}~\bibnamefont {Moille}}, \bibinfo {author} {\bibfnamefont
			{S.~P.}\ \bibnamefont {Yu}}, \bibinfo {author} {\bibfnamefont
			{X.}~\bibnamefont {Wang}}, \bibinfo {author} {\bibfnamefont {K.}~\bibnamefont
			{Srinivasan}}, \bibinfo {author} {\bibfnamefont {S.~B.}\ \bibnamefont
			{Papp}}, \bibinfo {author} {\bibfnamefont {K.}~\bibnamefont {Vahala}},\ and\
		\bibinfo {author} {\bibfnamefont {J.~E.}\ \bibnamefont {Bowers}},\ }\bibfield
	{title} {\emph {\bibinfo {title} {{\color{blue}Ultra-efficient frequency
					comb generation in {AlGaAs-on-insulator} microresonators}}},\ }\href@noop {}
	{\bibfield  {journal} {\bibinfo  {journal} {Nature Communications}\ }\textbf
		{\bibinfo {volume} {11}},\ \bibinfo {pages} {1} (\bibinfo {year}
		{2020})}\BibitemShut {NoStop}%
	\bibitem [{\citenamefont {Schwartz}\ \emph {et~al.}(2018)\citenamefont
		{Schwartz}, \citenamefont {Schmidt}, \citenamefont {Rengstl}, \citenamefont
		{Hornung}, \citenamefont {Hepp}, \citenamefont {Portalupi}, \citenamefont
		{llin}, \citenamefont {Jetter}, \citenamefont {Siegel},\ and\ \citenamefont
		{Michler}}]{Schwartz2018}%
	\BibitemOpen
	\bibfield  {author} {\bibinfo {author} {\bibfnamefont {M.}~\bibnamefont
			{Schwartz}}, \bibinfo {author} {\bibfnamefont {E.}~\bibnamefont {Schmidt}},
		\bibinfo {author} {\bibfnamefont {U.}~\bibnamefont {Rengstl}}, \bibinfo
		{author} {\bibfnamefont {F.}~\bibnamefont {Hornung}}, \bibinfo {author}
		{\bibfnamefont {S.}~\bibnamefont {Hepp}}, \bibinfo {author} {\bibfnamefont
			{S.~L.}\ \bibnamefont {Portalupi}}, \bibinfo {author} {\bibfnamefont
			{K.}~\bibnamefont {llin}}, \bibinfo {author} {\bibfnamefont {M.}~\bibnamefont
			{Jetter}}, \bibinfo {author} {\bibfnamefont {M.}~\bibnamefont {Siegel}},\
		and\ \bibinfo {author} {\bibfnamefont {P.}~\bibnamefont {Michler}},\
	}\bibfield  {title} {\emph {\bibinfo {title} {{\color{blue}Fully On-Chip
					Single-Photon Hanbury-Brown and Twiss Experiment on a Monolithic
					Semiconductor--Superconductor Platform}}},\ }\href@noop {} {\bibfield
		{journal} {\bibinfo  {journal} {Nano Letters}\ }\textbf {\bibinfo {volume}
			{18}},\ \bibinfo {pages} {6892} (\bibinfo {year} {2018})}\BibitemShut
	{NoStop}%
	\bibitem [{\citenamefont {Steiner}\ \emph {et~al.}(2021)\citenamefont
		{Steiner}, \citenamefont {Castro}, \citenamefont {Chang}, \citenamefont
		{Dang}, \citenamefont {Xie}, \citenamefont {Norman}, \citenamefont {Bowers},\
		and\ \citenamefont {Moody}}]{Steiner2021}%
	\BibitemOpen
	\bibfield  {author} {\bibinfo {author} {\bibfnamefont {T.~J.}\ \bibnamefont
			{Steiner}}, \bibinfo {author} {\bibfnamefont {J.~E.}\ \bibnamefont {Castro}},
		\bibinfo {author} {\bibfnamefont {L.}~\bibnamefont {Chang}}, \bibinfo
		{author} {\bibfnamefont {Q.}~\bibnamefont {Dang}}, \bibinfo {author}
		{\bibfnamefont {W.}~\bibnamefont {Xie}}, \bibinfo {author} {\bibfnamefont
			{J.}~\bibnamefont {Norman}}, \bibinfo {author} {\bibfnamefont {J.~E.}\
			\bibnamefont {Bowers}},\ and\ \bibinfo {author} {\bibfnamefont
			{G.}~\bibnamefont {Moody}},\ }\bibfield  {title} {\emph {\bibinfo {title}
			{{\color{blue}Ultrabright Entangled-Photon-Pair Generation from an
					{AlGaAs-On-Insulator} Microring Resonator}}},\ }\href@noop {} {\bibfield
		{journal} {\bibinfo  {journal} {PRX Quantum}\ }\textbf {\bibinfo {volume}
			{2}},\ \bibinfo {pages} {010337} (\bibinfo {year} {2021})}\BibitemShut
	{NoStop}%
	\bibitem [{\citenamefont {Castro}\ \emph {et~al.}(2022)\citenamefont {Castro},
		\citenamefont {Steiner}, \citenamefont {Thiel}, \citenamefont {Dinkelacker},
		\citenamefont {McDonald}, \citenamefont {Pintus}, \citenamefont {Chang},
		\citenamefont {Bowers},\ and\ \citenamefont {Moody}}]{CastroAlGaAsOI}%
	\BibitemOpen
	\bibfield  {author} {\bibinfo {author} {\bibfnamefont {J.~E.}\ \bibnamefont
			{Castro}}, \bibinfo {author} {\bibfnamefont {T.~J.}\ \bibnamefont {Steiner}},
		\bibinfo {author} {\bibfnamefont {L.}~\bibnamefont {Thiel}}, \bibinfo
		{author} {\bibfnamefont {A.}~\bibnamefont {Dinkelacker}}, \bibinfo {author}
		{\bibfnamefont {C.}~\bibnamefont {McDonald}}, \bibinfo {author}
		{\bibfnamefont {P.}~\bibnamefont {Pintus}}, \bibinfo {author} {\bibfnamefont
			{L.}~\bibnamefont {Chang}}, \bibinfo {author} {\bibfnamefont {J.~E.}\
			\bibnamefont {Bowers}},\ and\ \bibinfo {author} {\bibfnamefont
			{G.}~\bibnamefont {Moody}},\ }\bibfield  {title} {\emph {\bibinfo {title}
			{{\color{blue}Expanding the quantum photonic toolbox in {AlGaAsOI}}}},\
	}\href@noop {} {\bibfield  {journal} {\bibinfo  {journal} {APL Photonics}\
		}\textbf {\bibinfo {volume} {7}},\ \bibinfo {pages} {096103} (\bibinfo {year}
		{2022})}\BibitemShut {NoStop}%
	\bibitem [{\citenamefont {Boitier}\ \emph {et~al.}(2014)\citenamefont
		{Boitier}, \citenamefont {Orieux}, \citenamefont {Autebert}, \citenamefont
		{Lema\^{\i}tre}, \citenamefont {Galopin}, \citenamefont {Manquest},
		\citenamefont {Sirtori}, \citenamefont {Favero}, \citenamefont {Leo},\ and\
		\citenamefont {Ducci}}]{Boitier14}%
	\BibitemOpen
	\bibfield  {author} {\bibinfo {author} {\bibfnamefont {F.}~\bibnamefont
			{Boitier}}, \bibinfo {author} {\bibfnamefont {A.}~\bibnamefont {Orieux}},
		\bibinfo {author} {\bibfnamefont {C.}~\bibnamefont {Autebert}}, \bibinfo
		{author} {\bibfnamefont {A.}~\bibnamefont {Lema\^{\i}tre}}, \bibinfo {author}
		{\bibfnamefont {E.}~\bibnamefont {Galopin}}, \bibinfo {author} {\bibfnamefont
			{C.}~\bibnamefont {Manquest}}, \bibinfo {author} {\bibfnamefont
			{C.}~\bibnamefont {Sirtori}}, \bibinfo {author} {\bibfnamefont
			{I.}~\bibnamefont {Favero}}, \bibinfo {author} {\bibfnamefont
			{G.}~\bibnamefont {Leo}},\ and\ \bibinfo {author} {\bibfnamefont
			{S.}~\bibnamefont {Ducci}},\ }\bibfield  {title} {\emph {\bibinfo {title}
			{{\color{blue}Electrically injected photon-pair source at room
					temperature}}},\ }\href@noop {} {\bibfield  {journal} {\bibinfo  {journal}
			{Phys. Rev. Lett.}\ }\textbf {\bibinfo {volume} {112}},\ \bibinfo {pages}
		{183901} (\bibinfo {year} {2014})}\BibitemShut {NoStop}%
	\bibitem [{\citenamefont {Belhassen}\ \emph {et~al.}(2018)\citenamefont
		{Belhassen}, \citenamefont {Baboux}, \citenamefont {Yao}, \citenamefont
		{Amanti}, \citenamefont {Favero}, \citenamefont {Lema{\^\i}tre},
		\citenamefont {Kolthammer}, \citenamefont {Walmsley},\ and\ \citenamefont
		{Ducci}}]{Belhassen18}%
	\BibitemOpen
	\bibfield  {author} {\bibinfo {author} {\bibfnamefont {J.}~\bibnamefont
			{Belhassen}}, \bibinfo {author} {\bibfnamefont {F.}~\bibnamefont {Baboux}},
		\bibinfo {author} {\bibfnamefont {Q.}~\bibnamefont {Yao}}, \bibinfo {author}
		{\bibfnamefont {M.}~\bibnamefont {Amanti}}, \bibinfo {author} {\bibfnamefont
			{I.}~\bibnamefont {Favero}}, \bibinfo {author} {\bibfnamefont
			{A.}~\bibnamefont {Lema{\^\i}tre}}, \bibinfo {author} {\bibfnamefont
			{W.}~\bibnamefont {Kolthammer}}, \bibinfo {author} {\bibfnamefont
			{I.}~\bibnamefont {Walmsley}},\ and\ \bibinfo {author} {\bibfnamefont
			{S.}~\bibnamefont {Ducci}},\ }\bibfield  {title} {\emph {\bibinfo {title}
			{{\color{blue}On-chip {III-V} monolithic integration of heralded single
					photon sources and beamsplitters}}},\ }\href@noop {} {\bibfield  {journal}
		{\bibinfo  {journal} {Applied Physics Letters}\ }\textbf {\bibinfo {volume}
			{112}},\ \bibinfo {pages} {071105} (\bibinfo {year} {2018})}\BibitemShut
	{NoStop}%
	\bibitem [{\citenamefont {Wang}\ \emph {et~al.}(2014)\citenamefont {Wang},
		\citenamefont {Santamato}, \citenamefont {Jiang}, \citenamefont {Bonneau},
		\citenamefont {Engin}, \citenamefont {Silverstone}, \citenamefont {Lermer},
		\citenamefont {Beetz}, \citenamefont {Kamp}, \citenamefont {H{\"o}fling}
		\emph {et~al.}}]{Wang14}%
	\BibitemOpen
	\bibfield  {author} {\bibinfo {author} {\bibfnamefont {J.}~\bibnamefont
			{Wang}}, \bibinfo {author} {\bibfnamefont {A.}~\bibnamefont {Santamato}},
		\bibinfo {author} {\bibfnamefont {P.}~\bibnamefont {Jiang}}, \bibinfo
		{author} {\bibfnamefont {D.}~\bibnamefont {Bonneau}}, \bibinfo {author}
		{\bibfnamefont {E.}~\bibnamefont {Engin}}, \bibinfo {author} {\bibfnamefont
			{J.~W.}\ \bibnamefont {Silverstone}}, \bibinfo {author} {\bibfnamefont
			{M.}~\bibnamefont {Lermer}}, \bibinfo {author} {\bibfnamefont
			{J.}~\bibnamefont {Beetz}}, \bibinfo {author} {\bibfnamefont
			{M.}~\bibnamefont {Kamp}}, \bibinfo {author} {\bibfnamefont {S.}~\bibnamefont
			{H{\"o}fling}}, \emph {et~al.},\ }\bibfield  {title} {\emph {\bibinfo {title}
			{{\color{blue}Gallium arsenide quantum photonic waveguide circuits}}},\
	}\href@noop {} {\bibfield  {journal} {\bibinfo  {journal} {Optics
				Communications}\ }\textbf {\bibinfo {volume} {327}},\ \bibinfo {pages} {49}
		(\bibinfo {year} {2014})}\BibitemShut {NoStop}%
	\bibitem [{\citenamefont {Appas}\ \emph
		{et~al.}(2022{\natexlab{a}})\citenamefont {Appas}, \citenamefont {Meskine},
		\citenamefont {Lemaître}, \citenamefont {Morassi}, \citenamefont {Baboux},
		\citenamefont {Amanti},\ and\ \citenamefont {Ducci}}]{Appas2022}%
	\BibitemOpen
	\bibfield  {author} {\bibinfo {author} {\bibfnamefont {F.}~\bibnamefont
			{Appas}}, \bibinfo {author} {\bibfnamefont {O.}~\bibnamefont {Meskine}},
		\bibinfo {author} {\bibfnamefont {A.}~\bibnamefont {Lemaître}}, \bibinfo
		{author} {\bibfnamefont {M.}~\bibnamefont {Morassi}}, \bibinfo {author}
		{\bibfnamefont {F.}~\bibnamefont {Baboux}}, \bibinfo {author} {\bibfnamefont
			{M.~I.}\ \bibnamefont {Amanti}},\ and\ \bibinfo {author} {\bibfnamefont
			{S.}~\bibnamefont {Ducci}},\ }\bibfield  {title} {\emph {\bibinfo {title}
			{{\color{blue}Nonlinear Quantum Photonics With AlGaAs Bragg-Reflection
					Waveguides}}},\ }\href@noop {} {\bibfield  {journal} {\bibinfo  {journal}
			{Journal of Lightwave Technology}\ ,\ \bibinfo {pages} {1}} (\bibinfo {year}
		{2022}{\natexlab{a}})}\BibitemShut {NoStop}%
	\bibitem [{\citenamefont {Flamini}\ \emph
		{et~al.}(2018{\natexlab{b}})\citenamefont {Flamini}, \citenamefont
		{Spagnolo},\ and\ \citenamefont {Sciarrino}}]{Flamini18}%
	\BibitemOpen
	\bibfield  {author} {\bibinfo {author} {\bibfnamefont {F.}~\bibnamefont
			{Flamini}}, \bibinfo {author} {\bibfnamefont {N.}~\bibnamefont {Spagnolo}},\
		and\ \bibinfo {author} {\bibfnamefont {F.}~\bibnamefont {Sciarrino}},\
	}\bibfield  {title} {\emph {\bibinfo {title} {{\color{blue}Photonic quantum
					information processing: a review}}},\ }\href@noop {} {\bibfield  {journal}
		{\bibinfo  {journal} {Reports on Progress in Physics}\ }\textbf {\bibinfo
			{volume} {82}},\ \bibinfo {pages} {016001} (\bibinfo {year}
		{2018}{\natexlab{b}})}\BibitemShut {NoStop}%
	\bibitem [{\citenamefont {Bogdanov}\ \emph {et~al.}(2017)\citenamefont
		{Bogdanov}, \citenamefont {Shalaginov}, \citenamefont {Boltasseva},\ and\
		\citenamefont {Shalaev}}]{Bogdanov2017}%
	\BibitemOpen
	\bibfield  {author} {\bibinfo {author} {\bibfnamefont {S.}~\bibnamefont
			{Bogdanov}}, \bibinfo {author} {\bibfnamefont {M.~Y.}\ \bibnamefont
			{Shalaginov}}, \bibinfo {author} {\bibfnamefont {A.}~\bibnamefont
			{Boltasseva}},\ and\ \bibinfo {author} {\bibfnamefont {V.~M.}\ \bibnamefont
			{Shalaev}},\ }\bibfield  {title} {\emph {\bibinfo {title}
			{{\color{blue}Material platforms for integrated quantum photonics}}},\
	}\href@noop {} {\bibfield  {journal} {\bibinfo  {journal} {Opt. Mater.
				Express}\ }\textbf {\bibinfo {volume} {7}},\ \bibinfo {pages} {111} (\bibinfo
		{year} {2017})}\BibitemShut {NoStop}%
	\bibitem [{\citenamefont {Pelucchi}\ \emph {et~al.}(2022)\citenamefont
		{Pelucchi}, \citenamefont {Fagas}, \citenamefont {Aharonovich}, \citenamefont
		{Englund}, \citenamefont {Figueroa}, \citenamefont {Gong}, \citenamefont
		{Hannes}, \citenamefont {Liu}, \citenamefont {Lu}, \citenamefont {Matsuda},
		\citenamefont {Pan}, \citenamefont {Schreck}, \citenamefont {Sciarrino},
		\citenamefont {Silberhorn}, \citenamefont {Wang},\ and\ \citenamefont
		{J{\"o}ns}}]{Pelucchi2022}%
	\BibitemOpen
	\bibfield  {author} {\bibinfo {author} {\bibfnamefont {E.}~\bibnamefont
			{Pelucchi}}, \bibinfo {author} {\bibfnamefont {G.}~\bibnamefont {Fagas}},
		\bibinfo {author} {\bibfnamefont {I.}~\bibnamefont {Aharonovich}}, \bibinfo
		{author} {\bibfnamefont {D.}~\bibnamefont {Englund}}, \bibinfo {author}
		{\bibfnamefont {E.}~\bibnamefont {Figueroa}}, \bibinfo {author}
		{\bibfnamefont {Q.}~\bibnamefont {Gong}}, \bibinfo {author} {\bibfnamefont
			{H.}~\bibnamefont {Hannes}}, \bibinfo {author} {\bibfnamefont
			{J.}~\bibnamefont {Liu}}, \bibinfo {author} {\bibfnamefont {C.-Y.}\
			\bibnamefont {Lu}}, \bibinfo {author} {\bibfnamefont {N.}~\bibnamefont
			{Matsuda}}, \bibinfo {author} {\bibfnamefont {J.-W.}\ \bibnamefont {Pan}},
		\bibinfo {author} {\bibfnamefont {F.}~\bibnamefont {Schreck}}, \bibinfo
		{author} {\bibfnamefont {F.}~\bibnamefont {Sciarrino}}, \bibinfo {author}
		{\bibfnamefont {C.}~\bibnamefont {Silberhorn}}, \bibinfo {author}
		{\bibfnamefont {J.}~\bibnamefont {Wang}},\ and\ \bibinfo {author}
		{\bibfnamefont {K.~D.}\ \bibnamefont {J{\"o}ns}},\ }\bibfield  {title} {\emph
		{\bibinfo {title} {{\color{blue}The potential and global outlook of
					integrated photonics for quantum technologies}}},\ }\href@noop {} {\bibfield
		{journal} {\bibinfo  {journal} {Nature Reviews Physics}\ }\textbf {\bibinfo
			{volume} {4}},\ \bibinfo {pages} {194} (\bibinfo {year} {2022})}\BibitemShut
	{NoStop}%
	\bibitem [{\citenamefont {Wang}\ \emph {et~al.}(2020)\citenamefont {Wang},
		\citenamefont {Sciarrino}, \citenamefont {Laing},\ and\ \citenamefont
		{Thompson}}]{Wang20}%
	\BibitemOpen
	\bibfield  {author} {\bibinfo {author} {\bibfnamefont {J.}~\bibnamefont
			{Wang}}, \bibinfo {author} {\bibfnamefont {F.}~\bibnamefont {Sciarrino}},
		\bibinfo {author} {\bibfnamefont {A.}~\bibnamefont {Laing}},\ and\ \bibinfo
		{author} {\bibfnamefont {M.~G.}\ \bibnamefont {Thompson}},\ }\bibfield
	{title} {\emph {\bibinfo {title} {{\color{blue}Integrated photonic quantum
					technologies}}},\ }\href@noop {} {\bibfield  {journal} {\bibinfo  {journal}
			{Nature Photonics}\ }\textbf {\bibinfo {volume} {14}},\ \bibinfo {pages}
		{273} (\bibinfo {year} {2020})}\BibitemShut {NoStop}%
	\bibitem [{\citenamefont {Kim}\ \emph {et~al.}(2020)\citenamefont {Kim},
		\citenamefont {Aghaeimeibodi}, \citenamefont {Carolan}, \citenamefont
		{Englund},\ and\ \citenamefont {Waks}}]{Kim20}%
	\BibitemOpen
	\bibfield  {author} {\bibinfo {author} {\bibfnamefont {J.-H.}\ \bibnamefont
			{Kim}}, \bibinfo {author} {\bibfnamefont {S.}~\bibnamefont {Aghaeimeibodi}},
		\bibinfo {author} {\bibfnamefont {J.}~\bibnamefont {Carolan}}, \bibinfo
		{author} {\bibfnamefont {D.}~\bibnamefont {Englund}},\ and\ \bibinfo {author}
		{\bibfnamefont {E.}~\bibnamefont {Waks}},\ }\bibfield  {title} {\emph
		{\bibinfo {title} {{\color{blue}Hybrid integration methods for on-chip
					quantum photonics}}},\ }\href@noop {} {\bibfield  {journal} {\bibinfo
			{journal} {Optica}\ }\textbf {\bibinfo {volume} {7}},\ \bibinfo {pages} {291}
		(\bibinfo {year} {2020})}\BibitemShut {NoStop}%
	\bibitem [{\citenamefont {Dietrich}\ \emph {et~al.}(2016)\citenamefont
		{Dietrich}, \citenamefont {Fiore}, \citenamefont {Thompson}, \citenamefont
		{Kamp},\ and\ \citenamefont {H{\"o}fling}}]{Dietrich16}%
	\BibitemOpen
	\bibfield  {author} {\bibinfo {author} {\bibfnamefont {C.~P.}\ \bibnamefont
			{Dietrich}}, \bibinfo {author} {\bibfnamefont {A.}~\bibnamefont {Fiore}},
		\bibinfo {author} {\bibfnamefont {M.~G.}\ \bibnamefont {Thompson}}, \bibinfo
		{author} {\bibfnamefont {M.}~\bibnamefont {Kamp}},\ and\ \bibinfo {author}
		{\bibfnamefont {S.}~\bibnamefont {H{\"o}fling}},\ }\bibfield  {title} {\emph
		{\bibinfo {title} {{\color{blue}{GaAs} integrated quantum photonics: Towards
					compact and multi-functional quantum photonic integrated circuits}}},\
	}\href@noop {} {\bibfield  {journal} {\bibinfo  {journal} {Laser \& Photonics
				Reviews}\ }\textbf {\bibinfo {volume} {10}},\ \bibinfo {pages} {870}
		(\bibinfo {year} {2016})}\BibitemShut {NoStop}%
	\bibitem [{\citenamefont {Orieux}\ \emph {et~al.}(2017)\citenamefont {Orieux},
		\citenamefont {Versteegh}, \citenamefont {J{\"o}ns},\ and\ \citenamefont
		{Ducci}}]{Orieux2017}%
	\BibitemOpen
	\bibfield  {author} {\bibinfo {author} {\bibfnamefont {A.}~\bibnamefont
			{Orieux}}, \bibinfo {author} {\bibfnamefont {M.~A.~M.}\ \bibnamefont
			{Versteegh}}, \bibinfo {author} {\bibfnamefont {K.~D.}\ \bibnamefont
			{J{\"o}ns}},\ and\ \bibinfo {author} {\bibfnamefont {S.}~\bibnamefont
			{Ducci}},\ }\bibfield  {title} {\emph {\bibinfo {title}
			{{\color{blue}Semiconductor devices for entangled photon pair generation: a
					review}}},\ }\href@noop {} {\bibfield  {journal} {\bibinfo  {journal}
			{Reports on Progress in Physics}\ }\textbf {\bibinfo {volume} {80}},\
		\bibinfo {pages} {076001} (\bibinfo {year} {2017})}\BibitemShut {NoStop}%
	\bibitem [{\citenamefont {McDonald}\ \emph {et~al.}(2019)\citenamefont
		{McDonald}, \citenamefont {Moody}, \citenamefont {Nam}, \citenamefont
		{Mirin}, \citenamefont {Shainline}, \citenamefont {McCaughan}, \citenamefont
		{Buckley},\ and\ \citenamefont {Silverman}}]{mcdonald2019iii}%
	\BibitemOpen
	\bibfield  {author} {\bibinfo {author} {\bibfnamefont {C.}~\bibnamefont
			{McDonald}}, \bibinfo {author} {\bibfnamefont {G.}~\bibnamefont {Moody}},
		\bibinfo {author} {\bibfnamefont {S.~W.}\ \bibnamefont {Nam}}, \bibinfo
		{author} {\bibfnamefont {R.~P.}\ \bibnamefont {Mirin}}, \bibinfo {author}
		{\bibfnamefont {J.~M.}\ \bibnamefont {Shainline}}, \bibinfo {author}
		{\bibfnamefont {A.}~\bibnamefont {McCaughan}}, \bibinfo {author}
		{\bibfnamefont {S.}~\bibnamefont {Buckley}},\ and\ \bibinfo {author}
		{\bibfnamefont {K.~L.}\ \bibnamefont {Silverman}},\ }\bibfield  {title}
	{\emph {\bibinfo {title} {{\color{blue}{III-V} photonic integrated circuit
					with waveguide-coupled light-emitting diodes and {WSi} superconducting
					single-photon detectors}}},\ }\href@noop {} {\bibfield  {journal} {\bibinfo
			{journal} {Applied Physics Letters}\ }\textbf {\bibinfo {volume} {115}},\
		\bibinfo {pages} {081105} (\bibinfo {year} {2019})}\BibitemShut {NoStop}%
	\bibitem [{\citenamefont {Schimpf}\ \emph {et~al.}(2021)\citenamefont
		{Schimpf}, \citenamefont {Reindl}, \citenamefont {Basso~Basset},
		\citenamefont {J{\"o}ns}, \citenamefont {Trotta},\ and\ \citenamefont
		{Rastelli}}]{Schimpf2021}%
	\BibitemOpen
	\bibfield  {author} {\bibinfo {author} {\bibfnamefont {C.}~\bibnamefont
			{Schimpf}}, \bibinfo {author} {\bibfnamefont {M.}~\bibnamefont {Reindl}},
		\bibinfo {author} {\bibfnamefont {F.}~\bibnamefont {Basso~Basset}}, \bibinfo
		{author} {\bibfnamefont {K.~D.}\ \bibnamefont {J{\"o}ns}}, \bibinfo {author}
		{\bibfnamefont {R.}~\bibnamefont {Trotta}},\ and\ \bibinfo {author}
		{\bibfnamefont {A.}~\bibnamefont {Rastelli}},\ }\bibfield  {title} {\emph
		{\bibinfo {title} {{\color{blue}Quantum dots as potential sources of strongly
					entangled photons: Perspectives and challenges for applications in quantum
					networks}}},\ }\href@noop {} {\bibfield  {journal} {\bibinfo  {journal}
			{Applied Physics Letters}\ }\textbf {\bibinfo {volume} {118}},\ \bibinfo
		{pages} {100502} (\bibinfo {year} {2021})}\BibitemShut {NoStop}%
	\bibitem [{\citenamefont {Somaschi}\ \emph {et~al.}(2016)\citenamefont
		{Somaschi}, \citenamefont {Giesz}, \citenamefont {De~Santis}, \citenamefont
		{Loredo}, \citenamefont {Almeida}, \citenamefont {Hornecker}, \citenamefont
		{Portalupi}, \citenamefont {Grange}, \citenamefont {Ant{\'o}n}, \citenamefont
		{Demory}, \citenamefont {G{\'o}mez}, \citenamefont {Sagnes}, \citenamefont
		{Lanzillotti-Kimura}, \citenamefont {Lema{\'\i}tre}, \citenamefont
		{Auffeves}, \citenamefont {White}, \citenamefont {Lanco},\ and\ \citenamefont
		{Senellart}}]{Somaschi2016}%
	\BibitemOpen
	\bibfield  {author} {\bibinfo {author} {\bibfnamefont {N.}~\bibnamefont
			{Somaschi}}, \bibinfo {author} {\bibfnamefont {V.}~\bibnamefont {Giesz}},
		\bibinfo {author} {\bibfnamefont {L.}~\bibnamefont {De~Santis}}, \bibinfo
		{author} {\bibfnamefont {J.~C.}\ \bibnamefont {Loredo}}, \bibinfo {author}
		{\bibfnamefont {M.~P.}\ \bibnamefont {Almeida}}, \bibinfo {author}
		{\bibfnamefont {G.}~\bibnamefont {Hornecker}}, \bibinfo {author}
		{\bibfnamefont {S.~L.}\ \bibnamefont {Portalupi}}, \bibinfo {author}
		{\bibfnamefont {T.}~\bibnamefont {Grange}}, \bibinfo {author} {\bibfnamefont
			{C.}~\bibnamefont {Ant{\'o}n}}, \bibinfo {author} {\bibfnamefont
			{J.}~\bibnamefont {Demory}}, \bibinfo {author} {\bibfnamefont
			{C.}~\bibnamefont {G{\'o}mez}}, \bibinfo {author} {\bibfnamefont
			{I.}~\bibnamefont {Sagnes}}, \bibinfo {author} {\bibfnamefont {N.~D.}\
			\bibnamefont {Lanzillotti-Kimura}}, \bibinfo {author} {\bibfnamefont
			{A.}~\bibnamefont {Lema{\'\i}tre}}, \bibinfo {author} {\bibfnamefont
			{A.}~\bibnamefont {Auffeves}}, \bibinfo {author} {\bibfnamefont {A.~G.}\
			\bibnamefont {White}}, \bibinfo {author} {\bibfnamefont {L.}~\bibnamefont
			{Lanco}},\ and\ \bibinfo {author} {\bibfnamefont {P.}~\bibnamefont
			{Senellart}},\ }\bibfield  {title} {\emph {\bibinfo {title}
			{{\color{blue}Near-optimal single-photon sources in the solid state}}},\
	}\href@noop {} {\bibfield  {journal} {\bibinfo  {journal} {Nature Photonics}\
		}\textbf {\bibinfo {volume} {10}},\ \bibinfo {pages} {340} (\bibinfo {year}
		{2016})}\BibitemShut {NoStop}%
	\bibitem [{\citenamefont {Moody}\ \emph {et~al.}(2020)\citenamefont {Moody},
		\citenamefont {Chang}, \citenamefont {Steiner},\ and\ \citenamefont
		{Bowers}}]{moody2020}%
	\BibitemOpen
	\bibfield  {author} {\bibinfo {author} {\bibfnamefont {G.}~\bibnamefont
			{Moody}}, \bibinfo {author} {\bibfnamefont {L.}~\bibnamefont {Chang}},
		\bibinfo {author} {\bibfnamefont {T.~J.}\ \bibnamefont {Steiner}},\ and\
		\bibinfo {author} {\bibfnamefont {J.~E.}\ \bibnamefont {Bowers}},\ }\bibfield
	{title} {\emph {\bibinfo {title} {{\color{blue}{Chip-Scale Nonlinear
						Photonics for Quantum Light Generation}}}},\ }\href@noop {} {\bibfield
		{journal} {\bibinfo  {journal} {AVS Quantum Science}\ }\textbf {\bibinfo
			{volume} {2}},\ \bibinfo {pages} {041702} (\bibinfo {year}
		{2020})}\BibitemShut {NoStop}%
	\bibitem [{\citenamefont {Michler}(2017)}]{michler2017quantum}%
	\BibitemOpen
	\bibfield  {author} {\bibinfo {author} {\bibfnamefont {P.}~\bibnamefont
			{Michler}},\ }\href@noop {} {\emph {\bibinfo {title} {Quantum Dots for
				Quantum Information Technologies}}},\ Nano-Optics and Nanophotonics\
	(\bibinfo  {publisher} {Springer International Publishing},\ \bibinfo {year}
	{2017})\BibitemShut {NoStop}%
	\bibitem [{\citenamefont {Margalit}\ \emph {et~al.}(2021)\citenamefont
		{Margalit}, \citenamefont {Xiang}, \citenamefont {Bowers}, \citenamefont
		{Bjorlin}, \citenamefont {Blum},\ and\ \citenamefont
		{Bowers}}]{margalit2021perspective}%
	\BibitemOpen
	\bibfield  {author} {\bibinfo {author} {\bibfnamefont {N.}~\bibnamefont
			{Margalit}}, \bibinfo {author} {\bibfnamefont {C.}~\bibnamefont {Xiang}},
		\bibinfo {author} {\bibfnamefont {S.~M.}\ \bibnamefont {Bowers}}, \bibinfo
		{author} {\bibfnamefont {A.}~\bibnamefont {Bjorlin}}, \bibinfo {author}
		{\bibfnamefont {R.}~\bibnamefont {Blum}},\ and\ \bibinfo {author}
		{\bibfnamefont {J.~E.}\ \bibnamefont {Bowers}},\ }\bibfield  {title} {\emph
		{\bibinfo {title} {{\color{blue}Perspective on the future of silicon
					photonics and electronics}}},\ }\href@noop {} {\bibfield  {journal} {\bibinfo
			{journal} {Applied Physics Letters}\ }\textbf {\bibinfo {volume} {118}},\
		\bibinfo {pages} {220501} (\bibinfo {year} {2021})}\BibitemShut {NoStop}%
	\bibitem [{\citenamefont {Moody}\ and\ \citenamefont
		{Islam}(2022)}]{moody2022materials}%
	\BibitemOpen
	\bibfield  {author} {\bibinfo {author} {\bibfnamefont {G.}~\bibnamefont
			{Moody}}\ and\ \bibinfo {author} {\bibfnamefont {M.~S.}\ \bibnamefont
			{Islam}},\ }\bibfield  {title} {\emph {\bibinfo {title}
			{{\color{blue}Materials for ultra-efficient, high-speed optoelectronics}}},\
	}\href@noop {} {\bibfield  {journal} {\bibinfo  {journal} {MRS Bulletin}\
		}\textbf {\bibinfo {volume} {47}},\ \bibinfo {pages} {475} (\bibinfo {year}
		{2022})}\BibitemShut {NoStop}%
	\bibitem [{\citenamefont {Adcock}\ \emph {et~al.}(2020)\citenamefont {Adcock},
		\citenamefont {Bao}, \citenamefont {Chi}, \citenamefont {Chen}, \citenamefont
		{Bacco}, \citenamefont {Gong}, \citenamefont {Oxenlowe}, \citenamefont
		{Wang},\ and\ \citenamefont {Ding}}]{Adcock2020}%
	\BibitemOpen
	\bibfield  {author} {\bibinfo {author} {\bibfnamefont {J.~C.}\ \bibnamefont
			{Adcock}}, \bibinfo {author} {\bibfnamefont {J.}~\bibnamefont {Bao}},
		\bibinfo {author} {\bibfnamefont {Y.}~\bibnamefont {Chi}}, \bibinfo {author}
		{\bibfnamefont {X.}~\bibnamefont {Chen}}, \bibinfo {author} {\bibfnamefont
			{D.}~\bibnamefont {Bacco}}, \bibinfo {author} {\bibfnamefont
			{Q.}~\bibnamefont {Gong}}, \bibinfo {author} {\bibfnamefont {L.~K.}\
			\bibnamefont {Oxenlowe}}, \bibinfo {author} {\bibfnamefont {J.}~\bibnamefont
			{Wang}},\ and\ \bibinfo {author} {\bibfnamefont {Y.}~\bibnamefont {Ding}},\
	}\bibfield  {title} {\emph {\bibinfo {title} {{\color{blue}{Advances in
						Silicon Quantum Photonics}}}},\ }\href@noop {} {\bibfield  {journal}
		{\bibinfo  {journal} {IEEE Journal of Selected Topics in Quantum
				Electronics}\ }\textbf {\bibinfo {volume} {27}},\ \bibinfo {pages} {1}
		(\bibinfo {year} {2020})}\BibitemShut {NoStop}%
	\bibitem [{\citenamefont {Wang}\ \emph
		{et~al.}(2018{\natexlab{a}})\citenamefont {Wang}, \citenamefont {Paesani},
		\citenamefont {Ding}, \citenamefont {Santagati}, \citenamefont {Skrzypczyk},
		\citenamefont {Salavrakos}, \citenamefont {Tura}, \citenamefont {Augusiak},
		\citenamefont {Man{\v{c}}inska}, \citenamefont {Bacco}, \citenamefont
		{Bonneau}, \citenamefont {Silverstone}, \citenamefont {Gong}, \citenamefont
		{Ac{\'{i}}n}, \citenamefont {Rottwitt}, \citenamefont {Oxenl{\o}we},
		\citenamefont {O'Brien}, \citenamefont {Laing},\ and\ \citenamefont
		{Thompson}}]{Wang2018}%
	\BibitemOpen
	\bibfield  {author} {\bibinfo {author} {\bibfnamefont {J.}~\bibnamefont
			{Wang}}, \bibinfo {author} {\bibfnamefont {S.}~\bibnamefont {Paesani}},
		\bibinfo {author} {\bibfnamefont {Y.}~\bibnamefont {Ding}}, \bibinfo {author}
		{\bibfnamefont {R.}~\bibnamefont {Santagati}}, \bibinfo {author}
		{\bibfnamefont {P.}~\bibnamefont {Skrzypczyk}}, \bibinfo {author}
		{\bibfnamefont {A.}~\bibnamefont {Salavrakos}}, \bibinfo {author}
		{\bibfnamefont {J.}~\bibnamefont {Tura}}, \bibinfo {author} {\bibfnamefont
			{R.}~\bibnamefont {Augusiak}}, \bibinfo {author} {\bibfnamefont
			{L.}~\bibnamefont {Man{\v{c}}inska}}, \bibinfo {author} {\bibfnamefont
			{D.}~\bibnamefont {Bacco}}, \bibinfo {author} {\bibfnamefont
			{D.}~\bibnamefont {Bonneau}}, \bibinfo {author} {\bibfnamefont {J.~W.}\
			\bibnamefont {Silverstone}}, \bibinfo {author} {\bibfnamefont
			{Q.}~\bibnamefont {Gong}}, \bibinfo {author} {\bibfnamefont {A.}~\bibnamefont
			{Ac{\'{i}}n}}, \bibinfo {author} {\bibfnamefont {K.}~\bibnamefont
			{Rottwitt}}, \bibinfo {author} {\bibfnamefont {L.~K.}\ \bibnamefont
			{Oxenl{\o}we}}, \bibinfo {author} {\bibfnamefont {J.~L.}\ \bibnamefont
			{O'Brien}}, \bibinfo {author} {\bibfnamefont {A.}~\bibnamefont {Laing}},\
		and\ \bibinfo {author} {\bibfnamefont {M.~G.}\ \bibnamefont {Thompson}},\
	}\bibfield  {title} {\emph {\bibinfo {title} {{\color{blue}{Multidimensional
						quantum entanglement with large-scale integrated optics}}}},\ }\href@noop {}
	{\bibfield  {journal} {\bibinfo  {journal} {Science}\ }\textbf {\bibinfo
			{volume} {360}},\ \bibinfo {pages} {285} (\bibinfo {year}
		{2018}{\natexlab{a}})}\BibitemShut {NoStop}%
	\bibitem [{\citenamefont {Mobini}\ \emph {et~al.}(2022)\citenamefont {Mobini},
		\citenamefont {Espinosa}, \citenamefont {Vyas},\ and\ \citenamefont
		{Dolgaleva}}]{Mobini2022}%
	\BibitemOpen
	\bibfield  {author} {\bibinfo {author} {\bibfnamefont {E.}~\bibnamefont
			{Mobini}}, \bibinfo {author} {\bibfnamefont {D.~H.~G.}\ \bibnamefont
			{Espinosa}}, \bibinfo {author} {\bibfnamefont {K.}~\bibnamefont {Vyas}},\
		and\ \bibinfo {author} {\bibfnamefont {K.}~\bibnamefont {Dolgaleva}},\
	}\bibfield  {title} {\emph {\bibinfo {title} {{\color{blue}{AlGaAs} Nonlinear
					Integrated Photonics}}},\ }\href@noop {} {\bibfield  {journal} {\bibinfo
			{journal} {Micromachines}\ }\textbf {\bibinfo {volume} {13}} (\bibinfo {year}
		{2022})}\BibitemShut {NoStop}%
	\bibitem [{\citenamefont {Liao}\ and\ \citenamefont
		{Aitchison}(2017)}]{Liao17b}%
	\BibitemOpen
	\bibfield  {author} {\bibinfo {author} {\bibfnamefont {Z.}~\bibnamefont
			{Liao}}\ and\ \bibinfo {author} {\bibfnamefont {J.~S.}\ \bibnamefont
			{Aitchison}},\ }\bibfield  {title} {\emph {\bibinfo {title}
			{{\color{blue}Precision etching for multi-level {AlGaAs} waveguides}}},\
	}\href@noop {} {\bibfield  {journal} {\bibinfo  {journal} {Optical Materials
				Express}\ }\textbf {\bibinfo {volume} {7}},\ \bibinfo {pages} {895} (\bibinfo
		{year} {2017})}\BibitemShut {NoStop}%
	\bibitem [{\citenamefont {Ottaviano}\ \emph {et~al.}(2016)\citenamefont
		{Ottaviano}, \citenamefont {Pu}, \citenamefont {Semenova},\ and\
		\citenamefont {Yvind}}]{Semenova2016}%
	\BibitemOpen
	\bibfield  {author} {\bibinfo {author} {\bibfnamefont {L.}~\bibnamefont
			{Ottaviano}}, \bibinfo {author} {\bibfnamefont {M.}~\bibnamefont {Pu}},
		\bibinfo {author} {\bibfnamefont {E.}~\bibnamefont {Semenova}},\ and\
		\bibinfo {author} {\bibfnamefont {K.}~\bibnamefont {Yvind}},\ }\bibfield
	{title} {\emph {\bibinfo {title} {{\color{blue}Low-loss high-confinement
					waveguides and microring resonators in {AlGaAs}-on-insulator}}},\ }\href@noop
	{} {\bibfield  {journal} {\bibinfo  {journal} {Optics Letters}\ }\textbf
		{\bibinfo {volume} {41}},\ \bibinfo {pages} {3996} (\bibinfo {year}
		{2016})}\BibitemShut {NoStop}%
	\bibitem [{\citenamefont {Chang}\ \emph {et~al.}(2022)\citenamefont {Chang},
		\citenamefont {Cole}, \citenamefont {Moody},\ and\ \citenamefont
		{Bowers}}]{ChangCSOI}%
	\BibitemOpen
	\bibfield  {author} {\bibinfo {author} {\bibfnamefont {L.}~\bibnamefont
			{Chang}}, \bibinfo {author} {\bibfnamefont {G.~D.}\ \bibnamefont {Cole}},
		\bibinfo {author} {\bibfnamefont {G.}~\bibnamefont {Moody}},\ and\ \bibinfo
		{author} {\bibfnamefont {J.~E.}\ \bibnamefont {Bowers}},\ }\bibfield  {title}
	{\emph {\bibinfo {title} {{\color{blue}CSOI: Beyond Silicon-on-Insulator
					Photonics}}},\ }\href@noop {} {\bibfield  {journal} {\bibinfo  {journal}
			{Optics and Photonics News}\ }\textbf {\bibinfo {volume} {33}},\ \bibinfo
		{pages} {24} (\bibinfo {year} {2022})}\BibitemShut {NoStop}%
	\bibitem [{\citenamefont {Gao}\ \emph {et~al.}(2021)\citenamefont {Gao},
		\citenamefont {Zhang}, \citenamefont {Bo}, \citenamefont {Fang},
		\citenamefont {Hao}, \citenamefont {Yao}, \citenamefont {Lin}, \citenamefont
		{Guan}, \citenamefont {Deng}, \citenamefont {Wang} \emph
		{et~al.}}]{gao2021broadband}%
	\BibitemOpen
	\bibfield  {author} {\bibinfo {author} {\bibfnamefont {R.}~\bibnamefont
			{Gao}}, \bibinfo {author} {\bibfnamefont {H.}~\bibnamefont {Zhang}}, \bibinfo
		{author} {\bibfnamefont {F.}~\bibnamefont {Bo}}, \bibinfo {author}
		{\bibfnamefont {W.}~\bibnamefont {Fang}}, \bibinfo {author} {\bibfnamefont
			{Z.}~\bibnamefont {Hao}}, \bibinfo {author} {\bibfnamefont {N.}~\bibnamefont
			{Yao}}, \bibinfo {author} {\bibfnamefont {J.}~\bibnamefont {Lin}}, \bibinfo
		{author} {\bibfnamefont {J.}~\bibnamefont {Guan}}, \bibinfo {author}
		{\bibfnamefont {L.}~\bibnamefont {Deng}}, \bibinfo {author} {\bibfnamefont
			{M.}~\bibnamefont {Wang}}, \emph {et~al.},\ }\bibfield  {title} {\emph
		{\bibinfo {title} {{\color{blue}Broadband highly efficient nonlinear optical
					processes in on-chip integrated lithium niobate microdisk resonators of
					Q-factor above 108}}},\ }\href@noop {} {\bibfield  {journal} {\bibinfo
			{journal} {New Journal of Physics}\ }\textbf {\bibinfo {volume} {23}},\
		\bibinfo {pages} {123027} (\bibinfo {year} {2021})}\BibitemShut {NoStop}%
	\bibitem [{\citenamefont {Gao}\ \emph {et~al.}(2022)\citenamefont {Gao},
		\citenamefont {Yao}, \citenamefont {Guan}, \citenamefont {Deng},
		\citenamefont {Lin}, \citenamefont {Wang}, \citenamefont {Qiao},
		\citenamefont {Fang},\ and\ \citenamefont {Cheng}}]{gao2022lithium}%
	\BibitemOpen
	\bibfield  {author} {\bibinfo {author} {\bibfnamefont {R.}~\bibnamefont
			{Gao}}, \bibinfo {author} {\bibfnamefont {N.}~\bibnamefont {Yao}}, \bibinfo
		{author} {\bibfnamefont {J.}~\bibnamefont {Guan}}, \bibinfo {author}
		{\bibfnamefont {L.}~\bibnamefont {Deng}}, \bibinfo {author} {\bibfnamefont
			{J.}~\bibnamefont {Lin}}, \bibinfo {author} {\bibfnamefont {M.}~\bibnamefont
			{Wang}}, \bibinfo {author} {\bibfnamefont {L.}~\bibnamefont {Qiao}}, \bibinfo
		{author} {\bibfnamefont {W.}~\bibnamefont {Fang}},\ and\ \bibinfo {author}
		{\bibfnamefont {Y.}~\bibnamefont {Cheng}},\ }\bibfield  {title} {\emph
		{\bibinfo {title} {{\color{blue}Lithium niobate microring with ultra-high Q
					factor above 10 8}}},\ }\href@noop {} {\bibfield  {journal} {\bibinfo
			{journal} {Chinese Optics Letters}\ }\textbf {\bibinfo {volume} {20}},\
		\bibinfo {pages} {011902} (\bibinfo {year} {2022})}\BibitemShut {NoStop}%
	\bibitem [{\citenamefont {Liu}\ \emph {et~al.}(2022)\citenamefont {Liu},
		\citenamefont {Jin}, \citenamefont {Cheng}, \citenamefont {Chauhan},
		\citenamefont {Puckett}, \citenamefont {Nelson}, \citenamefont {Behunin},
		\citenamefont {Rakich},\ and\ \citenamefont {Blumenthal}}]{liu2022ultralow}%
	\BibitemOpen
	\bibfield  {author} {\bibinfo {author} {\bibfnamefont {K.}~\bibnamefont
			{Liu}}, \bibinfo {author} {\bibfnamefont {N.}~\bibnamefont {Jin}}, \bibinfo
		{author} {\bibfnamefont {H.}~\bibnamefont {Cheng}}, \bibinfo {author}
		{\bibfnamefont {N.}~\bibnamefont {Chauhan}}, \bibinfo {author} {\bibfnamefont
			{M.~W.}\ \bibnamefont {Puckett}}, \bibinfo {author} {\bibfnamefont {K.~D.}\
			\bibnamefont {Nelson}}, \bibinfo {author} {\bibfnamefont {R.~O.}\
			\bibnamefont {Behunin}}, \bibinfo {author} {\bibfnamefont {P.~T.}\
			\bibnamefont {Rakich}},\ and\ \bibinfo {author} {\bibfnamefont {D.~J.}\
			\bibnamefont {Blumenthal}},\ }\bibfield  {title} {\emph {\bibinfo {title}
			{{\color{blue}Ultralow 0.034 dB/m loss wafer-scale integrated photonics
					realizing 720 million Q and 380 $\mu$W threshold Brillouin lasing}}},\
	}\href@noop {} {\bibfield  {journal} {\bibinfo  {journal} {Optics letters}\
		}\textbf {\bibinfo {volume} {47}},\ \bibinfo {pages} {1855} (\bibinfo {year}
		{2022})}\BibitemShut {NoStop}%
	\bibitem [{\citenamefont {B{\"u}hler}\ \emph {et~al.}(2022)\citenamefont
		{B{\"u}hler}, \citenamefont {Wei{\ss}}, \citenamefont {Crespo-Poveda},
		\citenamefont {Nysten}, \citenamefont {Finley}, \citenamefont {M{\"u}ller},
		\citenamefont {Santos}, \citenamefont {Lima~Jr},\ and\ \citenamefont
		{Krenner}}]{buhler2022chip}%
	\BibitemOpen
	\bibfield  {author} {\bibinfo {author} {\bibfnamefont {D.~D.}\ \bibnamefont
			{B{\"u}hler}}, \bibinfo {author} {\bibfnamefont {M.}~\bibnamefont
			{Wei{\ss}}}, \bibinfo {author} {\bibfnamefont {A.}~\bibnamefont
			{Crespo-Poveda}}, \bibinfo {author} {\bibfnamefont {E.~D.}\ \bibnamefont
			{Nysten}}, \bibinfo {author} {\bibfnamefont {J.~J.}\ \bibnamefont {Finley}},
		\bibinfo {author} {\bibfnamefont {K.}~\bibnamefont {M{\"u}ller}}, \bibinfo
		{author} {\bibfnamefont {P.~V.}\ \bibnamefont {Santos}}, \bibinfo {author}
		{\bibfnamefont {M.~M.~d.}\ \bibnamefont {Lima~Jr}},\ and\ \bibinfo {author}
		{\bibfnamefont {H.~J.}\ \bibnamefont {Krenner}},\ }\bibfield  {title} {\emph
		{\bibinfo {title} {{\color{blue}On-chip piezo-optomechanical dynamic single
					photon routing and rotation of a photonic qubit}}},\ }\href@noop {}
	{\bibfield  {journal} {\bibinfo  {journal} {arXiv preprint arXiv:2202.10173}\
		} (\bibinfo {year} {2022})}\BibitemShut {NoStop}%
	\bibitem [{\citenamefont {Spickermann}\ \emph {et~al.}(1996)\citenamefont
		{Spickermann}, \citenamefont {Sakamoto}, \citenamefont {Peters},\ and\
		\citenamefont {Dagli}}]{Spickermann96}%
	\BibitemOpen
	\bibfield  {author} {\bibinfo {author} {\bibfnamefont {R.}~\bibnamefont
			{Spickermann}}, \bibinfo {author} {\bibfnamefont {S.}~\bibnamefont
			{Sakamoto}}, \bibinfo {author} {\bibfnamefont {M.}~\bibnamefont {Peters}},\
		and\ \bibinfo {author} {\bibfnamefont {N.}~\bibnamefont {Dagli}},\ }\bibfield
	{title} {\emph {\bibinfo {title} {{\color{blue}{GaAs/AlGaAs} travelling wave
					electro-optic modulator with an electrical bandwidth {$>$ 40 GHz}}}},\
	}\href@noop {} {\bibfield  {journal} {\bibinfo  {journal} {Electronics
				Letters}\ }\textbf {\bibinfo {volume} {32}},\ \bibinfo {pages} {1095}
		(\bibinfo {year} {1996})}\BibitemShut {NoStop}%
	\bibitem [{\citenamefont {Norman}\ \emph {et~al.}(2018)\citenamefont {Norman},
		\citenamefont {Jung}, \citenamefont {Wan},\ and\ \citenamefont
		{Bowers}}]{norman2018perspective}%
	\BibitemOpen
	\bibfield  {author} {\bibinfo {author} {\bibfnamefont {J.~C.}\ \bibnamefont
			{Norman}}, \bibinfo {author} {\bibfnamefont {D.}~\bibnamefont {Jung}},
		\bibinfo {author} {\bibfnamefont {Y.}~\bibnamefont {Wan}},\ and\ \bibinfo
		{author} {\bibfnamefont {J.~E.}\ \bibnamefont {Bowers}},\ }\bibfield  {title}
	{\emph {\bibinfo {title} {{\color{blue}Perspective: The future of quantum dot
					photonic integrated circuits}}},\ }\href@noop {} {\bibfield  {journal}
		{\bibinfo  {journal} {APL photonics}\ }\textbf {\bibinfo {volume} {3}},\
		\bibinfo {pages} {030901} (\bibinfo {year} {2018})}\BibitemShut {NoStop}%
	\bibitem [{\citenamefont {Liang}\ and\ \citenamefont
		{Bowers}(2021)}]{liang2021recent}%
	\BibitemOpen
	\bibfield  {author} {\bibinfo {author} {\bibfnamefont {D.}~\bibnamefont
			{Liang}}\ and\ \bibinfo {author} {\bibfnamefont {J.~E.}\ \bibnamefont
			{Bowers}},\ }\bibfield  {title} {\emph {\bibinfo {title} {{\color{blue}Recent
					progress in heterogeneous {III-V-on-silicon} photonic integration}}},\
	}\href@noop {} {\bibfield  {journal} {\bibinfo  {journal} {Light: Advanced
				Manufacturing}\ }\textbf {\bibinfo {volume} {2}},\ \bibinfo {pages} {59}
		(\bibinfo {year} {2021})}\BibitemShut {NoStop}%
	\bibitem [{\citenamefont {Senellart}\ \emph {et~al.}(2017)\citenamefont
		{Senellart}, \citenamefont {Solomon},\ and\ \citenamefont
		{White}}]{Senellart17}%
	\BibitemOpen
	\bibfield  {author} {\bibinfo {author} {\bibfnamefont {P.}~\bibnamefont
			{Senellart}}, \bibinfo {author} {\bibfnamefont {G.}~\bibnamefont {Solomon}},\
		and\ \bibinfo {author} {\bibfnamefont {A.}~\bibnamefont {White}},\ }\bibfield
	{title} {\emph {\bibinfo {title} {{\color{blue}High-performance
					semiconductor quantum-dot single-photon sources}}},\ }\href@noop {}
	{\bibfield  {journal} {\bibinfo  {journal} {Nature Nanotechnology}\ }\textbf
		{\bibinfo {volume} {12}},\ \bibinfo {pages} {1026} (\bibinfo {year}
		{2017})}\BibitemShut {NoStop}%
	\bibitem [{\citenamefont {Taillaert}\ \emph {et~al.}(2002)\citenamefont
		{Taillaert}, \citenamefont {Bogaerts}, \citenamefont {Bienstman},
		\citenamefont {Krauss}, \citenamefont {Van~Daele}, \citenamefont {Moerman},
		\citenamefont {Verstuyft}, \citenamefont {De~Mesel},\ and\ \citenamefont
		{Baets}}]{Taillaert02}%
	\BibitemOpen
	\bibfield  {author} {\bibinfo {author} {\bibfnamefont {D.}~\bibnamefont
			{Taillaert}}, \bibinfo {author} {\bibfnamefont {W.}~\bibnamefont {Bogaerts}},
		\bibinfo {author} {\bibfnamefont {P.}~\bibnamefont {Bienstman}}, \bibinfo
		{author} {\bibfnamefont {T.~F.}\ \bibnamefont {Krauss}}, \bibinfo {author}
		{\bibfnamefont {P.}~\bibnamefont {Van~Daele}}, \bibinfo {author}
		{\bibfnamefont {I.}~\bibnamefont {Moerman}}, \bibinfo {author} {\bibfnamefont
			{S.}~\bibnamefont {Verstuyft}}, \bibinfo {author} {\bibfnamefont
			{K.}~\bibnamefont {De~Mesel}},\ and\ \bibinfo {author} {\bibfnamefont
			{R.}~\bibnamefont {Baets}},\ }\bibfield  {title} {\emph {\bibinfo {title}
			{{\color{blue}An out-of-plane grating coupler for efficient butt-coupling
					between compact planar waveguides and single-mode fibers}}},\ }\href@noop {}
	{\bibfield  {journal} {\bibinfo  {journal} {IEEE Journal of Quantum
				Electronics}\ }\textbf {\bibinfo {volume} {38}},\ \bibinfo {pages} {949}
		(\bibinfo {year} {2002})}\BibitemShut {NoStop}%
	\bibitem [{\citenamefont {Almeida}\ \emph {et~al.}(2003)\citenamefont
		{Almeida}, \citenamefont {Panepucci},\ and\ \citenamefont
		{Lipson}}]{Almeida03}%
	\BibitemOpen
	\bibfield  {author} {\bibinfo {author} {\bibfnamefont {V.~R.}\ \bibnamefont
			{Almeida}}, \bibinfo {author} {\bibfnamefont {R.~R.}\ \bibnamefont
			{Panepucci}},\ and\ \bibinfo {author} {\bibfnamefont {M.}~\bibnamefont
			{Lipson}},\ }\bibfield  {title} {\emph {\bibinfo {title}
			{{\color{blue}Nanotaper for compact mode conversion}}},\ }\href@noop {}
	{\bibfield  {journal} {\bibinfo  {journal} {Optics letters}\ }\textbf
		{\bibinfo {volume} {28}},\ \bibinfo {pages} {1302} (\bibinfo {year}
		{2003})}\BibitemShut {NoStop}%
	\bibitem [{\citenamefont {Liao}\ \emph
		{et~al.}(2017{\natexlab{b}})\citenamefont {Liao}, \citenamefont {Wagner},
		\citenamefont {Alam}, \citenamefont {Tolstikhin},\ and\ \citenamefont
		{Aitchison}}]{Liao17}%
	\BibitemOpen
	\bibfield  {author} {\bibinfo {author} {\bibfnamefont {Z.}~\bibnamefont
			{Liao}}, \bibinfo {author} {\bibfnamefont {S.}~\bibnamefont {Wagner}},
		\bibinfo {author} {\bibfnamefont {M.}~\bibnamefont {Alam}}, \bibinfo {author}
		{\bibfnamefont {V.}~\bibnamefont {Tolstikhin}},\ and\ \bibinfo {author}
		{\bibfnamefont {J.~S.}\ \bibnamefont {Aitchison}},\ }\bibfield  {title}
	{\emph {\bibinfo {title} {{\color{blue}Vertically integrated spot-size
					converter in {AlGaAs-GaAs}}}},\ }\href@noop {} {\bibfield  {journal}
		{\bibinfo  {journal} {Optics Letters}\ }\textbf {\bibinfo {volume} {42}},\
		\bibinfo {pages} {4167} (\bibinfo {year} {2017}{\natexlab{b}})}\BibitemShut
	{NoStop}%
	\bibitem [{\citenamefont {Gr{\"o}blacher}\ \emph {et~al.}(2013)\citenamefont
		{Gr{\"o}blacher}, \citenamefont {Hill}, \citenamefont {Safavi-Naeini},
		\citenamefont {Chan},\ and\ \citenamefont {Painter}}]{Groblacher13}%
	\BibitemOpen
	\bibfield  {author} {\bibinfo {author} {\bibfnamefont {S.}~\bibnamefont
			{Gr{\"o}blacher}}, \bibinfo {author} {\bibfnamefont {J.~T.}\ \bibnamefont
			{Hill}}, \bibinfo {author} {\bibfnamefont {A.~H.}\ \bibnamefont
			{Safavi-Naeini}}, \bibinfo {author} {\bibfnamefont {J.}~\bibnamefont
			{Chan}},\ and\ \bibinfo {author} {\bibfnamefont {O.}~\bibnamefont
			{Painter}},\ }\bibfield  {title} {\emph {\bibinfo {title}
			{{\color{blue}Highly efficient coupling from an optical fiber to a nanoscale
					silicon optomechanical cavity}}},\ }\href@noop {} {\bibfield  {journal}
		{\bibinfo  {journal} {Applied Physics Letters}\ }\textbf {\bibinfo {volume}
			{103}},\ \bibinfo {pages} {181104} (\bibinfo {year} {2013})}\BibitemShut
	{NoStop}%
	\bibitem [{\citenamefont {Appas}\ \emph
		{et~al.}(2022{\natexlab{b}})\citenamefont {Appas}, \citenamefont {Meskine},
		\citenamefont {Lema{\^\i}tre}, \citenamefont {Palomo}, \citenamefont
		{Baboux}, \citenamefont {Amanti},\ and\ \citenamefont {Ducci}}]{Appas22}%
	\BibitemOpen
	\bibfield  {author} {\bibinfo {author} {\bibfnamefont {F.}~\bibnamefont
			{Appas}}, \bibinfo {author} {\bibfnamefont {O.}~\bibnamefont {Meskine}},
		\bibinfo {author} {\bibfnamefont {A.}~\bibnamefont {Lema{\^\i}tre}}, \bibinfo
		{author} {\bibfnamefont {J.}~\bibnamefont {Palomo}}, \bibinfo {author}
		{\bibfnamefont {F.}~\bibnamefont {Baboux}}, \bibinfo {author} {\bibfnamefont
			{M.~I.}\ \bibnamefont {Amanti}},\ and\ \bibinfo {author} {\bibfnamefont
			{S.}~\bibnamefont {Ducci}},\ }\bibfield  {title} {\emph {\bibinfo {title}
			{{\color{blue}Broadband biphoton generation and polarization splitting in a
					monolithic {AlGaAs} chip}}},\ }\href@noop {} {\bibfield  {journal} {\bibinfo
			{journal} {ArXiv:2208.14108}\ } (\bibinfo {year}
		{2022}{\natexlab{b}})}\BibitemShut {NoStop}%
	\bibitem [{\citenamefont {Mu}\ \emph {et~al.}(2020)\citenamefont {Mu},
		\citenamefont {Wu}, \citenamefont {Cheng},\ and\ \citenamefont
		{Fu}}]{Mu2020}%
	\BibitemOpen
	\bibfield  {author} {\bibinfo {author} {\bibfnamefont {X.}~\bibnamefont
			{Mu}}, \bibinfo {author} {\bibfnamefont {S.}~\bibnamefont {Wu}}, \bibinfo
		{author} {\bibfnamefont {L.}~\bibnamefont {Cheng}},\ and\ \bibinfo {author}
		{\bibfnamefont {H.~Y.}\ \bibnamefont {Fu}},\ }\bibfield  {title} {\emph
		{\bibinfo {title} {{\color{blue}{Edge couplers in silicon photonic integrated
						circuits: A review}}}},\ }\href@noop {} {\bibfield  {journal} {\bibinfo
			{journal} {Applied Sciences (Switzerland)}\ }\textbf {\bibinfo {volume}
			{10}},\ \bibinfo {pages} {1538} (\bibinfo {year} {2020})}\BibitemShut
	{NoStop}%
	\bibitem [{\citenamefont {Ramelow}\ \emph {et~al.}(2015)\citenamefont
		{Ramelow}, \citenamefont {Farsi}, \citenamefont {Clemmen}, \citenamefont
		{Orquiza}, \citenamefont {Luke}, \citenamefont {Lipson},\ and\ \citenamefont
		{Gaeta}}]{Ramelow2015}%
	\BibitemOpen
	\bibfield  {author} {\bibinfo {author} {\bibfnamefont {S.}~\bibnamefont
			{Ramelow}}, \bibinfo {author} {\bibfnamefont {A.}~\bibnamefont {Farsi}},
		\bibinfo {author} {\bibfnamefont {S.}~\bibnamefont {Clemmen}}, \bibinfo
		{author} {\bibfnamefont {D.}~\bibnamefont {Orquiza}}, \bibinfo {author}
		{\bibfnamefont {K.}~\bibnamefont {Luke}}, \bibinfo {author} {\bibfnamefont
			{M.}~\bibnamefont {Lipson}},\ and\ \bibinfo {author} {\bibfnamefont {A.~L.}\
			\bibnamefont {Gaeta}},\ }\bibfield  {title} {\emph {\bibinfo {title}
			{{\color{blue}Silicon-nitride platform for narrowband entangled photon
					generation}}},\ }\href@noop {} {\bibfield  {journal} {\bibinfo  {journal}
			{arXiv preprint arXiv:1508.04358}\ } (\bibinfo {year} {2015})}\BibitemShut
	{NoStop}%
	\bibitem [{\citenamefont {Chen}\ \emph {et~al.}(2022)\citenamefont {Chen},
		\citenamefont {Ruan}, \citenamefont {Fan}, \citenamefont {Wang},
		\citenamefont {Liu}, \citenamefont {Li}, \citenamefont {Chen},\ and\
		\citenamefont {Liu}}]{chen2022low}%
	\BibitemOpen
	\bibfield  {author} {\bibinfo {author} {\bibfnamefont {B.}~\bibnamefont
			{Chen}}, \bibinfo {author} {\bibfnamefont {Z.}~\bibnamefont {Ruan}}, \bibinfo
		{author} {\bibfnamefont {X.}~\bibnamefont {Fan}}, \bibinfo {author}
		{\bibfnamefont {Z.}~\bibnamefont {Wang}}, \bibinfo {author} {\bibfnamefont
			{J.}~\bibnamefont {Liu}}, \bibinfo {author} {\bibfnamefont {C.}~\bibnamefont
			{Li}}, \bibinfo {author} {\bibfnamefont {K.}~\bibnamefont {Chen}},\ and\
		\bibinfo {author} {\bibfnamefont {L.}~\bibnamefont {Liu}},\ }\bibfield
	{title} {\emph {\bibinfo {title} {{\color{blue}Low-loss fiber grating coupler
					on thin film lithium niobate platform}}},\ }\href@noop {} {\bibfield
		{journal} {\bibinfo  {journal} {APL Photonics}\ }\textbf {\bibinfo {volume}
			{7}},\ \bibinfo {pages} {076103} (\bibinfo {year} {2022})}\BibitemShut
	{NoStop}%
	\bibitem [{\citenamefont {Zhao}\ \emph {et~al.}(2020)\citenamefont {Zhao},
		\citenamefont {Kusolthossakul},\ and\ \citenamefont {Fang}}]{zhao2020high}%
	\BibitemOpen
	\bibfield  {author} {\bibinfo {author} {\bibfnamefont {M.}~\bibnamefont
			{Zhao}}, \bibinfo {author} {\bibfnamefont {W.}~\bibnamefont
			{Kusolthossakul}},\ and\ \bibinfo {author} {\bibfnamefont {K.}~\bibnamefont
			{Fang}},\ }\bibfield  {title} {\emph {\bibinfo {title}
			{{\color{blue}High-efficiency fiber-to-chip interface for aluminum nitride
					quantum photonics}}},\ }\href@noop {} {\bibfield  {journal} {\bibinfo
			{journal} {OSA Continuum}\ }\textbf {\bibinfo {volume} {3}},\ \bibinfo
		{pages} {952} (\bibinfo {year} {2020})}\BibitemShut {NoStop}%
	\bibitem [{\citenamefont {Sanchis}\ \emph {et~al.}(2009)\citenamefont
		{Sanchis}, \citenamefont {Villalba}, \citenamefont {Cuesta}, \citenamefont
		{H{\aa}kansson}, \citenamefont {Griol}, \citenamefont {Gal{\'{a}}n},
		\citenamefont {Brimont},\ and\ \citenamefont {Mart{\'{i}}}}]{Sanchis2009}%
	\BibitemOpen
	\bibfield  {author} {\bibinfo {author} {\bibfnamefont {P.}~\bibnamefont
			{Sanchis}}, \bibinfo {author} {\bibfnamefont {P.}~\bibnamefont {Villalba}},
		\bibinfo {author} {\bibfnamefont {F.}~\bibnamefont {Cuesta}}, \bibinfo
		{author} {\bibfnamefont {A.}~\bibnamefont {H{\aa}kansson}}, \bibinfo {author}
		{\bibfnamefont {A.}~\bibnamefont {Griol}}, \bibinfo {author} {\bibfnamefont
			{J.~V.}\ \bibnamefont {Gal{\'{a}}n}}, \bibinfo {author} {\bibfnamefont
			{A.}~\bibnamefont {Brimont}},\ and\ \bibinfo {author} {\bibfnamefont
			{J.}~\bibnamefont {Mart{\'{i}}}},\ }\bibfield  {title} {\emph {\bibinfo
			{title} {{\color{blue}{Highly efficient crossing structure for
						silicon-on-insulator waveguides}}}},\ }\href@noop {} {\bibfield  {journal}
		{\bibinfo  {journal} {Optics Letters}\ }\textbf {\bibinfo {volume} {34}},\
		\bibinfo {pages} {2760} (\bibinfo {year} {2009})}\BibitemShut {NoStop}%
	\bibitem [{\citenamefont {Yang}\ \emph {et~al.}(2019)\citenamefont {Yang},
		\citenamefont {Zheng}, \citenamefont {Hu}, \citenamefont {Zhang},
		\citenamefont {Yun},\ and\ \citenamefont {Cui}}]{Yang2019}%
	\BibitemOpen
	\bibfield  {author} {\bibinfo {author} {\bibfnamefont {H.}~\bibnamefont
			{Yang}}, \bibinfo {author} {\bibfnamefont {P.}~\bibnamefont {Zheng}},
		\bibinfo {author} {\bibfnamefont {G.}~\bibnamefont {Hu}}, \bibinfo {author}
		{\bibfnamefont {R.}~\bibnamefont {Zhang}}, \bibinfo {author} {\bibfnamefont
			{B.}~\bibnamefont {Yun}},\ and\ \bibinfo {author} {\bibfnamefont
			{Y.}~\bibnamefont {Cui}},\ }\bibfield  {title} {\emph {\bibinfo {title}
			{{\color{blue}{A broadband, low-crosstalk and low polarization dependent
						silicon nitride waveguide crossing based on the multimode-interference}}}},\
	}\href@noop {} {\bibfield  {journal} {\bibinfo  {journal} {Optics
				Communications}\ }\textbf {\bibinfo {volume} {450}},\ \bibinfo {pages} {28}
		(\bibinfo {year} {2019})}\BibitemShut {NoStop}%
	\bibitem [{\citenamefont {Zhang}\ \emph
		{et~al.}(2022{\natexlab{a}})\citenamefont {Zhang}, \citenamefont {Sun},
		\citenamefont {Chen}, \citenamefont {Feng}, \citenamefont {Zhang},
		\citenamefont {Chen},\ and\ \citenamefont {Wang}}]{zhang2022power}%
	\BibitemOpen
	\bibfield  {author} {\bibinfo {author} {\bibfnamefont {K.}~\bibnamefont
			{Zhang}}, \bibinfo {author} {\bibfnamefont {W.}~\bibnamefont {Sun}}, \bibinfo
		{author} {\bibfnamefont {Y.}~\bibnamefont {Chen}}, \bibinfo {author}
		{\bibfnamefont {H.}~\bibnamefont {Feng}}, \bibinfo {author} {\bibfnamefont
			{Y.}~\bibnamefont {Zhang}}, \bibinfo {author} {\bibfnamefont
			{Z.}~\bibnamefont {Chen}},\ and\ \bibinfo {author} {\bibfnamefont
			{C.}~\bibnamefont {Wang}},\ }\bibfield  {title} {\emph {\bibinfo {title}
			{{\color{blue}A power-efficient integrated lithium niobate electro-optic comb
					generator}}},\ }\href@noop {} {\bibfield  {journal} {\bibinfo  {journal}
			{arXiv preprint arXiv:2208.09603}\ } (\bibinfo {year}
		{2022}{\natexlab{a}})}\BibitemShut {NoStop}%
	\bibitem [{\citenamefont {Wang}\ \emph {et~al.}(2016)\citenamefont {Wang},
		\citenamefont {Bonneau}, \citenamefont {Villa}, \citenamefont {Silverstone},
		\citenamefont {Santagati}, \citenamefont {Miki}, \citenamefont {Yamashita},
		\citenamefont {Fujiwara}, \citenamefont {Sasaki}, \citenamefont {Terai},
		\citenamefont {Tanner}, \citenamefont {Natarajan}, \citenamefont {Hadfield},
		\citenamefont {O’Brien},\ and\ \citenamefont {Thompson}}]{Wang2016}%
	\BibitemOpen
	\bibfield  {author} {\bibinfo {author} {\bibfnamefont {J.}~\bibnamefont
			{Wang}}, \bibinfo {author} {\bibfnamefont {D.}~\bibnamefont {Bonneau}},
		\bibinfo {author} {\bibfnamefont {M.}~\bibnamefont {Villa}}, \bibinfo
		{author} {\bibfnamefont {J.~W.}\ \bibnamefont {Silverstone}}, \bibinfo
		{author} {\bibfnamefont {R.}~\bibnamefont {Santagati}}, \bibinfo {author}
		{\bibfnamefont {S.}~\bibnamefont {Miki}}, \bibinfo {author} {\bibfnamefont
			{T.}~\bibnamefont {Yamashita}}, \bibinfo {author} {\bibfnamefont
			{M.}~\bibnamefont {Fujiwara}}, \bibinfo {author} {\bibfnamefont
			{M.}~\bibnamefont {Sasaki}}, \bibinfo {author} {\bibfnamefont
			{H.}~\bibnamefont {Terai}}, \bibinfo {author} {\bibfnamefont {M.~G.}\
			\bibnamefont {Tanner}}, \bibinfo {author} {\bibfnamefont {C.~M.}\
			\bibnamefont {Natarajan}}, \bibinfo {author} {\bibfnamefont {R.~H.}\
			\bibnamefont {Hadfield}}, \bibinfo {author} {\bibfnamefont {J.~L.}\
			\bibnamefont {O’Brien}},\ and\ \bibinfo {author} {\bibfnamefont {M.~G.}\
			\bibnamefont {Thompson}},\ }\bibfield  {title} {\emph {\bibinfo {title}
			{{\color{blue}Chip-to-chip quantum photonic interconnect by path-polarization
					interconversion}}},\ }\href@noop {} {\bibfield  {journal} {\bibinfo
			{journal} {Optica}\ }\textbf {\bibinfo {volume} {3}},\ \bibinfo {pages} {407}
		(\bibinfo {year} {2016})}\BibitemShut {NoStop}%
	\bibitem [{\citenamefont {Rao}\ \emph {et~al.}(2021)\citenamefont {Rao},
		\citenamefont {Moille}, \citenamefont {Lu}, \citenamefont {Westly},
		\citenamefont {Geiselmann}, \citenamefont {Zervas},\ and\ \citenamefont
		{Srinivasan}}]{Rao2021}%
	\BibitemOpen
	\bibfield  {author} {\bibinfo {author} {\bibfnamefont {A.}~\bibnamefont
			{Rao}}, \bibinfo {author} {\bibfnamefont {G.}~\bibnamefont {Moille}},
		\bibinfo {author} {\bibfnamefont {X.}~\bibnamefont {Lu}}, \bibinfo {author}
		{\bibfnamefont {D.}~\bibnamefont {Westly}}, \bibinfo {author} {\bibfnamefont
			{M.}~\bibnamefont {Geiselmann}}, \bibinfo {author} {\bibfnamefont
			{M.}~\bibnamefont {Zervas}},\ and\ \bibinfo {author} {\bibfnamefont
			{K.}~\bibnamefont {Srinivasan}},\ }\bibfield  {title} {\emph {\bibinfo
			{title} {{\color{blue}Up to 50 dB Extinction in Broadband Single-Stage
					Thermo- Optic Mach-Zehnder Interferometers for Programmable Low-Loss Silicon
					Nitride Photonic Circuits}}},\ }\href@noop {} {\bibfield  {journal} {\bibinfo
			{journal} {Conference on Lasers and Electro-Optics (2021), paper SM1A.7}\ ,\
			\bibinfo {pages} {SM1A.7}} (\bibinfo {year} {2021})}\BibitemShut {NoStop}%
	\bibitem [{\citenamefont {He}\ \emph {et~al.}(2019)\citenamefont {He},
		\citenamefont {Xu}, \citenamefont {Ren}, \citenamefont {Jian}, \citenamefont
		{Ruan}, \citenamefont {Xu}, \citenamefont {Gao}, \citenamefont {Sun},
		\citenamefont {Wen}, \citenamefont {Zhou} \emph {et~al.}}]{he2019high}%
	\BibitemOpen
	\bibfield  {author} {\bibinfo {author} {\bibfnamefont {M.}~\bibnamefont
			{He}}, \bibinfo {author} {\bibfnamefont {M.}~\bibnamefont {Xu}}, \bibinfo
		{author} {\bibfnamefont {Y.}~\bibnamefont {Ren}}, \bibinfo {author}
		{\bibfnamefont {J.}~\bibnamefont {Jian}}, \bibinfo {author} {\bibfnamefont
			{Z.}~\bibnamefont {Ruan}}, \bibinfo {author} {\bibfnamefont {Y.}~\bibnamefont
			{Xu}}, \bibinfo {author} {\bibfnamefont {S.}~\bibnamefont {Gao}}, \bibinfo
		{author} {\bibfnamefont {S.}~\bibnamefont {Sun}}, \bibinfo {author}
		{\bibfnamefont {X.}~\bibnamefont {Wen}}, \bibinfo {author} {\bibfnamefont
			{L.}~\bibnamefont {Zhou}}, \emph {et~al.},\ }\bibfield  {title} {\emph
		{\bibinfo {title} {{\color{blue}High-performance hybrid silicon and lithium
					niobate Mach--Zehnder modulators for 100 Gbit s- 1 and beyond}}},\
	}\href@noop {} {\bibfield  {journal} {\bibinfo  {journal} {Nature Photonics}\
		}\textbf {\bibinfo {volume} {13}},\ \bibinfo {pages} {359} (\bibinfo {year}
		{2019})}\BibitemShut {NoStop}%
	\bibitem [{\citenamefont {Zhu}\ and\ \citenamefont
		{Lo}(2016)}]{zhu2016aluminum}%
	\BibitemOpen
	\bibfield  {author} {\bibinfo {author} {\bibfnamefont {S.}~\bibnamefont
			{Zhu}}\ and\ \bibinfo {author} {\bibfnamefont {G.-Q.}\ \bibnamefont {Lo}},\
	}\bibfield  {title} {\emph {\bibinfo {title} {{\color{blue}Aluminum nitride
					electro-optic phase shifter for backend integration on silicon}}},\
	}\href@noop {} {\bibfield  {journal} {\bibinfo  {journal} {Optics express}\
		}\textbf {\bibinfo {volume} {24}},\ \bibinfo {pages} {12501} (\bibinfo {year}
		{2016})}\BibitemShut {NoStop}%
	\bibitem [{\citenamefont {Lee}\ \emph {et~al.}(2019)\citenamefont {Lee},
		\citenamefont {Lee}, \citenamefont {Kim},\ and\ \citenamefont
		{Ju}}]{Lee2019}%
	\BibitemOpen
	\bibfield  {author} {\bibinfo {author} {\bibfnamefont {J.~M.}\ \bibnamefont
			{Lee}}, \bibinfo {author} {\bibfnamefont {W.~J.}\ \bibnamefont {Lee}},
		\bibinfo {author} {\bibfnamefont {M.~S.}\ \bibnamefont {Kim}},\ and\ \bibinfo
		{author} {\bibfnamefont {J.~J.}\ \bibnamefont {Ju}},\ }\bibfield  {title}
	{\emph {\bibinfo {title} {{\color{blue}Noise Filtering for Highly Correlated
					Photon Pairs from Silicon Waveguides}}},\ }\href@noop {} {\bibfield
		{journal} {\bibinfo  {journal} {Journal of Lightwave Technology}\ }\textbf
		{\bibinfo {volume} {37}},\ \bibinfo {pages} {5428} (\bibinfo {year}
		{2019})}\BibitemShut {NoStop}%
	\bibitem [{\citenamefont {Lee}\ \emph {et~al.}(2017)\citenamefont {Lee},
		\citenamefont {Kim}, \citenamefont {Ahn}, \citenamefont {Adelmini},
		\citenamefont {Fowler}, \citenamefont {Kopp}, \citenamefont {Oton},\ and\
		\citenamefont {Testa}}]{Lee2017}%
	\BibitemOpen
	\bibfield  {author} {\bibinfo {author} {\bibfnamefont {J.-M.}\ \bibnamefont
			{Lee}}, \bibinfo {author} {\bibfnamefont {M.-S.}\ \bibnamefont {Kim}},
		\bibinfo {author} {\bibfnamefont {J.~T.}\ \bibnamefont {Ahn}}, \bibinfo
		{author} {\bibfnamefont {L.}~\bibnamefont {Adelmini}}, \bibinfo {author}
		{\bibfnamefont {D.}~\bibnamefont {Fowler}}, \bibinfo {author} {\bibfnamefont
			{C.}~\bibnamefont {Kopp}}, \bibinfo {author} {\bibfnamefont {C.~J.}\
			\bibnamefont {Oton}},\ and\ \bibinfo {author} {\bibfnamefont
			{F.}~\bibnamefont {Testa}},\ }\bibfield  {title} {\emph {\bibinfo {title}
			{{\color{blue}Demonstration and fabrication tolerance study of
					temperature-insensitive silicon-photonic {MZI} tunable by a metal heater}}},\
	}\href@noop {} {\bibfield  {journal} {\bibinfo  {journal} {J. Lightwave
				Technol.}\ }\textbf {\bibinfo {volume} {35}},\ \bibinfo {pages} {4903}
		(\bibinfo {year} {2017})}\BibitemShut {NoStop}%
	\bibitem [{\citenamefont {Lee}\ \emph {et~al.}(2022)\citenamefont {Lee},
		\citenamefont {Lee}, \citenamefont {Kim}, \citenamefont {Cho}, \citenamefont
		{Ju}, \citenamefont {Navickaite},\ and\ \citenamefont {Fernandez}}]{Lee2022}%
	\BibitemOpen
	\bibfield  {author} {\bibinfo {author} {\bibfnamefont {J.~M.}\ \bibnamefont
			{Lee}}, \bibinfo {author} {\bibfnamefont {W.~J.}\ \bibnamefont {Lee}},
		\bibinfo {author} {\bibfnamefont {M.~S.}\ \bibnamefont {Kim}}, \bibinfo
		{author} {\bibfnamefont {S.~W.}\ \bibnamefont {Cho}}, \bibinfo {author}
		{\bibfnamefont {J.~J.}\ \bibnamefont {Ju}}, \bibinfo {author} {\bibfnamefont
			{G.}~\bibnamefont {Navickaite}},\ and\ \bibinfo {author} {\bibfnamefont
			{J.}~\bibnamefont {Fernandez}},\ }\bibfield  {title} {\emph {\bibinfo {title}
			{{\color{blue}Controlled-NOT operation of SiN-photonic circuit using photon
					pairs from silicon-photonic circuit}}},\ }\href@noop {} {\bibfield  {journal}
		{\bibinfo  {journal} {Optics Communications}\ }\textbf {\bibinfo {volume}
			{509}},\ \bibinfo {pages} {127863} (\bibinfo {year} {2022})}\BibitemShut
	{NoStop}%
	\bibitem [{\citenamefont {Liu}\ \emph {et~al.}(2020)\citenamefont {Liu},
		\citenamefont {Ying}, \citenamefont {Zhong}, \citenamefont {Xu},
		\citenamefont {Han}, \citenamefont {Yu},\ and\ \citenamefont
		{Cai}}]{liu2020highly}%
	\BibitemOpen
	\bibfield  {author} {\bibinfo {author} {\bibfnamefont {X.}~\bibnamefont
			{Liu}}, \bibinfo {author} {\bibfnamefont {P.}~\bibnamefont {Ying}}, \bibinfo
		{author} {\bibfnamefont {X.}~\bibnamefont {Zhong}}, \bibinfo {author}
		{\bibfnamefont {J.}~\bibnamefont {Xu}}, \bibinfo {author} {\bibfnamefont
			{Y.}~\bibnamefont {Han}}, \bibinfo {author} {\bibfnamefont {S.}~\bibnamefont
			{Yu}},\ and\ \bibinfo {author} {\bibfnamefont {X.}~\bibnamefont {Cai}},\
	}\bibfield  {title} {\emph {\bibinfo {title} {{\color{blue}Highly efficient
					thermo-optic tunable micro-ring resonator based on an {LNOI} platform}}},\
	}\href@noop {} {\bibfield  {journal} {\bibinfo  {journal} {Optics letters}\
		}\textbf {\bibinfo {volume} {45}},\ \bibinfo {pages} {6318} (\bibinfo {year}
		{2020})}\BibitemShut {NoStop}%
	\bibitem [{\citenamefont {Shin}\ \emph {et~al.}(2021)\citenamefont {Shin},
		\citenamefont {Sun}, \citenamefont {Soltani},\ and\ \citenamefont
		{Mi}}]{shin2021demonstration}%
	\BibitemOpen
	\bibfield  {author} {\bibinfo {author} {\bibfnamefont {W.}~\bibnamefont
			{Shin}}, \bibinfo {author} {\bibfnamefont {Y.}~\bibnamefont {Sun}}, \bibinfo
		{author} {\bibfnamefont {M.}~\bibnamefont {Soltani}},\ and\ \bibinfo {author}
		{\bibfnamefont {Z.}~\bibnamefont {Mi}},\ }\bibfield  {title} {\emph {\bibinfo
			{title} {{\color{blue}Demonstration of green and {UV} wavelength high {Q}
					aluminum nitride on sapphire microring resonators integrated with
					microheaters}}},\ }\href@noop {} {\bibfield  {journal} {\bibinfo  {journal}
			{Applied Physics Letters}\ }\textbf {\bibinfo {volume} {118}},\ \bibinfo
		{pages} {211103} (\bibinfo {year} {2021})}\BibitemShut {NoStop}%
	\bibitem [{\citenamefont {Carolan}\ \emph {et~al.}(2015)\citenamefont
		{Carolan}, \citenamefont {Harrold}, \citenamefont {Sparrow}, \citenamefont
		{Mart{\'\i}n-L{\'o}pez}, \citenamefont {Russell}, \citenamefont
		{Silverstone}, \citenamefont {Shadbolt}, \citenamefont {Matsuda},
		\citenamefont {Oguma}, \citenamefont {Itoh} \emph
		{et~al.}}]{carolan2015universal}%
	\BibitemOpen
	\bibfield  {author} {\bibinfo {author} {\bibfnamefont {J.}~\bibnamefont
			{Carolan}}, \bibinfo {author} {\bibfnamefont {C.}~\bibnamefont {Harrold}},
		\bibinfo {author} {\bibfnamefont {C.}~\bibnamefont {Sparrow}}, \bibinfo
		{author} {\bibfnamefont {E.}~\bibnamefont {Mart{\'\i}n-L{\'o}pez}}, \bibinfo
		{author} {\bibfnamefont {N.~J.}\ \bibnamefont {Russell}}, \bibinfo {author}
		{\bibfnamefont {J.~W.}\ \bibnamefont {Silverstone}}, \bibinfo {author}
		{\bibfnamefont {P.~J.}\ \bibnamefont {Shadbolt}}, \bibinfo {author}
		{\bibfnamefont {N.}~\bibnamefont {Matsuda}}, \bibinfo {author} {\bibfnamefont
			{M.}~\bibnamefont {Oguma}}, \bibinfo {author} {\bibfnamefont
			{M.}~\bibnamefont {Itoh}}, \emph {et~al.},\ }\bibfield  {title} {\emph
		{\bibinfo {title} {{\color{blue}Universal linear optics}}},\ }\href@noop {}
	{\bibfield  {journal} {\bibinfo  {journal} {Science}\ }\textbf {\bibinfo
			{volume} {349}},\ \bibinfo {pages} {711} (\bibinfo {year}
		{2015})}\BibitemShut {NoStop}%
	\bibitem [{\citenamefont {O'brien}(2007)}]{o2007optical}%
	\BibitemOpen
	\bibfield  {author} {\bibinfo {author} {\bibfnamefont {J.~L.}\ \bibnamefont
			{O'brien}},\ }\bibfield  {title} {\emph {\bibinfo {title}
			{{\color{blue}Optical quantum computing}}},\ }\href@noop {} {\bibfield
		{journal} {\bibinfo  {journal} {Science}\ }\textbf {\bibinfo {volume}
			{318}},\ \bibinfo {pages} {1567} (\bibinfo {year} {2007})}\BibitemShut
	{NoStop}%
	\bibitem [{\citenamefont {Stanton}\ \emph {et~al.}(2020)\citenamefont
		{Stanton}, \citenamefont {Chiles}, \citenamefont {Nader}, \citenamefont
		{Moody}, \citenamefont {Volet}, \citenamefont {Chang}, \citenamefont
		{Bowers}, \citenamefont {{Woo Nam}},\ and\ \citenamefont
		{Mirin}}]{Stanton2020}%
	\BibitemOpen
	\bibfield  {author} {\bibinfo {author} {\bibfnamefont {E.~J.}\ \bibnamefont
			{Stanton}}, \bibinfo {author} {\bibfnamefont {J.}~\bibnamefont {Chiles}},
		\bibinfo {author} {\bibfnamefont {N.}~\bibnamefont {Nader}}, \bibinfo
		{author} {\bibfnamefont {G.}~\bibnamefont {Moody}}, \bibinfo {author}
		{\bibfnamefont {N.}~\bibnamefont {Volet}}, \bibinfo {author} {\bibfnamefont
			{L.}~\bibnamefont {Chang}}, \bibinfo {author} {\bibfnamefont {J.~E.}\
			\bibnamefont {Bowers}}, \bibinfo {author} {\bibfnamefont {S.}~\bibnamefont
			{{Woo Nam}}},\ and\ \bibinfo {author} {\bibfnamefont {R.~P.}\ \bibnamefont
			{Mirin}},\ }\bibfield  {title} {\emph {\bibinfo {title}
			{{\color{blue}{Efficient second harmonic generation in nanophotonic
						GaAs-on-insulator waveguides}}}},\ }\href@noop {} {\bibfield  {journal}
		{\bibinfo  {journal} {Optics Express}\ }\textbf {\bibinfo {volume} {28}},\
		\bibinfo {pages} {9521} (\bibinfo {year} {2020})}\BibitemShut {NoStop}%
	\bibitem [{\citenamefont {Bienfang}\ \emph {et~al.}(2022)\citenamefont
		{Bienfang}, \citenamefont {Zwiller},\ and\ \citenamefont
		{Steinhauer}}]{bienfang2022materials}%
	\BibitemOpen
	\bibfield  {author} {\bibinfo {author} {\bibfnamefont {J.~C.}\ \bibnamefont
			{Bienfang}}, \bibinfo {author} {\bibfnamefont {V.}~\bibnamefont {Zwiller}},\
		and\ \bibinfo {author} {\bibfnamefont {S.}~\bibnamefont {Steinhauer}},\
	}\bibfield  {title} {\emph {\bibinfo {title} {{\color{blue}Materials,
					devices, and systems for high-speed single-photon counting}}},\ }\href@noop
	{} {\bibfield  {journal} {\bibinfo  {journal} {MRS Bulletin}\ ,\ \bibinfo
			{pages} {1}} (\bibinfo {year} {2022})}\BibitemShut {NoStop}%
	\bibitem [{\citenamefont {Khan}\ \emph {et~al.}(2020)\citenamefont {Khan},
		\citenamefont {Buckley}, \citenamefont {Chiles}, \citenamefont {Mirin},
		\citenamefont {Nam},\ and\ \citenamefont {Shainline}}]{khan2020low}%
	\BibitemOpen
	\bibfield  {author} {\bibinfo {author} {\bibfnamefont {S.}~\bibnamefont
			{Khan}}, \bibinfo {author} {\bibfnamefont {S.~M.}\ \bibnamefont {Buckley}},
		\bibinfo {author} {\bibfnamefont {J.}~\bibnamefont {Chiles}}, \bibinfo
		{author} {\bibfnamefont {R.~P.}\ \bibnamefont {Mirin}}, \bibinfo {author}
		{\bibfnamefont {S.~W.}\ \bibnamefont {Nam}},\ and\ \bibinfo {author}
		{\bibfnamefont {J.~M.}\ \bibnamefont {Shainline}},\ }\bibfield  {title}
	{\emph {\bibinfo {title} {{\color{blue}Low-loss, high-bandwidth fiber-to-chip
					coupling using capped adiabatic tapered fibers}}},\ }\href@noop {} {\bibfield
		{journal} {\bibinfo  {journal} {APL Photonics}\ }\textbf {\bibinfo {volume}
			{5}},\ \bibinfo {pages} {056101} (\bibinfo {year} {2020})}\BibitemShut
	{NoStop}%
	\bibitem [{\citenamefont {Shin}\ and\ \citenamefont
		{Dagli}(2013)}]{shin2013ultralow}%
	\BibitemOpen
	\bibfield  {author} {\bibinfo {author} {\bibfnamefont {J.~H.}\ \bibnamefont
			{Shin}}\ and\ \bibinfo {author} {\bibfnamefont {N.}~\bibnamefont {Dagli}},\
	}\bibfield  {title} {\emph {\bibinfo {title} {{\color{blue}Ultralow drive
					voltage substrate removed {GaAs/AlGaAs} electro-optic modulators at 1550
					nm}}},\ }\href@noop {} {\bibfield  {journal} {\bibinfo  {journal} {IEEE
				Journal of Selected Topics in Quantum Electronics}\ }\textbf {\bibinfo
			{volume} {19}},\ \bibinfo {pages} {150} (\bibinfo {year} {2013})}\BibitemShut
	{NoStop}%
	\bibitem [{\citenamefont {Pintus}\ \emph {et~al.}(2022)\citenamefont {Pintus},
		\citenamefont {Ranzani}, \citenamefont {Pinna}, \citenamefont {Huang},
		\citenamefont {Gustafsson}, \citenamefont {Karinou}, \citenamefont {Casula},
		\citenamefont {Shoji}, \citenamefont {Takamura}, \citenamefont {Mizumoto}
		\emph {et~al.}}]{pintus2022integrated}%
	\BibitemOpen
	\bibfield  {author} {\bibinfo {author} {\bibfnamefont {P.}~\bibnamefont
			{Pintus}}, \bibinfo {author} {\bibfnamefont {L.}~\bibnamefont {Ranzani}},
		\bibinfo {author} {\bibfnamefont {S.}~\bibnamefont {Pinna}}, \bibinfo
		{author} {\bibfnamefont {D.}~\bibnamefont {Huang}}, \bibinfo {author}
		{\bibfnamefont {M.~V.}\ \bibnamefont {Gustafsson}}, \bibinfo {author}
		{\bibfnamefont {F.}~\bibnamefont {Karinou}}, \bibinfo {author} {\bibfnamefont
			{G.~A.}\ \bibnamefont {Casula}}, \bibinfo {author} {\bibfnamefont
			{Y.}~\bibnamefont {Shoji}}, \bibinfo {author} {\bibfnamefont
			{Y.}~\bibnamefont {Takamura}}, \bibinfo {author} {\bibfnamefont
			{T.}~\bibnamefont {Mizumoto}}, \emph {et~al.},\ }\bibfield  {title} {\emph
		{\bibinfo {title} {{\color{blue}An integrated magneto-optic modulator for
					cryogenic applications}}},\ }\href@noop {} {\bibfield  {journal} {\bibinfo
			{journal} {Nature Electronics}\ ,\ \bibinfo {pages} {1}} (\bibinfo {year}
		{2022})}\BibitemShut {NoStop}%
	\bibitem [{\citenamefont {Khurana}\ \emph {et~al.}(2022)\citenamefont
		{Khurana}, \citenamefont {Jiang},\ and\ \citenamefont
		{Balram}}]{khurana2022piezo}%
	\BibitemOpen
	\bibfield  {author} {\bibinfo {author} {\bibfnamefont {A.}~\bibnamefont
			{Khurana}}, \bibinfo {author} {\bibfnamefont {P.}~\bibnamefont {Jiang}},\
		and\ \bibinfo {author} {\bibfnamefont {K.~C.}\ \bibnamefont {Balram}},\
	}\bibfield  {title} {\emph {\bibinfo {title}
			{{\color{blue}Piezo-optomechanical signal transduction using Lamb wave
					supermodes in a suspended Gallium Arsenide photonic integrated circuits
					platform}}},\ }\href@noop {} {\bibfield  {journal} {\bibinfo  {journal}
			{arXiv preprint arXiv:2203.09328}\ } (\bibinfo {year} {2022})}\BibitemShut
	{NoStop}%
	\bibitem [{\citenamefont {Dogru}\ \emph {et~al.}(2013)\citenamefont {Dogru},
		\citenamefont {Shin},\ and\ \citenamefont {Dagli}}]{dogru2013electrodes}%
	\BibitemOpen
	\bibfield  {author} {\bibinfo {author} {\bibfnamefont {S.}~\bibnamefont
			{Dogru}}, \bibinfo {author} {\bibfnamefont {J.~H.}\ \bibnamefont {Shin}},\
		and\ \bibinfo {author} {\bibfnamefont {N.}~\bibnamefont {Dagli}},\ }\bibfield
	{title} {\emph {\bibinfo {title} {{\color{blue}Electrodes for wide-bandwidth
					substrate-removed electro-optic modulators}}},\ }\href@noop {} {\bibfield
		{journal} {\bibinfo  {journal} {Optics Letters}\ }\textbf {\bibinfo {volume}
			{38}},\ \bibinfo {pages} {914} (\bibinfo {year} {2013})}\BibitemShut
	{NoStop}%
	\bibitem [{\citenamefont {Wang}\ \emph
		{et~al.}(2018{\natexlab{b}})\citenamefont {Wang}, \citenamefont {Zhang},
		\citenamefont {Chen}, \citenamefont {Bertrand}, \citenamefont {Shams-Ansari},
		\citenamefont {Chandrasekhar}, \citenamefont {Winzer},\ and\ \citenamefont
		{Lon{\v{c}}ar}}]{wang2018integrated}%
	\BibitemOpen
	\bibfield  {author} {\bibinfo {author} {\bibfnamefont {C.}~\bibnamefont
			{Wang}}, \bibinfo {author} {\bibfnamefont {M.}~\bibnamefont {Zhang}},
		\bibinfo {author} {\bibfnamefont {X.}~\bibnamefont {Chen}}, \bibinfo {author}
		{\bibfnamefont {M.}~\bibnamefont {Bertrand}}, \bibinfo {author}
		{\bibfnamefont {A.}~\bibnamefont {Shams-Ansari}}, \bibinfo {author}
		{\bibfnamefont {S.}~\bibnamefont {Chandrasekhar}}, \bibinfo {author}
		{\bibfnamefont {P.}~\bibnamefont {Winzer}},\ and\ \bibinfo {author}
		{\bibfnamefont {M.}~\bibnamefont {Lon{\v{c}}ar}},\ }\bibfield  {title} {\emph
		{\bibinfo {title} {{\color{blue}Integrated lithium niobate electro-optic
					modulators operating at {CMOS}-compatible voltages}}},\ }\href@noop {}
	{\bibfield  {journal} {\bibinfo  {journal} {Nature}\ }\textbf {\bibinfo
			{volume} {562}},\ \bibinfo {pages} {101} (\bibinfo {year}
		{2018}{\natexlab{b}})}\BibitemShut {NoStop}%
	\bibitem [{\citenamefont {Wang}\ \emph {et~al.}(2021)\citenamefont {Wang},
		\citenamefont {J{\"o}ns},\ and\ \citenamefont {Sun}}]{Wang2021}%
	\BibitemOpen
	\bibfield  {author} {\bibinfo {author} {\bibfnamefont {Y.}~\bibnamefont
			{Wang}}, \bibinfo {author} {\bibfnamefont {K.~D.}\ \bibnamefont {J{\"o}ns}},\
		and\ \bibinfo {author} {\bibfnamefont {Z.}~\bibnamefont {Sun}},\ }\bibfield
	{title} {\emph {\bibinfo {title} {{\color{blue}Integrated photon-pair sources
					with nonlinear optics}}},\ }\href@noop {} {\bibfield  {journal} {\bibinfo
			{journal} {Applied Physics Reviews}\ }\textbf {\bibinfo {volume} {8}},\
		\bibinfo {pages} {011314} (\bibinfo {year} {2021})}\BibitemShut {NoStop}%
	\bibitem [{\citenamefont {Midwinter}\ and\ \citenamefont
		{Warner}(1965)}]{Midwinter65}%
	\BibitemOpen
	\bibfield  {author} {\bibinfo {author} {\bibfnamefont {J.}~\bibnamefont
			{Midwinter}}\ and\ \bibinfo {author} {\bibfnamefont {J.}~\bibnamefont
			{Warner}},\ }\bibfield  {title} {\emph {\bibinfo {title} {{\color{blue}The
					effects of phase matching method and of uniaxial crystal symmetry on the
					polar distribution of second-order non-linear optical polarization}}},\
	}\href@noop {} {\bibfield  {journal} {\bibinfo  {journal} {British Journal of
				Applied Physics}\ }\textbf {\bibinfo {volume} {16}},\ \bibinfo {pages} {1135}
		(\bibinfo {year} {1965})}\BibitemShut {NoStop}%
	\bibitem [{\citenamefont {Ducci}\ \emph {et~al.}(2004)\citenamefont {Ducci},
		\citenamefont {Lanco}, \citenamefont {Berger}, \citenamefont {De~Rossi},
		\citenamefont {Ortiz},\ and\ \citenamefont {Calligaro}}]{Ducci04}%
	\BibitemOpen
	\bibfield  {author} {\bibinfo {author} {\bibfnamefont {S.}~\bibnamefont
			{Ducci}}, \bibinfo {author} {\bibfnamefont {L.}~\bibnamefont {Lanco}},
		\bibinfo {author} {\bibfnamefont {V.}~\bibnamefont {Berger}}, \bibinfo
		{author} {\bibfnamefont {A.}~\bibnamefont {De~Rossi}}, \bibinfo {author}
		{\bibfnamefont {V.}~\bibnamefont {Ortiz}},\ and\ \bibinfo {author}
		{\bibfnamefont {M.}~\bibnamefont {Calligaro}},\ }\bibfield  {title} {\emph
		{\bibinfo {title} {{\color{blue}Continuous-wave second-harmonic generation in
					modal phase matched semiconductor waveguides}}},\ }\href@noop {} {\bibfield
		{journal} {\bibinfo  {journal} {Applied Physics Letters}\ }\textbf {\bibinfo
			{volume} {84}},\ \bibinfo {pages} {2974} (\bibinfo {year}
		{2004})}\BibitemShut {NoStop}%
	\bibitem [{\citenamefont {Fiore}\ \emph {et~al.}(1998)\citenamefont {Fiore},
		\citenamefont {Janz}, \citenamefont {Delobel}, \citenamefont {Van~der Meer},
		\citenamefont {Bravetti}, \citenamefont {Berger}, \citenamefont {Rosencher},\
		and\ \citenamefont {Nagle}}]{Fiore98}%
	\BibitemOpen
	\bibfield  {author} {\bibinfo {author} {\bibfnamefont {A.}~\bibnamefont
			{Fiore}}, \bibinfo {author} {\bibfnamefont {S.}~\bibnamefont {Janz}},
		\bibinfo {author} {\bibfnamefont {L.}~\bibnamefont {Delobel}}, \bibinfo
		{author} {\bibfnamefont {P.}~\bibnamefont {Van~der Meer}}, \bibinfo {author}
		{\bibfnamefont {P.}~\bibnamefont {Bravetti}}, \bibinfo {author}
		{\bibfnamefont {V.}~\bibnamefont {Berger}}, \bibinfo {author} {\bibfnamefont
			{E.}~\bibnamefont {Rosencher}},\ and\ \bibinfo {author} {\bibfnamefont
			{J.}~\bibnamefont {Nagle}},\ }\bibfield  {title} {\emph {\bibinfo {title}
			{{\color{blue}Second-harmonic generation at $\lambda$= 1.6 $\mu$m in
					$AlGaAs/Al_2O_3$ waveguides using birefringence phase matching}}},\
	}\href@noop {} {\bibfield  {journal} {\bibinfo  {journal} {Applied Physics
				Letters}\ }\textbf {\bibinfo {volume} {72}},\ \bibinfo {pages} {2942}
		(\bibinfo {year} {1998})}\BibitemShut {NoStop}%
	\bibitem [{\citenamefont {Armstrong}\ \emph {et~al.}(1962)\citenamefont
		{Armstrong}, \citenamefont {Bloembergen}, \citenamefont {Ducuing},\ and\
		\citenamefont {Pershan}}]{Armstrong62}%
	\BibitemOpen
	\bibfield  {author} {\bibinfo {author} {\bibfnamefont {J.~A.}\ \bibnamefont
			{Armstrong}}, \bibinfo {author} {\bibfnamefont {N.}~\bibnamefont
			{Bloembergen}}, \bibinfo {author} {\bibfnamefont {J.}~\bibnamefont
			{Ducuing}},\ and\ \bibinfo {author} {\bibfnamefont {P.~S.}\ \bibnamefont
			{Pershan}},\ }\bibfield  {title} {\emph {\bibinfo {title}
			{{\color{blue}Interactions between light waves in a nonlinear dielectric}}},\
	}\href@noop {} {\bibfield  {journal} {\bibinfo  {journal} {Phys. Rev.}\
		}\textbf {\bibinfo {volume} {127}},\ \bibinfo {pages} {1918} (\bibinfo {year}
		{1962})}\BibitemShut {NoStop}%
	\bibitem [{\citenamefont {Tanzilli}\ \emph {et~al.}(2001)\citenamefont
		{Tanzilli}, \citenamefont {De~Riedmatten}, \citenamefont {Tittel},
		\citenamefont {Zbinden}, \citenamefont {Baldi}, \citenamefont {De~Micheli},
		\citenamefont {Ostrowsky},\ and\ \citenamefont {Gisin}}]{Tanzilli01}%
	\BibitemOpen
	\bibfield  {author} {\bibinfo {author} {\bibfnamefont {S.}~\bibnamefont
			{Tanzilli}}, \bibinfo {author} {\bibfnamefont {H.}~\bibnamefont
			{De~Riedmatten}}, \bibinfo {author} {\bibfnamefont {H.}~\bibnamefont
			{Tittel}}, \bibinfo {author} {\bibfnamefont {H.}~\bibnamefont {Zbinden}},
		\bibinfo {author} {\bibfnamefont {P.}~\bibnamefont {Baldi}}, \bibinfo
		{author} {\bibfnamefont {M.}~\bibnamefont {De~Micheli}}, \bibinfo {author}
		{\bibfnamefont {D.~B.}\ \bibnamefont {Ostrowsky}},\ and\ \bibinfo {author}
		{\bibfnamefont {N.}~\bibnamefont {Gisin}},\ }\bibfield  {title} {\emph
		{\bibinfo {title} {{\color{blue}Highly efficient photon-pair source using
					periodically poled lithium niobate waveguide}}},\ }\href@noop {} {\bibfield
		{journal} {\bibinfo  {journal} {Electronics Letters}\ }\textbf {\bibinfo
			{volume} {37}},\ \bibinfo {pages} {26} (\bibinfo {year} {2001})}\BibitemShut
	{NoStop}%
	\bibitem [{\citenamefont {Yoo}\ \emph {et~al.}(1995)\citenamefont {Yoo},
		\citenamefont {Bhat}, \citenamefont {Caneau},\ and\ \citenamefont
		{Koza}}]{Yoo95}%
	\BibitemOpen
	\bibfield  {author} {\bibinfo {author} {\bibfnamefont {S.}~\bibnamefont
			{Yoo}}, \bibinfo {author} {\bibfnamefont {R.}~\bibnamefont {Bhat}}, \bibinfo
		{author} {\bibfnamefont {C.}~\bibnamefont {Caneau}},\ and\ \bibinfo {author}
		{\bibfnamefont {M.}~\bibnamefont {Koza}},\ }\bibfield  {title} {\emph
		{\bibinfo {title} {{\color{blue}Quasi-phase-matched second-harmonic
					generation in AlGaAs waveguides with periodic domain inversion achieved by
					wafer-bonding}}},\ }\href@noop {} {\bibfield  {journal} {\bibinfo  {journal}
			{Applied Physics Letters}\ }\textbf {\bibinfo {volume} {66}},\ \bibinfo
		{pages} {3410} (\bibinfo {year} {1995})}\BibitemShut {NoStop}%
	\bibitem [{\citenamefont {Skauli}\ \emph {et~al.}(2002)\citenamefont {Skauli},
		\citenamefont {Vodopyanov}, \citenamefont {Pinguet}, \citenamefont {Schober},
		\citenamefont {Levi}, \citenamefont {Eyres}, \citenamefont {Fejer},
		\citenamefont {Harris}, \citenamefont {Gerard}, \citenamefont {Becouarn}
		\emph {et~al.}}]{Skauli02}%
	\BibitemOpen
	\bibfield  {author} {\bibinfo {author} {\bibfnamefont {T.}~\bibnamefont
			{Skauli}}, \bibinfo {author} {\bibfnamefont {K.}~\bibnamefont {Vodopyanov}},
		\bibinfo {author} {\bibfnamefont {T.}~\bibnamefont {Pinguet}}, \bibinfo
		{author} {\bibfnamefont {A.}~\bibnamefont {Schober}}, \bibinfo {author}
		{\bibfnamefont {O.}~\bibnamefont {Levi}}, \bibinfo {author} {\bibfnamefont
			{L.}~\bibnamefont {Eyres}}, \bibinfo {author} {\bibfnamefont
			{M.}~\bibnamefont {Fejer}}, \bibinfo {author} {\bibfnamefont
			{J.}~\bibnamefont {Harris}}, \bibinfo {author} {\bibfnamefont
			{B.}~\bibnamefont {Gerard}}, \bibinfo {author} {\bibfnamefont
			{L.}~\bibnamefont {Becouarn}}, \emph {et~al.},\ }\bibfield  {title} {\emph
		{\bibinfo {title} {{\color{blue}Measurement of the nonlinear coefficient of
					orientation-patterned {GaAs} and demonstration of highly efficient
					second-harmonic generation}}},\ }\href@noop {} {\bibfield  {journal}
		{\bibinfo  {journal} {Optics Letters}\ }\textbf {\bibinfo {volume} {27}},\
		\bibinfo {pages} {628} (\bibinfo {year} {2002})}\BibitemShut {NoStop}%
	\bibitem [{\citenamefont {Dumeige}\ and\ \citenamefont
		{Feron}(2006)}]{Dumeige06}%
	\BibitemOpen
	\bibfield  {author} {\bibinfo {author} {\bibfnamefont {Y.}~\bibnamefont
			{Dumeige}}\ and\ \bibinfo {author} {\bibfnamefont {P.}~\bibnamefont
			{Feron}},\ }\bibfield  {title} {\emph {\bibinfo {title}
			{{\color{blue}Whispering-gallery-mode analysis of phase-matched doubly
					resonant second-harmonic generation}}},\ }\href@noop {} {\bibfield  {journal}
		{\bibinfo  {journal} {Phys. Rev. A}\ }\textbf {\bibinfo {volume} {74}},\
		\bibinfo {pages} {063804} (\bibinfo {year} {2006})}\BibitemShut {NoStop}%
	\bibitem [{\citenamefont {Mariani}\ \emph {et~al.}(2014)\citenamefont
		{Mariani}, \citenamefont {Andronico}, \citenamefont {Lema{\^\i}tre},
		\citenamefont {Favero}, \citenamefont {Ducci},\ and\ \citenamefont
		{Leo}}]{Mariani14}%
	\BibitemOpen
	\bibfield  {author} {\bibinfo {author} {\bibfnamefont {S.}~\bibnamefont
			{Mariani}}, \bibinfo {author} {\bibfnamefont {A.}~\bibnamefont {Andronico}},
		\bibinfo {author} {\bibfnamefont {A.}~\bibnamefont {Lema{\^\i}tre}}, \bibinfo
		{author} {\bibfnamefont {I.}~\bibnamefont {Favero}}, \bibinfo {author}
		{\bibfnamefont {S.}~\bibnamefont {Ducci}},\ and\ \bibinfo {author}
		{\bibfnamefont {G.}~\bibnamefont {Leo}},\ }\bibfield  {title} {\emph
		{\bibinfo {title} {{\color{blue}Second-harmonic generation in {AlGaAs}
					microdisks in the telecom range}}},\ }\href@noop {} {\bibfield  {journal}
		{\bibinfo  {journal} {Optics Letters}\ }\textbf {\bibinfo {volume} {39}},\
		\bibinfo {pages} {3062} (\bibinfo {year} {2014})}\BibitemShut {NoStop}%
	\bibitem [{\citenamefont {Fu}\ \emph {et~al.}(2020)\citenamefont {Fu},
		\citenamefont {Xu}, \citenamefont {Liang}, \citenamefont {Shardlow},
		\citenamefont {Shepherd}, \citenamefont {Alam},\ and\ \citenamefont
		{Richardson}}]{Fu20}%
	\BibitemOpen
	\bibfield  {author} {\bibinfo {author} {\bibfnamefont {Q.}~\bibnamefont
			{Fu}}, \bibinfo {author} {\bibfnamefont {L.}~\bibnamefont {Xu}}, \bibinfo
		{author} {\bibfnamefont {S.}~\bibnamefont {Liang}}, \bibinfo {author}
		{\bibfnamefont {P.~C.}\ \bibnamefont {Shardlow}}, \bibinfo {author}
		{\bibfnamefont {D.~P.}\ \bibnamefont {Shepherd}}, \bibinfo {author}
		{\bibfnamefont {S.-u.}\ \bibnamefont {Alam}},\ and\ \bibinfo {author}
		{\bibfnamefont {D.~J.}\ \bibnamefont {Richardson}},\ }\bibfield  {title}
	{\emph {\bibinfo {title} {{\color{blue}High-average-power picosecond
					mid-infrared {OP-GaAs OPO}}}},\ }\href@noop {} {\bibfield  {journal}
		{\bibinfo  {journal} {Optics Express}\ }\textbf {\bibinfo {volume} {28}},\
		\bibinfo {pages} {5741} (\bibinfo {year} {2020})}\BibitemShut {NoStop}%
	\bibitem [{\citenamefont {Helmy}\ \emph {et~al.}(2011)\citenamefont {Helmy},
		\citenamefont {Abolghasem}, \citenamefont {Stewart~Aitchison}, \citenamefont
		{Bijlani}, \citenamefont {Han}, \citenamefont {Holmes}, \citenamefont
		{Hutchings}, \citenamefont {Younis},\ and\ \citenamefont
		{Wagner}}]{Helmy2011}%
	\BibitemOpen
	\bibfield  {author} {\bibinfo {author} {\bibfnamefont {A.}~\bibnamefont
			{Helmy}}, \bibinfo {author} {\bibfnamefont {P.}~\bibnamefont {Abolghasem}},
		\bibinfo {author} {\bibfnamefont {J.}~\bibnamefont {Stewart~Aitchison}},
		\bibinfo {author} {\bibfnamefont {B.}~\bibnamefont {Bijlani}}, \bibinfo
		{author} {\bibfnamefont {J.}~\bibnamefont {Han}}, \bibinfo {author}
		{\bibfnamefont {B.}~\bibnamefont {Holmes}}, \bibinfo {author} {\bibfnamefont
			{D.}~\bibnamefont {Hutchings}}, \bibinfo {author} {\bibfnamefont
			{U.}~\bibnamefont {Younis}},\ and\ \bibinfo {author} {\bibfnamefont
			{S.}~\bibnamefont {Wagner}},\ }\bibfield  {title} {\emph {\bibinfo {title}
			{{\color{blue}Recent advances in phase matching of second-order
					nonlinearities in monolithic semiconductor waveguides}}},\ }\href@noop {}
	{\bibfield  {journal} {\bibinfo  {journal} {Laser \& Photonics Reviews}\
		}\textbf {\bibinfo {volume} {5}},\ \bibinfo {pages} {272} (\bibinfo {year}
		{2011})}\BibitemShut {NoStop}%
	\bibitem [{\citenamefont {Abolghasem}\ \emph {et~al.}(2010)\citenamefont
		{Abolghasem}, \citenamefont {Han}, \citenamefont {Bijlani},\ and\
		\citenamefont {Helmy}}]{Abolghasem10}%
	\BibitemOpen
	\bibfield  {author} {\bibinfo {author} {\bibfnamefont {P.}~\bibnamefont
			{Abolghasem}}, \bibinfo {author} {\bibfnamefont {J.}~\bibnamefont {Han}},
		\bibinfo {author} {\bibfnamefont {B.~J.}\ \bibnamefont {Bijlani}},\ and\
		\bibinfo {author} {\bibfnamefont {A.~S.}\ \bibnamefont {Helmy}},\ }\bibfield
	{title} {\emph {\bibinfo {title} {{\color{blue}Type-0 second order nonlinear
					interaction in monolithic waveguides of isotropic semiconductors}}},\
	}\href@noop {} {\bibfield  {journal} {\bibinfo  {journal} {Optics Express}\
		}\textbf {\bibinfo {volume} {18}},\ \bibinfo {pages} {12681} (\bibinfo {year}
		{2010})}\BibitemShut {NoStop}%
	\bibitem [{\citenamefont {Horn}\ \emph {et~al.}(2012)\citenamefont {Horn},
		\citenamefont {Abolghasem}, \citenamefont {Bijlani}, \citenamefont {Kang},
		\citenamefont {Helmy},\ and\ \citenamefont {Weihs}}]{Horn12}%
	\BibitemOpen
	\bibfield  {author} {\bibinfo {author} {\bibfnamefont {R.}~\bibnamefont
			{Horn}}, \bibinfo {author} {\bibfnamefont {P.}~\bibnamefont {Abolghasem}},
		\bibinfo {author} {\bibfnamefont {B.~J.}\ \bibnamefont {Bijlani}}, \bibinfo
		{author} {\bibfnamefont {D.}~\bibnamefont {Kang}}, \bibinfo {author}
		{\bibfnamefont {A.}~\bibnamefont {Helmy}},\ and\ \bibinfo {author}
		{\bibfnamefont {G.}~\bibnamefont {Weihs}},\ }\bibfield  {title} {\emph
		{\bibinfo {title} {{\color{blue}Monolithic source of photon pairs}}},\
	}\href@noop {} {\bibfield  {journal} {\bibinfo  {journal} {Phys. Rev. Lett.}\
		}\textbf {\bibinfo {volume} {108}},\ \bibinfo {pages} {153605} (\bibinfo
		{year} {2012})}\BibitemShut {NoStop}%
	\bibitem [{\citenamefont {Zhukovsky}\ \emph {et~al.}(2012)\citenamefont
		{Zhukovsky}, \citenamefont {Helt}, \citenamefont {Abolghasem}, \citenamefont
		{Kang}, \citenamefont {Sipe},\ and\ \citenamefont {Helmy}}]{Zhukovsky2012}%
	\BibitemOpen
	\bibfield  {author} {\bibinfo {author} {\bibfnamefont {S.~V.}\ \bibnamefont
			{Zhukovsky}}, \bibinfo {author} {\bibfnamefont {L.~G.}\ \bibnamefont {Helt}},
		\bibinfo {author} {\bibfnamefont {P.}~\bibnamefont {Abolghasem}}, \bibinfo
		{author} {\bibfnamefont {D.}~\bibnamefont {Kang}}, \bibinfo {author}
		{\bibfnamefont {J.~E.}\ \bibnamefont {Sipe}},\ and\ \bibinfo {author}
		{\bibfnamefont {A.~S.}\ \bibnamefont {Helmy}},\ }\bibfield  {title} {\emph
		{\bibinfo {title} {{\color{blue}Bragg reflection waveguides as integrated
					sources of entangled photon pairs}}},\ }\href@noop {} {\bibfield  {journal}
		{\bibinfo  {journal} {J. Opt. Soc. Am. B}\ }\textbf {\bibinfo {volume}
			{29}},\ \bibinfo {pages} {2516} (\bibinfo {year} {2012})}\BibitemShut
	{NoStop}%
	\bibitem [{\citenamefont {Lanco}\ \emph {et~al.}(2006)\citenamefont {Lanco},
		\citenamefont {Ducci}, \citenamefont {Likforman}, \citenamefont {Marcadet},
		\citenamefont {van Houwelingen}, \citenamefont {Zbinden}, \citenamefont
		{Leo},\ and\ \citenamefont {Berger}}]{Lanco06}%
	\BibitemOpen
	\bibfield  {author} {\bibinfo {author} {\bibfnamefont {L.}~\bibnamefont
			{Lanco}}, \bibinfo {author} {\bibfnamefont {S.}~\bibnamefont {Ducci}},
		\bibinfo {author} {\bibfnamefont {J.-P.}\ \bibnamefont {Likforman}}, \bibinfo
		{author} {\bibfnamefont {X.}~\bibnamefont {Marcadet}}, \bibinfo {author}
		{\bibfnamefont {J.~A.~W.}\ \bibnamefont {van Houwelingen}}, \bibinfo {author}
		{\bibfnamefont {H.}~\bibnamefont {Zbinden}}, \bibinfo {author} {\bibfnamefont
			{G.}~\bibnamefont {Leo}},\ and\ \bibinfo {author} {\bibfnamefont
			{V.}~\bibnamefont {Berger}},\ }\bibfield  {title} {\emph {\bibinfo {title}
			{{\color{blue}Semiconductor waveguide source of counterpropagating twin
					photons}}},\ }\href@noop {} {\bibfield  {journal} {\bibinfo  {journal} {Phys.
				Rev. Lett.}\ }\textbf {\bibinfo {volume} {97}},\ \bibinfo {pages} {173901}
		(\bibinfo {year} {2006})}\BibitemShut {NoStop}%
	\bibitem [{\citenamefont {Orieux}\ \emph {et~al.}(2013)\citenamefont {Orieux},
		\citenamefont {Eckstein}, \citenamefont {Lema\^{\i}tre}, \citenamefont
		{Filloux}, \citenamefont {Favero}, \citenamefont {Leo}, \citenamefont
		{Coudreau}, \citenamefont {Keller}, \citenamefont {Milman},\ and\
		\citenamefont {Ducci}}]{Orieux13}%
	\BibitemOpen
	\bibfield  {author} {\bibinfo {author} {\bibfnamefont {A.}~\bibnamefont
			{Orieux}}, \bibinfo {author} {\bibfnamefont {A.}~\bibnamefont {Eckstein}},
		\bibinfo {author} {\bibfnamefont {A.}~\bibnamefont {Lema\^{\i}tre}}, \bibinfo
		{author} {\bibfnamefont {P.}~\bibnamefont {Filloux}}, \bibinfo {author}
		{\bibfnamefont {I.}~\bibnamefont {Favero}}, \bibinfo {author} {\bibfnamefont
			{G.}~\bibnamefont {Leo}}, \bibinfo {author} {\bibfnamefont {T.}~\bibnamefont
			{Coudreau}}, \bibinfo {author} {\bibfnamefont {A.}~\bibnamefont {Keller}},
		\bibinfo {author} {\bibfnamefont {P.}~\bibnamefont {Milman}},\ and\ \bibinfo
		{author} {\bibfnamefont {S.}~\bibnamefont {Ducci}},\ }\bibfield  {title}
	{\emph {\bibinfo {title} {{\color{blue}Direct {Bell} states generation on a
					{III-V} semiconductor chip at room temperature}}},\ }\href@noop {} {\bibfield
		{journal} {\bibinfo  {journal} {Phys. Rev. Lett.}\ }\textbf {\bibinfo
			{volume} {110}},\ \bibinfo {pages} {160502} (\bibinfo {year}
		{2013})}\BibitemShut {NoStop}%
	\bibitem [{\citenamefont {Francesconi}\ \emph {et~al.}(2020)\citenamefont
		{Francesconi}, \citenamefont {Baboux}, \citenamefont {Raymond}, \citenamefont
		{Fabre}, \citenamefont {Boucher}, \citenamefont {Lema{\^\i}tre},
		\citenamefont {Milman}, \citenamefont {Amanti},\ and\ \citenamefont
		{Ducci}}]{Francesconi20}%
	\BibitemOpen
	\bibfield  {author} {\bibinfo {author} {\bibfnamefont {S.}~\bibnamefont
			{Francesconi}}, \bibinfo {author} {\bibfnamefont {F.}~\bibnamefont {Baboux}},
		\bibinfo {author} {\bibfnamefont {A.}~\bibnamefont {Raymond}}, \bibinfo
		{author} {\bibfnamefont {N.}~\bibnamefont {Fabre}}, \bibinfo {author}
		{\bibfnamefont {G.}~\bibnamefont {Boucher}}, \bibinfo {author} {\bibfnamefont
			{A.}~\bibnamefont {Lema{\^\i}tre}}, \bibinfo {author} {\bibfnamefont
			{P.}~\bibnamefont {Milman}}, \bibinfo {author} {\bibfnamefont {M.~I.}\
			\bibnamefont {Amanti}},\ and\ \bibinfo {author} {\bibfnamefont
			{S.}~\bibnamefont {Ducci}},\ }\bibfield  {title} {\emph {\bibinfo {title}
			{{\color{blue}Engineering two-photon wavefunction and exchange statistics in
					a semiconductor chip}}},\ }\href@noop {} {\bibfield  {journal} {\bibinfo
			{journal} {Optica}\ }\textbf {\bibinfo {volume} {7}},\ \bibinfo {pages} {316}
		(\bibinfo {year} {2020})}\BibitemShut {NoStop}%
	\bibitem [{\citenamefont {G{\"u}nthner}\ \emph {et~al.}(2015)\citenamefont
		{G{\"u}nthner}, \citenamefont {Pressl}, \citenamefont {Laiho}, \citenamefont
		{Ge{\ss}ler}, \citenamefont {H{\"o}fling}, \citenamefont {Kamp},
		\citenamefont {Schneider},\ and\ \citenamefont {Weihs}}]{Gunthner15}%
	\BibitemOpen
	\bibfield  {author} {\bibinfo {author} {\bibfnamefont {T.}~\bibnamefont
			{G{\"u}nthner}}, \bibinfo {author} {\bibfnamefont {B.}~\bibnamefont
			{Pressl}}, \bibinfo {author} {\bibfnamefont {K.}~\bibnamefont {Laiho}},
		\bibinfo {author} {\bibfnamefont {J.}~\bibnamefont {Ge{\ss}ler}}, \bibinfo
		{author} {\bibfnamefont {S.}~\bibnamefont {H{\"o}fling}}, \bibinfo {author}
		{\bibfnamefont {M.}~\bibnamefont {Kamp}}, \bibinfo {author} {\bibfnamefont
			{C.}~\bibnamefont {Schneider}},\ and\ \bibinfo {author} {\bibfnamefont
			{G.}~\bibnamefont {Weihs}},\ }\bibfield  {title} {\emph {\bibinfo {title}
			{{\color{blue}Broadband indistinguishability from bright parametric
					downconversion in a semiconductor waveguide}}},\ }\href@noop {} {\bibfield
		{journal} {\bibinfo  {journal} {Journal of Optics}\ }\textbf {\bibinfo
			{volume} {17}},\ \bibinfo {pages} {125201} (\bibinfo {year}
		{2015})}\BibitemShut {NoStop}%
	\bibitem [{\citenamefont {Chen}\ \emph {et~al.}(2018)\citenamefont {Chen},
		\citenamefont {Auchter}, \citenamefont {Prilm{\"u}ller}, \citenamefont
		{Schlager}, \citenamefont {Kauten}, \citenamefont {Laiho}, \citenamefont
		{Pressl}, \citenamefont {Suchomel}, \citenamefont {Kamp}, \citenamefont
		{H{\"o}fling} \emph {et~al.}}]{Chen18}%
	\BibitemOpen
	\bibfield  {author} {\bibinfo {author} {\bibfnamefont {H.}~\bibnamefont
			{Chen}}, \bibinfo {author} {\bibfnamefont {S.}~\bibnamefont {Auchter}},
		\bibinfo {author} {\bibfnamefont {M.}~\bibnamefont {Prilm{\"u}ller}},
		\bibinfo {author} {\bibfnamefont {A.}~\bibnamefont {Schlager}}, \bibinfo
		{author} {\bibfnamefont {T.}~\bibnamefont {Kauten}}, \bibinfo {author}
		{\bibfnamefont {K.}~\bibnamefont {Laiho}}, \bibinfo {author} {\bibfnamefont
			{B.}~\bibnamefont {Pressl}}, \bibinfo {author} {\bibfnamefont
			{H.}~\bibnamefont {Suchomel}}, \bibinfo {author} {\bibfnamefont
			{M.}~\bibnamefont {Kamp}}, \bibinfo {author} {\bibfnamefont {S.}~\bibnamefont
			{H{\"o}fling}}, \emph {et~al.},\ }\bibfield  {title} {\emph {\bibinfo {title}
			{{\color{blue}Time-bin entangled photon pairs from {Bragg-reflection}
					waveguides}}},\ }\href@noop {} {\bibfield  {journal} {\bibinfo  {journal}
			{APL Photonics}\ }\textbf {\bibinfo {volume} {3}},\ \bibinfo {pages} {080804}
		(\bibinfo {year} {2018})}\BibitemShut {NoStop}%
	\bibitem [{\citenamefont {Vall{\'e}s}\ \emph {et~al.}(2013)\citenamefont
		{Vall{\'e}s}, \citenamefont {Hendrych}, \citenamefont {Svozilik},
		\citenamefont {Machulka}, \citenamefont {Abolghasem}, \citenamefont {Kang},
		\citenamefont {Bijlani}, \citenamefont {Helmy},\ and\ \citenamefont
		{Torres}}]{Valles13}%
	\BibitemOpen
	\bibfield  {author} {\bibinfo {author} {\bibfnamefont {A.}~\bibnamefont
			{Vall{\'e}s}}, \bibinfo {author} {\bibfnamefont {M.}~\bibnamefont
			{Hendrych}}, \bibinfo {author} {\bibfnamefont {J.}~\bibnamefont {Svozilik}},
		\bibinfo {author} {\bibfnamefont {R.}~\bibnamefont {Machulka}}, \bibinfo
		{author} {\bibfnamefont {P.}~\bibnamefont {Abolghasem}}, \bibinfo {author}
		{\bibfnamefont {D.}~\bibnamefont {Kang}}, \bibinfo {author} {\bibfnamefont
			{B.}~\bibnamefont {Bijlani}}, \bibinfo {author} {\bibfnamefont
			{A.}~\bibnamefont {Helmy}},\ and\ \bibinfo {author} {\bibfnamefont
			{J.}~\bibnamefont {Torres}},\ }\bibfield  {title} {\emph {\bibinfo {title}
			{{\color{blue}Generation of polarization-entangled photon pairs in a Bragg
					reflection waveguide}}},\ }\href@noop {} {\bibfield  {journal} {\bibinfo
			{journal} {Optics Express}\ }\textbf {\bibinfo {volume} {21}},\ \bibinfo
		{pages} {10841} (\bibinfo {year} {2013})}\BibitemShut {NoStop}%
	\bibitem [{\citenamefont {Horn}\ \emph {et~al.}(2013)\citenamefont {Horn},
		\citenamefont {Kolenderski}, \citenamefont {Kang}, \citenamefont
		{Abolghasem}, \citenamefont {Scarcella}, \citenamefont {Della~Frera},
		\citenamefont {Tosi}, \citenamefont {Helt}, \citenamefont {Zhukovsky},
		\citenamefont {Sipe} \emph {et~al.}}]{Horn13}%
	\BibitemOpen
	\bibfield  {author} {\bibinfo {author} {\bibfnamefont {R.~T.}\ \bibnamefont
			{Horn}}, \bibinfo {author} {\bibfnamefont {P.}~\bibnamefont {Kolenderski}},
		\bibinfo {author} {\bibfnamefont {D.}~\bibnamefont {Kang}}, \bibinfo {author}
		{\bibfnamefont {P.}~\bibnamefont {Abolghasem}}, \bibinfo {author}
		{\bibfnamefont {C.}~\bibnamefont {Scarcella}}, \bibinfo {author}
		{\bibfnamefont {A.}~\bibnamefont {Della~Frera}}, \bibinfo {author}
		{\bibfnamefont {A.}~\bibnamefont {Tosi}}, \bibinfo {author} {\bibfnamefont
			{L.~G.}\ \bibnamefont {Helt}}, \bibinfo {author} {\bibfnamefont {S.~V.}\
			\bibnamefont {Zhukovsky}}, \bibinfo {author} {\bibfnamefont {J.~E.}\
			\bibnamefont {Sipe}}, \emph {et~al.},\ }\bibfield  {title} {\emph {\bibinfo
			{title} {{\color{blue}Inherent polarization entanglement generated from a
					monolithic semiconductor chip}}},\ }\href@noop {} {\bibfield  {journal}
		{\bibinfo  {journal} {Scientific Reports}\ }\textbf {\bibinfo {volume} {3}}
		(\bibinfo {year} {2013})}\BibitemShut {NoStop}%
	\bibitem [{\citenamefont {Kang}\ \emph {et~al.}(2016)\citenamefont {Kang},
		\citenamefont {Anirban},\ and\ \citenamefont {Helmy}}]{Kang16}%
	\BibitemOpen
	\bibfield  {author} {\bibinfo {author} {\bibfnamefont {D.}~\bibnamefont
			{Kang}}, \bibinfo {author} {\bibfnamefont {A.}~\bibnamefont {Anirban}},\ and\
		\bibinfo {author} {\bibfnamefont {A.~S.}\ \bibnamefont {Helmy}},\ }\bibfield
	{title} {\emph {\bibinfo {title} {{\color{blue}Monolithic semiconductor chips
					as a source for broadband wavelength-multiplexed polarization entangled
					photons}}},\ }\href@noop {} {\bibfield  {journal} {\bibinfo  {journal}
			{Optics Express}\ }\textbf {\bibinfo {volume} {24}},\ \bibinfo {pages}
		{15160} (\bibinfo {year} {2016})}\BibitemShut {NoStop}%
	\bibitem [{\citenamefont {Appas}\ \emph {et~al.}(2021)\citenamefont {Appas},
		\citenamefont {Baboux}, \citenamefont {Amanti}, \citenamefont
		{Lema{\^\i}tre}, \citenamefont {Boitier}, \citenamefont {Diamanti},\ and\
		\citenamefont {Ducci}}]{Appas21}%
	\BibitemOpen
	\bibfield  {author} {\bibinfo {author} {\bibfnamefont {F.}~\bibnamefont
			{Appas}}, \bibinfo {author} {\bibfnamefont {F.}~\bibnamefont {Baboux}},
		\bibinfo {author} {\bibfnamefont {M.}~\bibnamefont {Amanti}}, \bibinfo
		{author} {\bibfnamefont {A.}~\bibnamefont {Lema{\^\i}tre}}, \bibinfo {author}
		{\bibfnamefont {F.}~\bibnamefont {Boitier}}, \bibinfo {author} {\bibfnamefont
			{E.}~\bibnamefont {Diamanti}},\ and\ \bibinfo {author} {\bibfnamefont
			{S.}~\bibnamefont {Ducci}},\ }\bibfield  {title} {\emph {\bibinfo {title}
			{{\color{blue}{Flexible entanglement-distribution network with an AlGaAs chip
						for secure communications}}}},\ }\href@noop {} {\bibfield  {journal}
		{\bibinfo  {journal} {npj Quantum Information}\ }\textbf {\bibinfo {volume}
			{7}},\ \bibinfo {pages} {1} (\bibinfo {year} {2021})}\BibitemShut {NoStop}%
	\bibitem [{\citenamefont {Francesconi}\ \emph {et~al.}(2021)\citenamefont
		{Francesconi}, \citenamefont {Raymond}, \citenamefont {Fabre}, \citenamefont
		{Lema{\^\i}tre}, \citenamefont {Amanti}, \citenamefont {Milman},
		\citenamefont {Baboux},\ and\ \citenamefont {Ducci}}]{Francesconi21}%
	\BibitemOpen
	\bibfield  {author} {\bibinfo {author} {\bibfnamefont {S.}~\bibnamefont
			{Francesconi}}, \bibinfo {author} {\bibfnamefont {A.}~\bibnamefont
			{Raymond}}, \bibinfo {author} {\bibfnamefont {N.}~\bibnamefont {Fabre}},
		\bibinfo {author} {\bibfnamefont {A.}~\bibnamefont {Lema{\^\i}tre}}, \bibinfo
		{author} {\bibfnamefont {M.~I.}\ \bibnamefont {Amanti}}, \bibinfo {author}
		{\bibfnamefont {P.}~\bibnamefont {Milman}}, \bibinfo {author} {\bibfnamefont
			{F.}~\bibnamefont {Baboux}},\ and\ \bibinfo {author} {\bibfnamefont
			{S.}~\bibnamefont {Ducci}},\ }\bibfield  {title} {\emph {\bibinfo {title}
			{{\color{blue}Anyonic two-photon statistics with a semiconductor chip}}},\
	}\href@noop {} {\bibfield  {journal} {\bibinfo  {journal} {ACS Photonics}\
		}\textbf {\bibinfo {volume} {8}},\ \bibinfo {pages} {2764} (\bibinfo {year}
		{2021})}\BibitemShut {NoStop}%
	\bibitem [{\citenamefont {Francesconi}\ \emph {et~al.}(2022)\citenamefont
		{Francesconi}, \citenamefont {Raymond}, \citenamefont {Duhamel},
		\citenamefont {Filloux}, \citenamefont {Lema{\^\i}tre}, \citenamefont
		{Milman}, \citenamefont {Amanti}, \citenamefont {Baboux},\ and\ \citenamefont
		{Ducci}}]{Francesconi22}%
	\BibitemOpen
	\bibfield  {author} {\bibinfo {author} {\bibfnamefont {S.}~\bibnamefont
			{Francesconi}}, \bibinfo {author} {\bibfnamefont {A.}~\bibnamefont
			{Raymond}}, \bibinfo {author} {\bibfnamefont {R.}~\bibnamefont {Duhamel}},
		\bibinfo {author} {\bibfnamefont {P.}~\bibnamefont {Filloux}}, \bibinfo
		{author} {\bibfnamefont {A.}~\bibnamefont {Lema{\^\i}tre}}, \bibinfo {author}
		{\bibfnamefont {P.}~\bibnamefont {Milman}}, \bibinfo {author} {\bibfnamefont
			{M.~I.}\ \bibnamefont {Amanti}}, \bibinfo {author} {\bibfnamefont
			{F.}~\bibnamefont {Baboux}},\ and\ \bibinfo {author} {\bibfnamefont
			{S.}~\bibnamefont {Ducci}},\ }\bibfield  {title} {\emph {\bibinfo {title}
			{{\color{blue}On-chip generation of hybrid polarization-frequency entangled
					biphoton states}}},\ }\href@noop {} {\bibfield  {journal} {\bibinfo
			{journal} {arXiv 2207.10943}\ } (\bibinfo {year} {2022})}\BibitemShut
	{NoStop}%
	\bibitem [{\citenamefont {Sarrafi}\ \emph {et~al.}(2013)\citenamefont
		{Sarrafi}, \citenamefont {Zhu}, \citenamefont {Dolgaleva}, \citenamefont
		{Holmes}, \citenamefont {Hutchings}, \citenamefont {Aitchison},\ and\
		\citenamefont {Qian}}]{Sarrafi13}%
	\BibitemOpen
	\bibfield  {author} {\bibinfo {author} {\bibfnamefont {P.}~\bibnamefont
			{Sarrafi}}, \bibinfo {author} {\bibfnamefont {E.~Y.}\ \bibnamefont {Zhu}},
		\bibinfo {author} {\bibfnamefont {K.}~\bibnamefont {Dolgaleva}}, \bibinfo
		{author} {\bibfnamefont {B.~M.}\ \bibnamefont {Holmes}}, \bibinfo {author}
		{\bibfnamefont {D.~C.}\ \bibnamefont {Hutchings}}, \bibinfo {author}
		{\bibfnamefont {J.~S.}\ \bibnamefont {Aitchison}},\ and\ \bibinfo {author}
		{\bibfnamefont {L.}~\bibnamefont {Qian}},\ }\bibfield  {title} {\emph
		{\bibinfo {title} {{\color{blue}Continuous-wave quasi-phase-matched waveguide
					correlated photon pair source on a {III--V} chip}}},\ }\href@noop {}
	{\bibfield  {journal} {\bibinfo  {journal} {Applied Physics Letters}\
		}\textbf {\bibinfo {volume} {103}},\ \bibinfo {pages} {251115} (\bibinfo
		{year} {2013})}\BibitemShut {NoStop}%
	\bibitem [{\citenamefont {Sarrafi}\ \emph {et~al.}(2014)\citenamefont
		{Sarrafi}, \citenamefont {Zhu}, \citenamefont {Holmes}, \citenamefont
		{Hutchings}, \citenamefont {Aitchison},\ and\ \citenamefont
		{Qian}}]{Sarrafi14}%
	\BibitemOpen
	\bibfield  {author} {\bibinfo {author} {\bibfnamefont {P.}~\bibnamefont
			{Sarrafi}}, \bibinfo {author} {\bibfnamefont {E.~Y.}\ \bibnamefont {Zhu}},
		\bibinfo {author} {\bibfnamefont {B.~M.}\ \bibnamefont {Holmes}}, \bibinfo
		{author} {\bibfnamefont {D.~C.}\ \bibnamefont {Hutchings}}, \bibinfo {author}
		{\bibfnamefont {S.}~\bibnamefont {Aitchison}},\ and\ \bibinfo {author}
		{\bibfnamefont {L.}~\bibnamefont {Qian}},\ }\bibfield  {title} {\emph
		{\bibinfo {title} {{\color{blue}High-visibility two-photon interference of
					frequency--time entangled photons generated in a quasi-phase-matched {AlGaAs}
					waveguide}}},\ }\href@noop {} {\bibfield  {journal} {\bibinfo  {journal}
			{Optics Letters}\ }\textbf {\bibinfo {volume} {39}},\ \bibinfo {pages} {5188}
		(\bibinfo {year} {2014})}\BibitemShut {NoStop}%
	\bibitem [{\citenamefont {Kultavewuti}\ \emph {et~al.}(2016)\citenamefont
		{Kultavewuti}, \citenamefont {Zhu}, \citenamefont {Qian}, \citenamefont
		{Pusino}, \citenamefont {Sorel},\ and\ \citenamefont
		{Aitchison}}]{Kultavewuti2016}%
	\BibitemOpen
	\bibfield  {author} {\bibinfo {author} {\bibfnamefont {P.}~\bibnamefont
			{Kultavewuti}}, \bibinfo {author} {\bibfnamefont {E.~Y.}\ \bibnamefont
			{Zhu}}, \bibinfo {author} {\bibfnamefont {L.}~\bibnamefont {Qian}}, \bibinfo
		{author} {\bibfnamefont {V.}~\bibnamefont {Pusino}}, \bibinfo {author}
		{\bibfnamefont {M.}~\bibnamefont {Sorel}},\ and\ \bibinfo {author}
		{\bibfnamefont {J.~S.}\ \bibnamefont {Aitchison}},\ }\bibfield  {title}
	{\emph {\bibinfo {title} {{\color{blue}Correlated photon pair generation in
					{AlGaAs} nanowaveguides via spontaneous four-wave mixing}}},\ }\href@noop {}
	{\bibfield  {journal} {\bibinfo  {journal} {Opt. Express}\ }\textbf {\bibinfo
			{volume} {24}},\ \bibinfo {pages} {3365} (\bibinfo {year}
		{2016})}\BibitemShut {NoStop}%
	\bibitem [{\citenamefont {Mahmudlu}\ \emph {et~al.}(2021)\citenamefont
		{Mahmudlu}, \citenamefont {May}, \citenamefont {Angulo}, \citenamefont
		{Sorel},\ and\ \citenamefont {Kues}}]{Mahmudlu2021}%
	\BibitemOpen
	\bibfield  {author} {\bibinfo {author} {\bibfnamefont {H.}~\bibnamefont
			{Mahmudlu}}, \bibinfo {author} {\bibfnamefont {S.}~\bibnamefont {May}},
		\bibinfo {author} {\bibfnamefont {A.}~\bibnamefont {Angulo}}, \bibinfo
		{author} {\bibfnamefont {M.}~\bibnamefont {Sorel}},\ and\ \bibinfo {author}
		{\bibfnamefont {M.}~\bibnamefont {Kues}},\ }\bibfield  {title} {\emph
		{\bibinfo {title} {{\color{blue}{AlGaAs-on-insulator} waveguide for highly
					efficient photon-pair generation via spontaneous four-wave mixing}}},\
	}\href@noop {} {\bibfield  {journal} {\bibinfo  {journal} {Opt. Lett.}\
		}\textbf {\bibinfo {volume} {46}},\ \bibinfo {pages} {1061} (\bibinfo {year}
		{2021})}\BibitemShut {NoStop}%
	\bibitem [{\citenamefont {Pu}\ \emph {et~al.}(2016)\citenamefont {Pu},
		\citenamefont {Ottaviano}, \citenamefont {Semenova},\ and\ \citenamefont
		{Yvind}}]{pu2016efficient}%
	\BibitemOpen
	\bibfield  {author} {\bibinfo {author} {\bibfnamefont {M.}~\bibnamefont
			{Pu}}, \bibinfo {author} {\bibfnamefont {L.}~\bibnamefont {Ottaviano}},
		\bibinfo {author} {\bibfnamefont {E.}~\bibnamefont {Semenova}},\ and\
		\bibinfo {author} {\bibfnamefont {K.}~\bibnamefont {Yvind}},\ }\bibfield
	{title} {\emph {\bibinfo {title} {{\color{blue}Efficient frequency comb
					generation in AlGaAs-on-insulator}}},\ }\href@noop {} {\bibfield  {journal}
		{\bibinfo  {journal} {Optica}\ }\textbf {\bibinfo {volume} {3}},\ \bibinfo
		{pages} {823} (\bibinfo {year} {2016})}\BibitemShut {NoStop}%
	\bibitem [{\citenamefont {Kuyken}\ \emph {et~al.}(2020)\citenamefont {Kuyken},
		\citenamefont {Billet}, \citenamefont {Leo}, \citenamefont {Yvind},\ and\
		\citenamefont {Pu}}]{kuyken2020octave}%
	\BibitemOpen
	\bibfield  {author} {\bibinfo {author} {\bibfnamefont {B.}~\bibnamefont
			{Kuyken}}, \bibinfo {author} {\bibfnamefont {M.}~\bibnamefont {Billet}},
		\bibinfo {author} {\bibfnamefont {F.}~\bibnamefont {Leo}}, \bibinfo {author}
		{\bibfnamefont {K.}~\bibnamefont {Yvind}},\ and\ \bibinfo {author}
		{\bibfnamefont {M.}~\bibnamefont {Pu}},\ }\bibfield  {title} {\emph {\bibinfo
			{title} {{\color{blue}Octave-spanning coherent supercontinuum generation in
					an {AlGaAs-on-insulator} waveguide}}},\ }\href@noop {} {\bibfield  {journal}
		{\bibinfo  {journal} {Optics Letters}\ }\textbf {\bibinfo {volume} {45}},\
		\bibinfo {pages} {603} (\bibinfo {year} {2020})}\BibitemShut {NoStop}%
	\bibitem [{\citenamefont {Zhang}\ \emph
		{et~al.}(2022{\natexlab{b}})\citenamefont {Zhang}, \citenamefont {Yuan},
		\citenamefont {Cheng}, \citenamefont {Mei}, \citenamefont {Lai},
		\citenamefont {Zhou}, \citenamefont {Wu}, \citenamefont {Yan}, \citenamefont
		{Wang}, \citenamefont {Yu} \emph {et~al.}}]{zhang2022polarization}%
	\BibitemOpen
	\bibfield  {author} {\bibinfo {author} {\bibfnamefont {L.}~\bibnamefont
			{Zhang}}, \bibinfo {author} {\bibfnamefont {J.}~\bibnamefont {Yuan}},
		\bibinfo {author} {\bibfnamefont {Y.}~\bibnamefont {Cheng}}, \bibinfo
		{author} {\bibfnamefont {C.}~\bibnamefont {Mei}}, \bibinfo {author}
		{\bibfnamefont {J.}~\bibnamefont {Lai}}, \bibinfo {author} {\bibfnamefont
			{X.}~\bibnamefont {Zhou}}, \bibinfo {author} {\bibfnamefont {Q.}~\bibnamefont
			{Wu}}, \bibinfo {author} {\bibfnamefont {B.}~\bibnamefont {Yan}}, \bibinfo
		{author} {\bibfnamefont {K.}~\bibnamefont {Wang}}, \bibinfo {author}
		{\bibfnamefont {C.}~\bibnamefont {Yu}}, \emph {et~al.},\ }\bibfield  {title}
	{\emph {\bibinfo {title} {{\color{blue}Polarization-insensitive reverse-ridge
					{AlGaAs} waveguide for the mid-infrared supercontinuum generation}}},\
	}\href@noop {} {\bibfield  {journal} {\bibinfo  {journal} {Optics
				Communications}\ }\textbf {\bibinfo {volume} {502}},\ \bibinfo {pages}
		{127407} (\bibinfo {year} {2022}{\natexlab{b}})}\BibitemShut {NoStop}%
	\bibitem [{\citenamefont {Chiles}\ \emph {et~al.}(2019)\citenamefont {Chiles},
		\citenamefont {Nader}, \citenamefont {Stanton}, \citenamefont {Herman},
		\citenamefont {Moody}, \citenamefont {Zhu}, \citenamefont {Skehan},
		\citenamefont {Guha}, \citenamefont {Kowligy}, \citenamefont {Gopinath} \emph
		{et~al.}}]{chiles2019multifunctional}%
	\BibitemOpen
	\bibfield  {author} {\bibinfo {author} {\bibfnamefont {J.}~\bibnamefont
			{Chiles}}, \bibinfo {author} {\bibfnamefont {N.}~\bibnamefont {Nader}},
		\bibinfo {author} {\bibfnamefont {E.~J.}\ \bibnamefont {Stanton}}, \bibinfo
		{author} {\bibfnamefont {D.}~\bibnamefont {Herman}}, \bibinfo {author}
		{\bibfnamefont {G.}~\bibnamefont {Moody}}, \bibinfo {author} {\bibfnamefont
			{J.}~\bibnamefont {Zhu}}, \bibinfo {author} {\bibfnamefont {J.~C.}\
			\bibnamefont {Skehan}}, \bibinfo {author} {\bibfnamefont {B.}~\bibnamefont
			{Guha}}, \bibinfo {author} {\bibfnamefont {A.}~\bibnamefont {Kowligy}},
		\bibinfo {author} {\bibfnamefont {J.~T.}\ \bibnamefont {Gopinath}}, \emph
		{et~al.},\ }\bibfield  {title} {\emph {\bibinfo {title}
			{{\color{blue}Multifunctional integrated photonics in the mid-infrared with
					suspended {AlGaAs} on silicon}}},\ }\href@noop {} {\bibfield  {journal}
		{\bibinfo  {journal} {Optica}\ }\textbf {\bibinfo {volume} {6}},\ \bibinfo
		{pages} {1246} (\bibinfo {year} {2019})}\BibitemShut {NoStop}%
	\bibitem [{\citenamefont {Parker}\ \emph {et~al.}(2000)\citenamefont {Parker},
		\citenamefont {Bose},\ and\ \citenamefont {Plenio}}]{Parker2000}%
	\BibitemOpen
	\bibfield  {author} {\bibinfo {author} {\bibfnamefont {S.}~\bibnamefont
			{Parker}}, \bibinfo {author} {\bibfnamefont {S.}~\bibnamefont {Bose}},\ and\
		\bibinfo {author} {\bibfnamefont {M.~B.}\ \bibnamefont {Plenio}},\ }\bibfield
	{title} {\emph {\bibinfo {title} {{\color{blue}Entanglement quantification
					and purification in continuous-variable systems}}},\ }\href@noop {}
	{\bibfield  {journal} {\bibinfo  {journal} {Phys. Rev. A}\ }\textbf {\bibinfo
			{volume} {61}},\ \bibinfo {pages} {032305} (\bibinfo {year}
		{2000})}\BibitemShut {NoStop}%
	\bibitem [{\citenamefont {Zielnicki}\ \emph {et~al.}(2018)\citenamefont
		{Zielnicki}, \citenamefont {Garay-Palmett}, \citenamefont {Cruz-Delgado},
		\citenamefont {Cruz-Ramirez}, \citenamefont {O'Boyle}, \citenamefont {Fang},
		\citenamefont {Lorenz}, \citenamefont {U'Ren},\ and\ \citenamefont
		{Kwiat}}]{Zielnicki2018}%
	\BibitemOpen
	\bibfield  {author} {\bibinfo {author} {\bibfnamefont {K.}~\bibnamefont
			{Zielnicki}}, \bibinfo {author} {\bibfnamefont {K.}~\bibnamefont
			{Garay-Palmett}}, \bibinfo {author} {\bibfnamefont {D.}~\bibnamefont
			{Cruz-Delgado}}, \bibinfo {author} {\bibfnamefont {H.}~\bibnamefont
			{Cruz-Ramirez}}, \bibinfo {author} {\bibfnamefont {M.~F.}\ \bibnamefont
			{O'Boyle}}, \bibinfo {author} {\bibfnamefont {B.}~\bibnamefont {Fang}},
		\bibinfo {author} {\bibfnamefont {V.~O.}\ \bibnamefont {Lorenz}}, \bibinfo
		{author} {\bibfnamefont {A.~B.}\ \bibnamefont {U'Ren}},\ and\ \bibinfo
		{author} {\bibfnamefont {P.~G.}\ \bibnamefont {Kwiat}},\ }\bibfield  {title}
	{\emph {\bibinfo {title} {{\color{blue}Joint spectral characterization of
					photon-pair sources}}},\ }\href@noop {} {\bibfield  {journal} {\bibinfo
			{journal} {Journal of Modern Optics}\ }\textbf {\bibinfo {volume} {65}},\
		\bibinfo {pages} {1141} (\bibinfo {year} {2018})}\BibitemShut {NoStop}%
	\bibitem [{\citenamefont {Hong}\ \emph {et~al.}(1987)\citenamefont {Hong},
		\citenamefont {Ou},\ and\ \citenamefont {Mandel}}]{Hong87}%
	\BibitemOpen
	\bibfield  {author} {\bibinfo {author} {\bibfnamefont {C.~K.}\ \bibnamefont
			{Hong}}, \bibinfo {author} {\bibfnamefont {Z.~Y.}\ \bibnamefont {Ou}},\ and\
		\bibinfo {author} {\bibfnamefont {L.}~\bibnamefont {Mandel}},\ }\bibfield
	{title} {\emph {\bibinfo {title} {{\color{blue}Measurement of subpicosecond
					time intervals between two photons by interference}}},\ }\href@noop {}
	{\bibfield  {journal} {\bibinfo  {journal} {Phys. Rev. Lett.}\ }\textbf
		{\bibinfo {volume} {59}},\ \bibinfo {pages} {2044} (\bibinfo {year}
		{1987})}\BibitemShut {NoStop}%
	\bibitem [{\citenamefont {Fedrizzi}\ \emph {et~al.}(2009)\citenamefont
		{Fedrizzi}, \citenamefont {Herbst}, \citenamefont {Aspelmeyer}, \citenamefont
		{Barbieri}, \citenamefont {Jennewein},\ and\ \citenamefont
		{Zeilinger}}]{Fedrizzi09}%
	\BibitemOpen
	\bibfield  {author} {\bibinfo {author} {\bibfnamefont {A.}~\bibnamefont
			{Fedrizzi}}, \bibinfo {author} {\bibfnamefont {T.}~\bibnamefont {Herbst}},
		\bibinfo {author} {\bibfnamefont {M.}~\bibnamefont {Aspelmeyer}}, \bibinfo
		{author} {\bibfnamefont {M.}~\bibnamefont {Barbieri}}, \bibinfo {author}
		{\bibfnamefont {T.}~\bibnamefont {Jennewein}},\ and\ \bibinfo {author}
		{\bibfnamefont {A.}~\bibnamefont {Zeilinger}},\ }\bibfield  {title} {\emph
		{\bibinfo {title} {{\color{blue}Anti-symmetrization reveals hidden
					entanglement}}},\ }\href@noop {} {\bibfield  {journal} {\bibinfo  {journal}
			{New Journal of Physics}\ }\textbf {\bibinfo {volume} {11}},\ \bibinfo
		{pages} {103052} (\bibinfo {year} {2009})}\BibitemShut {NoStop}%
	\bibitem [{\citenamefont {Douce}\ \emph {et~al.}(2013)\citenamefont {Douce},
		\citenamefont {Eckstein}, \citenamefont {Walborn}, \citenamefont {Khoury},
		\citenamefont {Ducci}, \citenamefont {Keller}, \citenamefont {Coudreau},\
		and\ \citenamefont {Milman}}]{Douce13}%
	\BibitemOpen
	\bibfield  {author} {\bibinfo {author} {\bibfnamefont {T.}~\bibnamefont
			{Douce}}, \bibinfo {author} {\bibfnamefont {A.}~\bibnamefont {Eckstein}},
		\bibinfo {author} {\bibfnamefont {S.~P.}\ \bibnamefont {Walborn}}, \bibinfo
		{author} {\bibfnamefont {A.~Z.}\ \bibnamefont {Khoury}}, \bibinfo {author}
		{\bibfnamefont {S.}~\bibnamefont {Ducci}}, \bibinfo {author} {\bibfnamefont
			{A.}~\bibnamefont {Keller}}, \bibinfo {author} {\bibfnamefont
			{T.}~\bibnamefont {Coudreau}},\ and\ \bibinfo {author} {\bibfnamefont
			{P.}~\bibnamefont {Milman}},\ }\bibfield  {title} {\emph {\bibinfo {title}
			{{\color{blue}Direct measurement of the biphoton {Wigner} function through
					two-photon interference}}},\ }\href@noop {} {\bibfield  {journal} {\bibinfo
			{journal} {Scientific Reports}\ }\textbf {\bibinfo {volume} {3}},\ \bibinfo
		{pages} {3530} (\bibinfo {year} {2013})}\BibitemShut {NoStop}%
	\bibitem [{\citenamefont {James}\ \emph {et~al.}(2001)\citenamefont {James},
		\citenamefont {Kwiat}, \citenamefont {Munro},\ and\ \citenamefont
		{White}}]{James2001}%
	\BibitemOpen
	\bibfield  {author} {\bibinfo {author} {\bibfnamefont {D.~F.~V.}\
			\bibnamefont {James}}, \bibinfo {author} {\bibfnamefont {P.~G.}\ \bibnamefont
			{Kwiat}}, \bibinfo {author} {\bibfnamefont {W.~J.}\ \bibnamefont {Munro}},\
		and\ \bibinfo {author} {\bibfnamefont {A.~G.}\ \bibnamefont {White}},\
	}\bibfield  {title} {\emph {\bibinfo {title} {{\color{blue}Measurement of
					qubits}}},\ }\href@noop {} {\bibfield  {journal} {\bibinfo  {journal} {Phys.
				Rev. A}\ }\textbf {\bibinfo {volume} {64}},\ \bibinfo {pages} {052312}
		(\bibinfo {year} {2001})}\BibitemShut {NoStop}%
	\bibitem [{\citenamefont {Franson}(1989)}]{Franson1989}%
	\BibitemOpen
	\bibfield  {author} {\bibinfo {author} {\bibfnamefont {J.~D.}\ \bibnamefont
			{Franson}},\ }\bibfield  {title} {\emph {\bibinfo {title} {{\color{blue}Bell
					inequality for position and time}}},\ }\href@noop {} {\bibfield  {journal}
		{\bibinfo  {journal} {Phys. Rev. Lett.}\ }\textbf {\bibinfo {volume} {62}},\
		\bibinfo {pages} {2205} (\bibinfo {year} {1989})}\BibitemShut {NoStop}%
	\bibitem [{\citenamefont {Clauser}\ \emph {et~al.}(1969)\citenamefont
		{Clauser}, \citenamefont {Horne}, \citenamefont {Shimony},\ and\
		\citenamefont {Holt}}]{CHSH1969}%
	\BibitemOpen
	\bibfield  {author} {\bibinfo {author} {\bibfnamefont {J.~F.}\ \bibnamefont
			{Clauser}}, \bibinfo {author} {\bibfnamefont {M.~A.}\ \bibnamefont {Horne}},
		\bibinfo {author} {\bibfnamefont {A.}~\bibnamefont {Shimony}},\ and\ \bibinfo
		{author} {\bibfnamefont {R.~A.}\ \bibnamefont {Holt}},\ }\bibfield  {title}
	{\emph {\bibinfo {title} {{\color{blue}Proposed Experiment to Test Local
					Hidden-Variable Theories}}},\ }\href@noop {} {\bibfield  {journal} {\bibinfo
			{journal} {Phys. Rev. Lett.}\ }\textbf {\bibinfo {volume} {23}},\ \bibinfo
		{pages} {880} (\bibinfo {year} {1969})}\BibitemShut {NoStop}%
	\bibitem [{\citenamefont {Hutchings}\ \emph {et~al.}(2010)\citenamefont
		{Hutchings}, \citenamefont {Wagner}, \citenamefont {Holmes}, \citenamefont
		{Younis}, \citenamefont {Helmy},\ and\ \citenamefont
		{Aitchison}}]{Hutchings10}%
	\BibitemOpen
	\bibfield  {author} {\bibinfo {author} {\bibfnamefont {D.~C.}\ \bibnamefont
			{Hutchings}}, \bibinfo {author} {\bibfnamefont {S.~J.}\ \bibnamefont
			{Wagner}}, \bibinfo {author} {\bibfnamefont {B.~M.}\ \bibnamefont {Holmes}},
		\bibinfo {author} {\bibfnamefont {U.}~\bibnamefont {Younis}}, \bibinfo
		{author} {\bibfnamefont {A.~S.}\ \bibnamefont {Helmy}},\ and\ \bibinfo
		{author} {\bibfnamefont {J.~S.}\ \bibnamefont {Aitchison}},\ }\bibfield
	{title} {\emph {\bibinfo {title} {{\color{blue}{Type-II} quasi phase matching
					in periodically intermixed semiconductor superlattice waveguides}}},\
	}\href@noop {} {\bibfield  {journal} {\bibinfo  {journal} {Optics Letters}\
		}\textbf {\bibinfo {volume} {35}},\ \bibinfo {pages} {1299} (\bibinfo {year}
		{2010})}\BibitemShut {NoStop}%
	\bibitem [{\citenamefont {Caillet}\ \emph {et~al.}(2010)\citenamefont
		{Caillet}, \citenamefont {Orieux}, \citenamefont {Lema{\^\i}tre},
		\citenamefont {Filloux}, \citenamefont {Favero}, \citenamefont {Leo},\ and\
		\citenamefont {Ducci}}]{Caillet10}%
	\BibitemOpen
	\bibfield  {author} {\bibinfo {author} {\bibfnamefont {X.}~\bibnamefont
			{Caillet}}, \bibinfo {author} {\bibfnamefont {A.}~\bibnamefont {Orieux}},
		\bibinfo {author} {\bibfnamefont {A.}~\bibnamefont {Lema{\^\i}tre}}, \bibinfo
		{author} {\bibfnamefont {P.}~\bibnamefont {Filloux}}, \bibinfo {author}
		{\bibfnamefont {I.}~\bibnamefont {Favero}}, \bibinfo {author} {\bibfnamefont
			{G.}~\bibnamefont {Leo}},\ and\ \bibinfo {author} {\bibfnamefont
			{S.}~\bibnamefont {Ducci}},\ }\bibfield  {title} {\emph {\bibinfo {title}
			{{\color{blue}Two-photon interference with a semiconductor integrated source
					at room temperature}}},\ }\href@noop {} {\bibfield  {journal} {\bibinfo
			{journal} {Optics Express}\ }\textbf {\bibinfo {volume} {18}},\ \bibinfo
		{pages} {9967} (\bibinfo {year} {2010})}\BibitemShut {NoStop}%
	\bibitem [{\citenamefont {Aspect}\ \emph {et~al.}(1981)\citenamefont {Aspect},
		\citenamefont {Grangier},\ and\ \citenamefont {Roger}}]{Aspect81}%
	\BibitemOpen
	\bibfield  {author} {\bibinfo {author} {\bibfnamefont {A.}~\bibnamefont
			{Aspect}}, \bibinfo {author} {\bibfnamefont {P.}~\bibnamefont {Grangier}},\
		and\ \bibinfo {author} {\bibfnamefont {G.}~\bibnamefont {Roger}},\ }\bibfield
	{title} {\emph {\bibinfo {title} {{\color{blue}Experimental tests of
					realistic local theories via Bell's theorem}}},\ }\href@noop {} {\bibfield
		{journal} {\bibinfo  {journal} {Phys. Rev. Lett.}\ }\textbf {\bibinfo
			{volume} {47}},\ \bibinfo {pages} {460} (\bibinfo {year} {1981})}\BibitemShut
	{NoStop}%
	\bibitem [{\citenamefont {Shor}(1997)}]{Shor97}%
	\BibitemOpen
	\bibfield  {author} {\bibinfo {author} {\bibfnamefont {P.~W.}\ \bibnamefont
			{Shor}},\ }\bibfield  {title} {\emph {\bibinfo {title}
			{{\color{blue}Polynomial-time algorithms for prime factorization and discrete
					logarithms on a quantum computer}}},\ }\href@noop {} {\bibfield  {journal}
		{\bibinfo  {journal} {SIAM Review}\ }\textbf {\bibinfo {volume} {26}},\
		\bibinfo {pages} {1484} (\bibinfo {year} {1997})}\BibitemShut {NoStop}%
	\bibitem [{\citenamefont {Ekert}(1991)}]{Ekert91}%
	\BibitemOpen
	\bibfield  {author} {\bibinfo {author} {\bibfnamefont {A.~K.}\ \bibnamefont
			{Ekert}},\ }\bibfield  {title} {\emph {\bibinfo {title} {{\color{blue}Quantum
					cryptography based on Bell's theorem}}},\ }\href@noop {} {\bibfield
		{journal} {\bibinfo  {journal} {Phys. Rev. Lett.}\ }\textbf {\bibinfo
			{volume} {67}},\ \bibinfo {pages} {661} (\bibinfo {year} {1991})}\BibitemShut
	{NoStop}%
	\bibitem [{\citenamefont {Bouwmeester}\ \emph {et~al.}(1997)\citenamefont
		{Bouwmeester}, \citenamefont {Pan}, \citenamefont {Mattle}, \citenamefont
		{Eibl}, \citenamefont {Weinfurter},\ and\ \citenamefont
		{Zeilinger}}]{Bouwmeester97}%
	\BibitemOpen
	\bibfield  {author} {\bibinfo {author} {\bibfnamefont {D.}~\bibnamefont
			{Bouwmeester}}, \bibinfo {author} {\bibfnamefont {J.-W.}\ \bibnamefont
			{Pan}}, \bibinfo {author} {\bibfnamefont {K.}~\bibnamefont {Mattle}},
		\bibinfo {author} {\bibfnamefont {M.}~\bibnamefont {Eibl}}, \bibinfo {author}
		{\bibfnamefont {H.}~\bibnamefont {Weinfurter}},\ and\ \bibinfo {author}
		{\bibfnamefont {A.}~\bibnamefont {Zeilinger}},\ }\bibfield  {title} {\emph
		{\bibinfo {title} {{\color{blue}Experimental quantum teleportation}}},\
	}\href@noop {} {\bibfield  {journal} {\bibinfo  {journal} {Nature}\ }\textbf
		{\bibinfo {volume} {390}},\ \bibinfo {pages} {575} (\bibinfo {year}
		{1997})}\BibitemShut {NoStop}%
	\bibitem [{\citenamefont {Kwiat}\ \emph {et~al.}(1995)\citenamefont {Kwiat},
		\citenamefont {Mattle}, \citenamefont {Weinfurter}, \citenamefont
		{Zeilinger}, \citenamefont {Sergienko},\ and\ \citenamefont
		{Shih}}]{Kwiat95}%
	\BibitemOpen
	\bibfield  {author} {\bibinfo {author} {\bibfnamefont {P.~G.}\ \bibnamefont
			{Kwiat}}, \bibinfo {author} {\bibfnamefont {K.}~\bibnamefont {Mattle}},
		\bibinfo {author} {\bibfnamefont {H.}~\bibnamefont {Weinfurter}}, \bibinfo
		{author} {\bibfnamefont {A.}~\bibnamefont {Zeilinger}}, \bibinfo {author}
		{\bibfnamefont {A.~V.}\ \bibnamefont {Sergienko}},\ and\ \bibinfo {author}
		{\bibfnamefont {Y.}~\bibnamefont {Shih}},\ }\bibfield  {title} {\emph
		{\bibinfo {title} {{\color{blue}New high-intensity source of
					polarization-entangled photon pairs}}},\ }\href@noop {} {\bibfield  {journal}
		{\bibinfo  {journal} {Phys. Rev. Lett.}\ }\textbf {\bibinfo {volume} {75}},\
		\bibinfo {pages} {4337} (\bibinfo {year} {1995})}\BibitemShut {NoStop}%
	\bibitem [{\citenamefont {Shi}\ and\ \citenamefont {Tomita}(2004)}]{Shi04}%
	\BibitemOpen
	\bibfield  {author} {\bibinfo {author} {\bibfnamefont {B.-S.}\ \bibnamefont
			{Shi}}\ and\ \bibinfo {author} {\bibfnamefont {A.}~\bibnamefont {Tomita}},\
	}\bibfield  {title} {\emph {\bibinfo {title} {{\color{blue}Generation of a
					pulsed polarization entangled photon pair using a {Sagnac}
					interferometer}}},\ }\href@noop {} {\bibfield  {journal} {\bibinfo  {journal}
			{Phys. Rev. A}\ }\textbf {\bibinfo {volume} {69}},\ \bibinfo {pages} {013803}
		(\bibinfo {year} {2004})}\BibitemShut {NoStop}%
	\bibitem [{\citenamefont {Kim}\ \emph {et~al.}(2006)\citenamefont {Kim},
		\citenamefont {Fiorentino},\ and\ \citenamefont {Wong}}]{Kim06}%
	\BibitemOpen
	\bibfield  {author} {\bibinfo {author} {\bibfnamefont {T.}~\bibnamefont
			{Kim}}, \bibinfo {author} {\bibfnamefont {M.}~\bibnamefont {Fiorentino}},\
		and\ \bibinfo {author} {\bibfnamefont {F.~N.~C.}\ \bibnamefont {Wong}},\
	}\bibfield  {title} {\emph {\bibinfo {title} {{\color{blue}Phase-stable
					source of polarization-entangled photons using a polarization {Sagnac}
					interferometer}}},\ }\href@noop {} {\bibfield  {journal} {\bibinfo  {journal}
			{Phys. Rev. A}\ }\textbf {\bibinfo {volume} {73}},\ \bibinfo {pages} {012316}
		(\bibinfo {year} {2006})}\BibitemShut {NoStop}%
	\bibitem [{\citenamefont {Wengerowsky}\ \emph {et~al.}(2018)\citenamefont
		{Wengerowsky}, \citenamefont {Joshi}, \citenamefont {Steinlechner},
		\citenamefont {H{\"u}bel},\ and\ \citenamefont {Ursin}}]{Wengerowsky18}%
	\BibitemOpen
	\bibfield  {author} {\bibinfo {author} {\bibfnamefont {S.}~\bibnamefont
			{Wengerowsky}}, \bibinfo {author} {\bibfnamefont {S.~K.}\ \bibnamefont
			{Joshi}}, \bibinfo {author} {\bibfnamefont {F.}~\bibnamefont {Steinlechner}},
		\bibinfo {author} {\bibfnamefont {H.}~\bibnamefont {H{\"u}bel}},\ and\
		\bibinfo {author} {\bibfnamefont {R.}~\bibnamefont {Ursin}},\ }\bibfield
	{title} {\emph {\bibinfo {title} {{\color{blue}An entanglement-based
					wavelength-multiplexed quantum communication network}}},\ }\href@noop {}
	{\bibfield  {journal} {\bibinfo  {journal} {Nature}\ }\textbf {\bibinfo
			{volume} {564}},\ \bibinfo {pages} {225} (\bibinfo {year}
		{2018})}\BibitemShut {NoStop}%
	\bibitem [{\citenamefont {Kang}\ \emph {et~al.}(2015)\citenamefont {Kang},
		\citenamefont {Kim}, \citenamefont {He},\ and\ \citenamefont
		{Helmy}}]{Kang15}%
	\BibitemOpen
	\bibfield  {author} {\bibinfo {author} {\bibfnamefont {D.}~\bibnamefont
			{Kang}}, \bibinfo {author} {\bibfnamefont {M.}~\bibnamefont {Kim}}, \bibinfo
		{author} {\bibfnamefont {H.}~\bibnamefont {He}},\ and\ \bibinfo {author}
		{\bibfnamefont {A.~S.}\ \bibnamefont {Helmy}},\ }\bibfield  {title} {\emph
		{\bibinfo {title} {{\color{blue}Two polarization-entangled sources from the
					same semiconductor chip}}},\ }\href@noop {} {\bibfield  {journal} {\bibinfo
			{journal} {Physical Review A}\ }\textbf {\bibinfo {volume} {92}},\ \bibinfo
		{pages} {013821} (\bibinfo {year} {2015})}\BibitemShut {NoStop}%
	\bibitem [{\citenamefont {Schlager}\ \emph {et~al.}(2017)\citenamefont
		{Schlager}, \citenamefont {Pressl}, \citenamefont {Laiho}, \citenamefont
		{Suchomel}, \citenamefont {Kamp}, \citenamefont {H{\"o}fling}, \citenamefont
		{Schneider},\ and\ \citenamefont {Weihs}}]{Schlager17}%
	\BibitemOpen
	\bibfield  {author} {\bibinfo {author} {\bibfnamefont {A.}~\bibnamefont
			{Schlager}}, \bibinfo {author} {\bibfnamefont {B.}~\bibnamefont {Pressl}},
		\bibinfo {author} {\bibfnamefont {K.}~\bibnamefont {Laiho}}, \bibinfo
		{author} {\bibfnamefont {H.}~\bibnamefont {Suchomel}}, \bibinfo {author}
		{\bibfnamefont {M.}~\bibnamefont {Kamp}}, \bibinfo {author} {\bibfnamefont
			{S.}~\bibnamefont {H{\"o}fling}}, \bibinfo {author} {\bibfnamefont
			{C.}~\bibnamefont {Schneider}},\ and\ \bibinfo {author} {\bibfnamefont
			{G.}~\bibnamefont {Weihs}},\ }\bibfield  {title} {\emph {\bibinfo {title}
			{{\color{blue}Temporally versatile polarization entanglement from {Bragg}
					reflection waveguides}}},\ }\href@noop {} {\bibfield  {journal} {\bibinfo
			{journal} {Optics Letters}\ }\textbf {\bibinfo {volume} {42}},\ \bibinfo
		{pages} {2102} (\bibinfo {year} {2017})}\BibitemShut {NoStop}%
	\bibitem [{\citenamefont {Martin}\ \emph {et~al.}(2010)\citenamefont {Martin},
		\citenamefont {Issautier}, \citenamefont {Herrmann}, \citenamefont {Sohler},
		\citenamefont {Ostrowsky}, \citenamefont {Alibart},\ and\ \citenamefont
		{Tanzilli}}]{Martin10}%
	\BibitemOpen
	\bibfield  {author} {\bibinfo {author} {\bibfnamefont {A.}~\bibnamefont
			{Martin}}, \bibinfo {author} {\bibfnamefont {A.}~\bibnamefont {Issautier}},
		\bibinfo {author} {\bibfnamefont {H.}~\bibnamefont {Herrmann}}, \bibinfo
		{author} {\bibfnamefont {W.}~\bibnamefont {Sohler}}, \bibinfo {author}
		{\bibfnamefont {D.~B.}\ \bibnamefont {Ostrowsky}}, \bibinfo {author}
		{\bibfnamefont {O.}~\bibnamefont {Alibart}},\ and\ \bibinfo {author}
		{\bibfnamefont {S.}~\bibnamefont {Tanzilli}},\ }\bibfield  {title} {\emph
		{\bibinfo {title} {{\color{blue}A polarization entangled photon-pair source
					based on a {type-II PPLN} waveguide emitting at a telecom wavelength}}},\
	}\href@noop {} {\bibfield  {journal} {\bibinfo  {journal} {New Journal of
				Physics}\ }\textbf {\bibinfo {volume} {12}},\ \bibinfo {pages} {103005}
		(\bibinfo {year} {2010})}\BibitemShut {NoStop}%
	\bibitem [{\citenamefont {Matsuda}\ \emph {et~al.}(2012)\citenamefont
		{Matsuda}, \citenamefont {Le~Jeannic}, \citenamefont {Fukuda}, \citenamefont
		{Tsuchizawa}, \citenamefont {Munro}, \citenamefont {Shimizu}, \citenamefont
		{Yamada}, \citenamefont {Tokura},\ and\ \citenamefont {Takesue}}]{Matsuda12}%
	\BibitemOpen
	\bibfield  {author} {\bibinfo {author} {\bibfnamefont {N.}~\bibnamefont
			{Matsuda}}, \bibinfo {author} {\bibfnamefont {H.}~\bibnamefont {Le~Jeannic}},
		\bibinfo {author} {\bibfnamefont {H.}~\bibnamefont {Fukuda}}, \bibinfo
		{author} {\bibfnamefont {T.}~\bibnamefont {Tsuchizawa}}, \bibinfo {author}
		{\bibfnamefont {W.~J.}\ \bibnamefont {Munro}}, \bibinfo {author}
		{\bibfnamefont {K.}~\bibnamefont {Shimizu}}, \bibinfo {author} {\bibfnamefont
			{K.}~\bibnamefont {Yamada}}, \bibinfo {author} {\bibfnamefont
			{Y.}~\bibnamefont {Tokura}},\ and\ \bibinfo {author} {\bibfnamefont
			{H.}~\bibnamefont {Takesue}},\ }\bibfield  {title} {\emph {\bibinfo {title}
			{{\color{blue}A monolithically integrated polarization entangled photon pair
					source on a silicon chip}}},\ }\href@noop {} {\bibfield  {journal} {\bibinfo
			{journal} {Scientific Reports}\ }\textbf {\bibinfo {volume} {2}},\ \bibinfo
		{pages} {1} (\bibinfo {year} {2012})}\BibitemShut {NoStop}%
	\bibitem [{\citenamefont {Dada}\ \emph {et~al.}(2011)\citenamefont {Dada},
		\citenamefont {Leach}, \citenamefont {Buller}, \citenamefont {Padgett},\ and\
		\citenamefont {Andersson}}]{Dada11}%
	\BibitemOpen
	\bibfield  {author} {\bibinfo {author} {\bibfnamefont {A.~C.}\ \bibnamefont
			{Dada}}, \bibinfo {author} {\bibfnamefont {J.}~\bibnamefont {Leach}},
		\bibinfo {author} {\bibfnamefont {G.~S.}\ \bibnamefont {Buller}}, \bibinfo
		{author} {\bibfnamefont {M.~J.}\ \bibnamefont {Padgett}},\ and\ \bibinfo
		{author} {\bibfnamefont {E.}~\bibnamefont {Andersson}},\ }\bibfield  {title}
	{\emph {\bibinfo {title} {{\color{blue}Experimental high-dimensional
					two-photon entanglement and violations of generalized {Bell}
					inequalities}}},\ }\href@noop {} {\bibfield  {journal} {\bibinfo  {journal}
			{Nature Physics}\ }\textbf {\bibinfo {volume} {7}},\ \bibinfo {pages} {677}
		(\bibinfo {year} {2011})}\BibitemShut {NoStop}%
	\bibitem [{\citenamefont {Barreiro}\ \emph {et~al.}(2008)\citenamefont
		{Barreiro}, \citenamefont {Wei},\ and\ \citenamefont {Kwiat}}]{Barreiro08}%
	\BibitemOpen
	\bibfield  {author} {\bibinfo {author} {\bibfnamefont {J.~T.}\ \bibnamefont
			{Barreiro}}, \bibinfo {author} {\bibfnamefont {T.-C.}\ \bibnamefont {Wei}},\
		and\ \bibinfo {author} {\bibfnamefont {P.~G.}\ \bibnamefont {Kwiat}},\
	}\bibfield  {title} {\emph {\bibinfo {title} {{\color{blue}Beating the
					channel capacity limit for linear photonic superdense coding}}},\ }\href@noop
	{} {\bibfield  {journal} {\bibinfo  {journal} {Nature Physics}\ }\textbf
		{\bibinfo {volume} {4}},\ \bibinfo {pages} {282} (\bibinfo {year}
		{2008})}\BibitemShut {NoStop}%
	\bibitem [{\citenamefont {Lanyon}\ \emph {et~al.}(2009)\citenamefont {Lanyon},
		\citenamefont {Barbieri}, \citenamefont {Almeida}, \citenamefont {Jennewein},
		\citenamefont {Ralph}, \citenamefont {Resch}, \citenamefont {Pryde},
		\citenamefont {O’brien}, \citenamefont {Gilchrist},\ and\ \citenamefont
		{White}}]{Lanyon09}%
	\BibitemOpen
	\bibfield  {author} {\bibinfo {author} {\bibfnamefont {B.~P.}\ \bibnamefont
			{Lanyon}}, \bibinfo {author} {\bibfnamefont {M.}~\bibnamefont {Barbieri}},
		\bibinfo {author} {\bibfnamefont {M.~P.}\ \bibnamefont {Almeida}}, \bibinfo
		{author} {\bibfnamefont {T.}~\bibnamefont {Jennewein}}, \bibinfo {author}
		{\bibfnamefont {T.~C.}\ \bibnamefont {Ralph}}, \bibinfo {author}
		{\bibfnamefont {K.~J.}\ \bibnamefont {Resch}}, \bibinfo {author}
		{\bibfnamefont {G.~J.}\ \bibnamefont {Pryde}}, \bibinfo {author}
		{\bibfnamefont {J.~L.}\ \bibnamefont {O’brien}}, \bibinfo {author}
		{\bibfnamefont {A.}~\bibnamefont {Gilchrist}},\ and\ \bibinfo {author}
		{\bibfnamefont {A.~G.}\ \bibnamefont {White}},\ }\bibfield  {title} {\emph
		{\bibinfo {title} {{\color{blue}Simplifying quantum logic using
					higher-dimensional {Hilbert} spaces}}},\ }\href@noop {} {\bibfield  {journal}
		{\bibinfo  {journal} {Nature Physics}\ }\textbf {\bibinfo {volume} {5}},\
		\bibinfo {pages} {134} (\bibinfo {year} {2009})}\BibitemShut {NoStop}%
	\bibitem [{\citenamefont {Ansari}\ \emph
		{et~al.}(2018{\natexlab{a}})\citenamefont {Ansari}, \citenamefont {Donohue},
		\citenamefont {Brecht},\ and\ \citenamefont {Silberhorn}}]{Ansari18b}%
	\BibitemOpen
	\bibfield  {author} {\bibinfo {author} {\bibfnamefont {V.}~\bibnamefont
			{Ansari}}, \bibinfo {author} {\bibfnamefont {J.~M.}\ \bibnamefont {Donohue}},
		\bibinfo {author} {\bibfnamefont {B.}~\bibnamefont {Brecht}},\ and\ \bibinfo
		{author} {\bibfnamefont {C.}~\bibnamefont {Silberhorn}},\ }\bibfield  {title}
	{\emph {\bibinfo {title} {{\color{blue}Tailoring nonlinear processes for
					quantum optics with pulsed temporal-mode encodings}}},\ }\href@noop {}
	{\bibfield  {journal} {\bibinfo  {journal} {Optica}\ }\textbf {\bibinfo
			{volume} {5}},\ \bibinfo {pages} {534} (\bibinfo {year}
		{2018}{\natexlab{a}})}\BibitemShut {NoStop}%
	\bibitem [{\citenamefont {Kues}\ \emph {et~al.}(2019)\citenamefont {Kues},
		\citenamefont {Reimer}, \citenamefont {Lukens}, \citenamefont {Munro},
		\citenamefont {Weiner}, \citenamefont {Moss},\ and\ \citenamefont
		{Morandotti}}]{Kues19}%
	\BibitemOpen
	\bibfield  {author} {\bibinfo {author} {\bibfnamefont {M.}~\bibnamefont
			{Kues}}, \bibinfo {author} {\bibfnamefont {C.}~\bibnamefont {Reimer}},
		\bibinfo {author} {\bibfnamefont {J.~M.}\ \bibnamefont {Lukens}}, \bibinfo
		{author} {\bibfnamefont {W.~J.}\ \bibnamefont {Munro}}, \bibinfo {author}
		{\bibfnamefont {A.~M.}\ \bibnamefont {Weiner}}, \bibinfo {author}
		{\bibfnamefont {D.~J.}\ \bibnamefont {Moss}},\ and\ \bibinfo {author}
		{\bibfnamefont {R.}~\bibnamefont {Morandotti}},\ }\bibfield  {title} {\emph
		{\bibinfo {title} {{\color{blue}Quantum optical microcombs}}},\ }\href@noop
	{} {\bibfield  {journal} {\bibinfo  {journal} {Nature Photonics}\ }\textbf
		{\bibinfo {volume} {13}},\ \bibinfo {pages} {170} (\bibinfo {year}
		{2019})}\BibitemShut {NoStop}%
	\bibitem [{\citenamefont {Ansari}\ \emph
		{et~al.}(2018{\natexlab{b}})\citenamefont {Ansari}, \citenamefont {Donohue},
		\citenamefont {Allgaier}, \citenamefont {Sansoni}, \citenamefont {Brecht},
		\citenamefont {Roslund}, \citenamefont {Treps}, \citenamefont {Harder},\ and\
		\citenamefont {Silberhorn}}]{Ansari18}%
	\BibitemOpen
	\bibfield  {author} {\bibinfo {author} {\bibfnamefont {V.}~\bibnamefont
			{Ansari}}, \bibinfo {author} {\bibfnamefont {J.~M.}\ \bibnamefont {Donohue}},
		\bibinfo {author} {\bibfnamefont {M.}~\bibnamefont {Allgaier}}, \bibinfo
		{author} {\bibfnamefont {L.}~\bibnamefont {Sansoni}}, \bibinfo {author}
		{\bibfnamefont {B.}~\bibnamefont {Brecht}}, \bibinfo {author} {\bibfnamefont
			{J.}~\bibnamefont {Roslund}}, \bibinfo {author} {\bibfnamefont
			{N.}~\bibnamefont {Treps}}, \bibinfo {author} {\bibfnamefont
			{G.}~\bibnamefont {Harder}},\ and\ \bibinfo {author} {\bibfnamefont
			{C.}~\bibnamefont {Silberhorn}},\ }\bibfield  {title} {\emph {\bibinfo
			{title} {{\color{blue}Tomography and purification of the temporal-mode
					structure of quantum light}}},\ }\href@noop {} {\bibfield  {journal}
		{\bibinfo  {journal} {Phys. Rev. Lett.}\ }\textbf {\bibinfo {volume} {120}},\
		\bibinfo {pages} {213601} (\bibinfo {year} {2018}{\natexlab{b}})}\BibitemShut
	{NoStop}%
	\bibitem [{\citenamefont {Ansari}\ \emph
		{et~al.}(2018{\natexlab{c}})\citenamefont {Ansari}, \citenamefont {Roccia},
		\citenamefont {Santandrea}, \citenamefont {Doostdar}, \citenamefont {Eigner},
		\citenamefont {Padberg}, \citenamefont {Gianani}, \citenamefont {Sbroscia},
		\citenamefont {Donohue}, \citenamefont {Mancino} \emph {et~al.}}]{Ansari18c}%
	\BibitemOpen
	\bibfield  {author} {\bibinfo {author} {\bibfnamefont {V.}~\bibnamefont
			{Ansari}}, \bibinfo {author} {\bibfnamefont {E.}~\bibnamefont {Roccia}},
		\bibinfo {author} {\bibfnamefont {M.}~\bibnamefont {Santandrea}}, \bibinfo
		{author} {\bibfnamefont {M.}~\bibnamefont {Doostdar}}, \bibinfo {author}
		{\bibfnamefont {C.}~\bibnamefont {Eigner}}, \bibinfo {author} {\bibfnamefont
			{L.}~\bibnamefont {Padberg}}, \bibinfo {author} {\bibfnamefont
			{I.}~\bibnamefont {Gianani}}, \bibinfo {author} {\bibfnamefont
			{M.}~\bibnamefont {Sbroscia}}, \bibinfo {author} {\bibfnamefont {J.~M.}\
			\bibnamefont {Donohue}}, \bibinfo {author} {\bibfnamefont {L.}~\bibnamefont
			{Mancino}}, \emph {et~al.},\ }\bibfield  {title} {\emph {\bibinfo {title}
			{{\color{blue}Heralded generation of high-purity ultrashort single photons in
					programmable temporal shapes}}},\ }\href@noop {} {\bibfield  {journal}
		{\bibinfo  {journal} {Optics Express}\ }\textbf {\bibinfo {volume} {26}},\
		\bibinfo {pages} {2764} (\bibinfo {year} {2018}{\natexlab{c}})}\BibitemShut
	{NoStop}%
	\bibitem [{\citenamefont {Graffitti}\ \emph {et~al.}(2018)\citenamefont
		{Graffitti}, \citenamefont {Barrow}, \citenamefont {Proietti}, \citenamefont
		{Kundys},\ and\ \citenamefont {Fedrizzi}}]{Graffitti18}%
	\BibitemOpen
	\bibfield  {author} {\bibinfo {author} {\bibfnamefont {F.}~\bibnamefont
			{Graffitti}}, \bibinfo {author} {\bibfnamefont {P.}~\bibnamefont {Barrow}},
		\bibinfo {author} {\bibfnamefont {M.}~\bibnamefont {Proietti}}, \bibinfo
		{author} {\bibfnamefont {D.}~\bibnamefont {Kundys}},\ and\ \bibinfo {author}
		{\bibfnamefont {A.}~\bibnamefont {Fedrizzi}},\ }\bibfield  {title} {\emph
		{\bibinfo {title} {{\color{blue}Independent high-purity photons created in
					domain-engineered crystals}}},\ }\href@noop {} {\bibfield  {journal}
		{\bibinfo  {journal} {Optica}\ }\textbf {\bibinfo {volume} {5}},\ \bibinfo
		{pages} {514} (\bibinfo {year} {2018})}\BibitemShut {NoStop}%
	\bibitem [{\citenamefont {Graffitti}\ \emph {et~al.}(2020)\citenamefont
		{Graffitti}, \citenamefont {Barrow}, \citenamefont {Pickston}, \citenamefont
		{Bra\ifmmode~\acute{n}\else \'{n}\fi{}czyk},\ and\ \citenamefont
		{Fedrizzi}}]{Graffitti20}%
	\BibitemOpen
	\bibfield  {author} {\bibinfo {author} {\bibfnamefont {F.}~\bibnamefont
			{Graffitti}}, \bibinfo {author} {\bibfnamefont {P.}~\bibnamefont {Barrow}},
		\bibinfo {author} {\bibfnamefont {A.}~\bibnamefont {Pickston}}, \bibinfo
		{author} {\bibfnamefont {A.~M.}\ \bibnamefont {Bra\ifmmode~\acute{n}\else
				\'{n}\fi{}czyk}},\ and\ \bibinfo {author} {\bibfnamefont {A.}~\bibnamefont
			{Fedrizzi}},\ }\bibfield  {title} {\emph {\bibinfo {title}
			{{\color{blue}Direct generation of tailored pulse-mode entanglement}}},\
	}\href@noop {} {\bibfield  {journal} {\bibinfo  {journal} {Phys. Rev. Lett.}\
		}\textbf {\bibinfo {volume} {124}},\ \bibinfo {pages} {053603} (\bibinfo
		{year} {2020})}\BibitemShut {NoStop}%
	\bibitem [{\citenamefont {Giovannetti}\ \emph {et~al.}(2001)\citenamefont
		{Giovannetti}, \citenamefont {Lloyd}, \citenamefont {Maccone},\ and\
		\citenamefont {Wong}}]{Giovannetti01_PRL}%
	\BibitemOpen
	\bibfield  {author} {\bibinfo {author} {\bibfnamefont {V.}~\bibnamefont
			{Giovannetti}}, \bibinfo {author} {\bibfnamefont {S.}~\bibnamefont {Lloyd}},
		\bibinfo {author} {\bibfnamefont {L.}~\bibnamefont {Maccone}},\ and\ \bibinfo
		{author} {\bibfnamefont {F.~N.~C.}\ \bibnamefont {Wong}},\ }\bibfield
	{title} {\emph {\bibinfo {title} {{\color{blue}Clock synchronization with
					dispersion cancellation}}},\ }\href@noop {} {\bibfield  {journal} {\bibinfo
			{journal} {Phys. Rev. Lett.}\ }\textbf {\bibinfo {volume} {87}},\ \bibinfo
		{pages} {117902} (\bibinfo {year} {2001})}\BibitemShut {NoStop}%
	\bibitem [{\citenamefont {Erdmann}\ \emph {et~al.}(2000)\citenamefont
		{Erdmann}, \citenamefont {Branning}, \citenamefont {Grice},\ and\
		\citenamefont {Walmsley}}]{Erdmann00}%
	\BibitemOpen
	\bibfield  {author} {\bibinfo {author} {\bibfnamefont {R.}~\bibnamefont
			{Erdmann}}, \bibinfo {author} {\bibfnamefont {D.}~\bibnamefont {Branning}},
		\bibinfo {author} {\bibfnamefont {W.}~\bibnamefont {Grice}},\ and\ \bibinfo
		{author} {\bibfnamefont {I.~A.}\ \bibnamefont {Walmsley}},\ }\bibfield
	{title} {\emph {\bibinfo {title} {{\color{blue}Restoring dispersion
					cancellation for entangled photons produced by ultrashort pulses}}},\
	}\href@noop {} {\bibfield  {journal} {\bibinfo  {journal} {Phys. Rev. A}\
		}\textbf {\bibinfo {volume} {62}},\ \bibinfo {pages} {053810} (\bibinfo
		{year} {2000})}\BibitemShut {NoStop}%
	\bibitem [{\citenamefont {Fabre}\ \emph
		{et~al.}(2022{\natexlab{a}})\citenamefont {Fabre}, \citenamefont {Amanti},
		\citenamefont {Baboux}, \citenamefont {Keller}, \citenamefont {Ducci},\ and\
		\citenamefont {Milman}}]{Fabre22_HOM}%
	\BibitemOpen
	\bibfield  {author} {\bibinfo {author} {\bibfnamefont {N.}~\bibnamefont
			{Fabre}}, \bibinfo {author} {\bibfnamefont {M.}~\bibnamefont {Amanti}},
		\bibinfo {author} {\bibfnamefont {F.}~\bibnamefont {Baboux}}, \bibinfo
		{author} {\bibfnamefont {A.}~\bibnamefont {Keller}}, \bibinfo {author}
		{\bibfnamefont {S.}~\bibnamefont {Ducci}},\ and\ \bibinfo {author}
		{\bibfnamefont {P.}~\bibnamefont {Milman}},\ }\bibfield  {title} {\emph
		{\bibinfo {title} {{\color{blue}The {Hong--Ou--Mandel} experiment: from
					photon indistinguishability to continuous-variable quantum computing}}},\
	}\href@noop {} {\bibfield  {journal} {\bibinfo  {journal} {The European
				Physical Journal D}\ }\textbf {\bibinfo {volume} {76}},\ \bibinfo {pages}
		{196} (\bibinfo {year} {2022}{\natexlab{a}})}\BibitemShut {NoStop}%
	\bibitem [{\citenamefont {Halperin}(1984)}]{Halperin84}%
	\BibitemOpen
	\bibfield  {author} {\bibinfo {author} {\bibfnamefont {B.~I.}\ \bibnamefont
			{Halperin}},\ }\bibfield  {title} {\emph {\bibinfo {title}
			{{\color{blue}Statistics of quasiparticles and the hierarchy of fractional
					quantized {Hall} states}}},\ }\href@noop {} {\bibfield  {journal} {\bibinfo
			{journal} {Phys. Rev. Lett.}\ }\textbf {\bibinfo {volume} {52}},\ \bibinfo
		{pages} {1583} (\bibinfo {year} {1984})}\BibitemShut {NoStop}%
	\bibitem [{\citenamefont {Bartolomei}\ \emph {et~al.}(2020)\citenamefont
		{Bartolomei}, \citenamefont {Kumar}, \citenamefont {Bisognin}, \citenamefont
		{Marguerite}, \citenamefont {Berroir}, \citenamefont {Bocquillon},
		\citenamefont {Placais}, \citenamefont {Cavanna}, \citenamefont {Dong},
		\citenamefont {Gennser} \emph {et~al.}}]{Bartolomei20}%
	\BibitemOpen
	\bibfield  {author} {\bibinfo {author} {\bibfnamefont {H.}~\bibnamefont
			{Bartolomei}}, \bibinfo {author} {\bibfnamefont {M.}~\bibnamefont {Kumar}},
		\bibinfo {author} {\bibfnamefont {R.}~\bibnamefont {Bisognin}}, \bibinfo
		{author} {\bibfnamefont {A.}~\bibnamefont {Marguerite}}, \bibinfo {author}
		{\bibfnamefont {J.-M.}\ \bibnamefont {Berroir}}, \bibinfo {author}
		{\bibfnamefont {E.}~\bibnamefont {Bocquillon}}, \bibinfo {author}
		{\bibfnamefont {B.}~\bibnamefont {Placais}}, \bibinfo {author} {\bibfnamefont
			{A.}~\bibnamefont {Cavanna}}, \bibinfo {author} {\bibfnamefont
			{Q.}~\bibnamefont {Dong}}, \bibinfo {author} {\bibfnamefont {U.}~\bibnamefont
			{Gennser}}, \emph {et~al.},\ }\bibfield  {title} {\emph {\bibinfo {title}
			{{\color{blue}Fractional statistics in anyon collisions}}},\ }\href@noop {}
	{\bibfield  {journal} {\bibinfo  {journal} {Science}\ }\textbf {\bibinfo
			{volume} {368}},\ \bibinfo {pages} {173} (\bibinfo {year}
		{2020})}\BibitemShut {NoStop}%
	\bibitem [{\citenamefont {Kitaev}(2003)}]{Kitaev03}%
	\BibitemOpen
	\bibfield  {author} {\bibinfo {author} {\bibfnamefont {A.~Y.}\ \bibnamefont
			{Kitaev}},\ }\bibfield  {title} {\emph {\bibinfo {title}
			{{\color{blue}Fault-tolerant quantum computation by anyons}}},\ }\href@noop
	{} {\bibfield  {journal} {\bibinfo  {journal} {Annals of Physics}\ }\textbf
		{\bibinfo {volume} {303}},\ \bibinfo {pages} {2} (\bibinfo {year}
		{2003})}\BibitemShut {NoStop}%
	\bibitem [{\citenamefont {Maltese}\ \emph {et~al.}(2020)\citenamefont
		{Maltese}, \citenamefont {Amanti}, \citenamefont {Appas}, \citenamefont
		{Sinnl}, \citenamefont {Lema{\^\i}tre}, \citenamefont {Milman}, \citenamefont
		{Baboux},\ and\ \citenamefont {Ducci}}]{Maltese20}%
	\BibitemOpen
	\bibfield  {author} {\bibinfo {author} {\bibfnamefont {G.}~\bibnamefont
			{Maltese}}, \bibinfo {author} {\bibfnamefont {M.}~\bibnamefont {Amanti}},
		\bibinfo {author} {\bibfnamefont {F.}~\bibnamefont {Appas}}, \bibinfo
		{author} {\bibfnamefont {G.}~\bibnamefont {Sinnl}}, \bibinfo {author}
		{\bibfnamefont {A.}~\bibnamefont {Lema{\^\i}tre}}, \bibinfo {author}
		{\bibfnamefont {P.}~\bibnamefont {Milman}}, \bibinfo {author} {\bibfnamefont
			{F.}~\bibnamefont {Baboux}},\ and\ \bibinfo {author} {\bibfnamefont
			{S.}~\bibnamefont {Ducci}},\ }\bibfield  {title} {\emph {\bibinfo {title}
			{{\color{blue}Generation and symmetry control of quantum frequency combs}}},\
	}\href@noop {} {\bibfield  {journal} {\bibinfo  {journal} {npj Quantum
				Information}\ }\textbf {\bibinfo {volume} {6}},\ \bibinfo {pages} {13}
		(\bibinfo {year} {2020})}\BibitemShut {NoStop}%
	\bibitem [{\citenamefont {Fabre}\ \emph {et~al.}(2020)\citenamefont {Fabre},
		\citenamefont {Maltese}, \citenamefont {Appas}, \citenamefont {Felicetti},
		\citenamefont {Ketterer}, \citenamefont {Keller}, \citenamefont {Coudreau},
		\citenamefont {Baboux}, \citenamefont {Amanti}, \citenamefont {Ducci} \emph
		{et~al.}}]{Fabre20}%
	\BibitemOpen
	\bibfield  {author} {\bibinfo {author} {\bibfnamefont {N.}~\bibnamefont
			{Fabre}}, \bibinfo {author} {\bibfnamefont {G.}~\bibnamefont {Maltese}},
		\bibinfo {author} {\bibfnamefont {F.}~\bibnamefont {Appas}}, \bibinfo
		{author} {\bibfnamefont {S.}~\bibnamefont {Felicetti}}, \bibinfo {author}
		{\bibfnamefont {A.}~\bibnamefont {Ketterer}}, \bibinfo {author}
		{\bibfnamefont {A.}~\bibnamefont {Keller}}, \bibinfo {author} {\bibfnamefont
			{T.}~\bibnamefont {Coudreau}}, \bibinfo {author} {\bibfnamefont
			{F.}~\bibnamefont {Baboux}}, \bibinfo {author} {\bibfnamefont
			{M.}~\bibnamefont {Amanti}}, \bibinfo {author} {\bibfnamefont
			{S.}~\bibnamefont {Ducci}}, \emph {et~al.},\ }\bibfield  {title} {\emph
		{\bibinfo {title} {{\color{blue}Generation of a time-frequency grid state
					with integrated biphoton frequency combs}}},\ }\href@noop {} {\bibfield
		{journal} {\bibinfo  {journal} {Phys. Rev. A}\ }\textbf {\bibinfo {volume}
			{102}},\ \bibinfo {pages} {012607} (\bibinfo {year} {2020})}\BibitemShut
	{NoStop}%
	\bibitem [{\citenamefont {Crespi}\ \emph {et~al.}(2013)\citenamefont {Crespi},
		\citenamefont {Osellame}, \citenamefont {Ramponi}, \citenamefont {Brod},
		\citenamefont {Galvao}, \citenamefont {Spagnolo}, \citenamefont {Vitelli},
		\citenamefont {Maiorino}, \citenamefont {Mataloni},\ and\ \citenamefont
		{Sciarrino}}]{Crespi13}%
	\BibitemOpen
	\bibfield  {author} {\bibinfo {author} {\bibfnamefont {A.}~\bibnamefont
			{Crespi}}, \bibinfo {author} {\bibfnamefont {R.}~\bibnamefont {Osellame}},
		\bibinfo {author} {\bibfnamefont {R.}~\bibnamefont {Ramponi}}, \bibinfo
		{author} {\bibfnamefont {D.~J.}\ \bibnamefont {Brod}}, \bibinfo {author}
		{\bibfnamefont {E.~F.}\ \bibnamefont {Galvao}}, \bibinfo {author}
		{\bibfnamefont {N.}~\bibnamefont {Spagnolo}}, \bibinfo {author}
		{\bibfnamefont {C.}~\bibnamefont {Vitelli}}, \bibinfo {author} {\bibfnamefont
			{E.}~\bibnamefont {Maiorino}}, \bibinfo {author} {\bibfnamefont
			{P.}~\bibnamefont {Mataloni}},\ and\ \bibinfo {author} {\bibfnamefont
			{F.}~\bibnamefont {Sciarrino}},\ }\bibfield  {title} {\emph {\bibinfo {title}
			{{\color{blue}Integrated multimode interferometers with arbitrary designs for
					photonic boson sampling}}},\ }\href@noop {} {\bibfield  {journal} {\bibinfo
			{journal} {Nature Photonics}\ }\textbf {\bibinfo {volume} {7}},\ \bibinfo
		{pages} {545} (\bibinfo {year} {2013})}\BibitemShut {NoStop}%
	\bibitem [{\citenamefont {Matthews}\ \emph {et~al.}(2013)\citenamefont
		{Matthews}, \citenamefont {Poulios}, \citenamefont {Meinecke}, \citenamefont
		{Politi}, \citenamefont {Peruzzo}, \citenamefont {Ismail}, \citenamefont
		{W{\"o}rhoff}, \citenamefont {Thompson},\ and\ \citenamefont
		{O'Brien}}]{Matthews13}%
	\BibitemOpen
	\bibfield  {author} {\bibinfo {author} {\bibfnamefont {J.~C.}\ \bibnamefont
			{Matthews}}, \bibinfo {author} {\bibfnamefont {K.}~\bibnamefont {Poulios}},
		\bibinfo {author} {\bibfnamefont {J.~D.}\ \bibnamefont {Meinecke}}, \bibinfo
		{author} {\bibfnamefont {A.}~\bibnamefont {Politi}}, \bibinfo {author}
		{\bibfnamefont {A.}~\bibnamefont {Peruzzo}}, \bibinfo {author} {\bibfnamefont
			{N.}~\bibnamefont {Ismail}}, \bibinfo {author} {\bibfnamefont
			{K.}~\bibnamefont {W{\"o}rhoff}}, \bibinfo {author} {\bibfnamefont {M.~G.}\
			\bibnamefont {Thompson}},\ and\ \bibinfo {author} {\bibfnamefont {J.~L.}\
			\bibnamefont {O'Brien}},\ }\bibfield  {title} {\emph {\bibinfo {title}
			{{\color{blue}Observing fermionic statistics with photons in arbitrary
					processes}}},\ }\href@noop {} {\bibfield  {journal} {\bibinfo  {journal}
			{Scientific Reports}\ }\textbf {\bibinfo {volume} {3}},\ \bibinfo {pages}
		{1539} (\bibinfo {year} {2013})}\BibitemShut {NoStop}%
	\bibitem [{\citenamefont {Crespi}\ \emph {et~al.}(2015)\citenamefont {Crespi},
		\citenamefont {Sansoni}, \citenamefont {Della~Valle}, \citenamefont {Ciamei},
		\citenamefont {Ramponi}, \citenamefont {Sciarrino}, \citenamefont {Mataloni},
		\citenamefont {Longhi},\ and\ \citenamefont {Osellame}}]{Crespi15}%
	\BibitemOpen
	\bibfield  {author} {\bibinfo {author} {\bibfnamefont {A.}~\bibnamefont
			{Crespi}}, \bibinfo {author} {\bibfnamefont {L.}~\bibnamefont {Sansoni}},
		\bibinfo {author} {\bibfnamefont {G.}~\bibnamefont {Della~Valle}}, \bibinfo
		{author} {\bibfnamefont {A.}~\bibnamefont {Ciamei}}, \bibinfo {author}
		{\bibfnamefont {R.}~\bibnamefont {Ramponi}}, \bibinfo {author} {\bibfnamefont
			{F.}~\bibnamefont {Sciarrino}}, \bibinfo {author} {\bibfnamefont
			{P.}~\bibnamefont {Mataloni}}, \bibinfo {author} {\bibfnamefont
			{S.}~\bibnamefont {Longhi}},\ and\ \bibinfo {author} {\bibfnamefont
			{R.}~\bibnamefont {Osellame}},\ }\bibfield  {title} {\emph {\bibinfo {title}
			{{\color{blue}Particle statistics affects quantum decay and {Fano}
					interference}}},\ }\href@noop {} {\bibfield  {journal} {\bibinfo  {journal}
			{Phys. Rev. Lett.}\ }\textbf {\bibinfo {volume} {114}},\ \bibinfo {pages}
		{090201} (\bibinfo {year} {2015})}\BibitemShut {NoStop}%
	\bibitem [{\citenamefont {Jex}\ \emph {et~al.}(2003)\citenamefont {Jex},
		\citenamefont {Alber}, \citenamefont {Barnett},\ and\ \citenamefont
		{Delgado}}]{Jex03}%
	\BibitemOpen
	\bibfield  {author} {\bibinfo {author} {\bibfnamefont {I.}~\bibnamefont
			{Jex}}, \bibinfo {author} {\bibfnamefont {G.}~\bibnamefont {Alber}}, \bibinfo
		{author} {\bibfnamefont {S.}~\bibnamefont {Barnett}},\ and\ \bibinfo {author}
		{\bibfnamefont {A.}~\bibnamefont {Delgado}},\ }\bibfield  {title} {\emph
		{\bibinfo {title} {{\color{blue}Antisymmetric multi-partite quantum states
					and their applications}}},\ }\href@noop {} {\bibfield  {journal} {\bibinfo
			{journal} {Fortschritte der Physik: Progress of Physics}\ }\textbf {\bibinfo
			{volume} {51}},\ \bibinfo {pages} {172} (\bibinfo {year} {2003})}\BibitemShut
	{NoStop}%
	\bibitem [{\citenamefont {Goyal}\ \emph {et~al.}(2014)\citenamefont {Goyal},
		\citenamefont {Boukama-Dzoussi}, \citenamefont {Ghosh}, \citenamefont
		{Roux},\ and\ \citenamefont {Konrad}}]{Goyal14}%
	\BibitemOpen
	\bibfield  {author} {\bibinfo {author} {\bibfnamefont {S.~K.}\ \bibnamefont
			{Goyal}}, \bibinfo {author} {\bibfnamefont {P.~E.}\ \bibnamefont
			{Boukama-Dzoussi}}, \bibinfo {author} {\bibfnamefont {S.}~\bibnamefont
			{Ghosh}}, \bibinfo {author} {\bibfnamefont {F.~S.}\ \bibnamefont {Roux}},\
		and\ \bibinfo {author} {\bibfnamefont {T.}~\bibnamefont {Konrad}},\
	}\bibfield  {title} {\emph {\bibinfo {title}
			{{\color{blue}Qudit-teleportation for photons with linear optics}}},\
	}\href@noop {} {\bibfield  {journal} {\bibinfo  {journal} {Scientific
				Reports}\ }\textbf {\bibinfo {volume} {4}},\ \bibinfo {pages} {4543}
		(\bibinfo {year} {2014})}\BibitemShut {NoStop}%
	\bibitem [{\citenamefont {Santiago-Cruz}\ \emph {et~al.}(2022)\citenamefont
		{Santiago-Cruz}, \citenamefont {Gennaro}, \citenamefont {Mitrofanov},
		\citenamefont {Addamane}, \citenamefont {Reno}, \citenamefont {Brener},\ and\
		\citenamefont {Chekhova}}]{Santiago2022}%
	\BibitemOpen
	\bibfield  {author} {\bibinfo {author} {\bibfnamefont {T.}~\bibnamefont
			{Santiago-Cruz}}, \bibinfo {author} {\bibfnamefont {S.~D.}\ \bibnamefont
			{Gennaro}}, \bibinfo {author} {\bibfnamefont {O.}~\bibnamefont {Mitrofanov}},
		\bibinfo {author} {\bibfnamefont {S.}~\bibnamefont {Addamane}}, \bibinfo
		{author} {\bibfnamefont {J.}~\bibnamefont {Reno}}, \bibinfo {author}
		{\bibfnamefont {I.}~\bibnamefont {Brener}},\ and\ \bibinfo {author}
		{\bibfnamefont {M.~V.}\ \bibnamefont {Chekhova}},\ }\bibfield  {title} {\emph
		{\bibinfo {title} {{\color{blue}Resonant metasurfaces for generating complex
					quantum states}}},\ }\href@noop {} {\bibfield  {journal} {\bibinfo  {journal}
			{Science}\ }\textbf {\bibinfo {volume} {377}},\ \bibinfo {pages} {991}
		(\bibinfo {year} {2022})}\BibitemShut {NoStop}%
	\bibitem [{\citenamefont {Spring}\ \emph {et~al.}(2017)\citenamefont {Spring},
		\citenamefont {Mennea}, \citenamefont {Metcalf}, \citenamefont {Humphreys},
		\citenamefont {Gates}, \citenamefont {Rogers}, \citenamefont {S{\"o}ller},
		\citenamefont {Smith}, \citenamefont {Kolthammer}, \citenamefont {Smith}
		\emph {et~al.}}]{Spring17}%
	\BibitemOpen
	\bibfield  {author} {\bibinfo {author} {\bibfnamefont {J.~B.}\ \bibnamefont
			{Spring}}, \bibinfo {author} {\bibfnamefont {P.~L.}\ \bibnamefont {Mennea}},
		\bibinfo {author} {\bibfnamefont {B.~J.}\ \bibnamefont {Metcalf}}, \bibinfo
		{author} {\bibfnamefont {P.~C.}\ \bibnamefont {Humphreys}}, \bibinfo {author}
		{\bibfnamefont {J.~C.}\ \bibnamefont {Gates}}, \bibinfo {author}
		{\bibfnamefont {H.~L.}\ \bibnamefont {Rogers}}, \bibinfo {author}
		{\bibfnamefont {C.}~\bibnamefont {S{\"o}ller}}, \bibinfo {author}
		{\bibfnamefont {B.~J.}\ \bibnamefont {Smith}}, \bibinfo {author}
		{\bibfnamefont {W.~S.}\ \bibnamefont {Kolthammer}}, \bibinfo {author}
		{\bibfnamefont {P.~G.}\ \bibnamefont {Smith}}, \emph {et~al.},\ }\bibfield
	{title} {\emph {\bibinfo {title} {{\color{blue}Chip-based array of
					near-identical, pure, heralded single-photon sources}}},\ }\href@noop {}
	{\bibfield  {journal} {\bibinfo  {journal} {Optica}\ }\textbf {\bibinfo
			{volume} {4}},\ \bibinfo {pages} {90} (\bibinfo {year} {2017})}\BibitemShut
	{NoStop}%
	\bibitem [{\citenamefont {Paesani}\ \emph {et~al.}(2019)\citenamefont
		{Paesani}, \citenamefont {Ding}, \citenamefont {Santagati}, \citenamefont
		{Chakhmakhchyan}, \citenamefont {Vigliar}, \citenamefont {Rottwitt},
		\citenamefont {Oxenl{\o}we}, \citenamefont {Wang}, \citenamefont {Thompson},\
		and\ \citenamefont {Laing}}]{Paesani19}%
	\BibitemOpen
	\bibfield  {author} {\bibinfo {author} {\bibfnamefont {S.}~\bibnamefont
			{Paesani}}, \bibinfo {author} {\bibfnamefont {Y.}~\bibnamefont {Ding}},
		\bibinfo {author} {\bibfnamefont {R.}~\bibnamefont {Santagati}}, \bibinfo
		{author} {\bibfnamefont {L.}~\bibnamefont {Chakhmakhchyan}}, \bibinfo
		{author} {\bibfnamefont {C.}~\bibnamefont {Vigliar}}, \bibinfo {author}
		{\bibfnamefont {K.}~\bibnamefont {Rottwitt}}, \bibinfo {author}
		{\bibfnamefont {L.~K.}\ \bibnamefont {Oxenl{\o}we}}, \bibinfo {author}
		{\bibfnamefont {J.}~\bibnamefont {Wang}}, \bibinfo {author} {\bibfnamefont
			{M.~G.}\ \bibnamefont {Thompson}},\ and\ \bibinfo {author} {\bibfnamefont
			{A.}~\bibnamefont {Laing}},\ }\bibfield  {title} {\emph {\bibinfo {title}
			{{\color{blue}{Generation and sampling of quantum states of light in a
						silicon chip}}}},\ }\href@noop {} {\bibfield  {journal} {\bibinfo  {journal}
			{Nature Physics}\ }\textbf {\bibinfo {volume} {15}},\ \bibinfo {pages} {925}
		(\bibinfo {year} {2019})}\BibitemShut {NoStop}%
	\bibitem [{\citenamefont {Raymond}\ \emph {et~al.}(2024)\citenamefont
		{Raymond}, \citenamefont {Zecchetto}, \citenamefont {Palomo}, \citenamefont
		{Morassi}, \citenamefont {Lema\^{\i}tre}, \citenamefont {Raineri},
		\citenamefont {Amanti}, \citenamefont {Ducci},\ and\ \citenamefont
		{Baboux}}]{Raymond24}%
	\BibitemOpen
	\bibfield  {author} {\bibinfo {author} {\bibfnamefont {A.}~\bibnamefont
			{Raymond}}, \bibinfo {author} {\bibfnamefont {A.}~\bibnamefont {Zecchetto}},
		\bibinfo {author} {\bibfnamefont {J.}~\bibnamefont {Palomo}}, \bibinfo
		{author} {\bibfnamefont {M.}~\bibnamefont {Morassi}}, \bibinfo {author}
		{\bibfnamefont {A.}~\bibnamefont {Lema\^{\i}tre}}, \bibinfo {author}
		{\bibfnamefont {F.}~\bibnamefont {Raineri}}, \bibinfo {author} {\bibfnamefont
			{M.~I.}\ \bibnamefont {Amanti}}, \bibinfo {author} {\bibfnamefont
			{S.}~\bibnamefont {Ducci}},\ and\ \bibinfo {author} {\bibfnamefont
			{F.}~\bibnamefont {Baboux}},\ }\bibfield  {title} {\emph {\bibinfo {title}
			{{\color{blue}Tunable Generation of Spatial Entanglement in Nonlinear
					Waveguide Arrays}}},\ }\href@noop {} {\bibfield  {journal} {\bibinfo
			{journal} {Phys. Rev. Lett.}\ }\textbf {\bibinfo {volume} {133}},\ \bibinfo
		{pages} {233602} (\bibinfo {year} {2024})}\BibitemShut {NoStop}%
	\bibitem [{\citenamefont {Raymond}\ \emph {et~al.}(2025)\citenamefont
		{Raymond}, \citenamefont {Cathala}, \citenamefont {Morassi}, \citenamefont
		{Lema\^{i}tre}, \citenamefont {Raineri}, \citenamefont {Ducci},\ and\
		\citenamefont {Baboux}}]{Raymond25}%
	\BibitemOpen
	\bibfield  {author} {\bibinfo {author} {\bibfnamefont {A.}~\bibnamefont
			{Raymond}}, \bibinfo {author} {\bibfnamefont {P.}~\bibnamefont {Cathala}},
		\bibinfo {author} {\bibfnamefont {M.}~\bibnamefont {Morassi}}, \bibinfo
		{author} {\bibfnamefont {A.}~\bibnamefont {Lema\^{i}tre}}, \bibinfo {author}
		{\bibfnamefont {F.}~\bibnamefont {Raineri}}, \bibinfo {author} {\bibfnamefont
			{S.}~\bibnamefont {Ducci}},\ and\ \bibinfo {author} {\bibfnamefont
			{F.}~\bibnamefont {Baboux}},\ }\bibfield  {title} {\emph {\bibinfo {title}
			{{\color{blue}Tailoring quantum walks in integrated photonic lattices}}},\
	}\href@noop {} {\bibfield  {journal} {\bibinfo  {journal} {Opt. Express}\
		}\textbf {\bibinfo {volume} {33}},\ \bibinfo {pages} {45869} (\bibinfo {year}
		{2025})}\BibitemShut {NoStop}%
	\bibitem [{\citenamefont {Peruzzo}\ \emph {et~al.}(2010)\citenamefont
		{Peruzzo}, \citenamefont {Lobino}, \citenamefont {Matthews}, \citenamefont
		{Matsuda}, \citenamefont {Politi}, \citenamefont {Poulios}, \citenamefont
		{Zhou}, \citenamefont {Lahini}, \citenamefont {Ismail}, \citenamefont
		{W{\"o}rhoff} \emph {et~al.}}]{Peruzzo10}%
	\BibitemOpen
	\bibfield  {author} {\bibinfo {author} {\bibfnamefont {A.}~\bibnamefont
			{Peruzzo}}, \bibinfo {author} {\bibfnamefont {M.}~\bibnamefont {Lobino}},
		\bibinfo {author} {\bibfnamefont {J.~C.}\ \bibnamefont {Matthews}}, \bibinfo
		{author} {\bibfnamefont {N.}~\bibnamefont {Matsuda}}, \bibinfo {author}
		{\bibfnamefont {A.}~\bibnamefont {Politi}}, \bibinfo {author} {\bibfnamefont
			{K.}~\bibnamefont {Poulios}}, \bibinfo {author} {\bibfnamefont {X.-Q.}\
			\bibnamefont {Zhou}}, \bibinfo {author} {\bibfnamefont {Y.}~\bibnamefont
			{Lahini}}, \bibinfo {author} {\bibfnamefont {N.}~\bibnamefont {Ismail}},
		\bibinfo {author} {\bibfnamefont {K.}~\bibnamefont {W{\"o}rhoff}}, \emph
		{et~al.},\ }\bibfield  {title} {\emph {\bibinfo {title} {{\color{blue}Quantum
					walks of correlated photons}}},\ }\href@noop {} {\bibfield  {journal}
		{\bibinfo  {journal} {Science}\ }\textbf {\bibinfo {volume} {329}},\ \bibinfo
		{pages} {1500} (\bibinfo {year} {2010})}\BibitemShut {NoStop}%
	\bibitem [{\citenamefont {Solntsev}\ \emph {et~al.}(2014)\citenamefont
		{Solntsev}, \citenamefont {Setzpfandt}, \citenamefont {Clark}, \citenamefont
		{Wu}, \citenamefont {Collins}, \citenamefont {Xiong}, \citenamefont
		{Schreiber}, \citenamefont {Katzschmann}, \citenamefont {Eilenberger},
		\citenamefont {Schiek}, \citenamefont {Sohler}, \citenamefont {Mitchell},
		\citenamefont {Silberhorn}, \citenamefont {Eggleton}, \citenamefont
		{Pertsch}, \citenamefont {Sukhorukov}, \citenamefont {Neshev},\ and\
		\citenamefont {Kivshar}}]{Solntsev14}%
	\BibitemOpen
	\bibfield  {author} {\bibinfo {author} {\bibfnamefont {A.~S.}\ \bibnamefont
			{Solntsev}}, \bibinfo {author} {\bibfnamefont {F.}~\bibnamefont
			{Setzpfandt}}, \bibinfo {author} {\bibfnamefont {A.~S.}\ \bibnamefont
			{Clark}}, \bibinfo {author} {\bibfnamefont {C.~W.}\ \bibnamefont {Wu}},
		\bibinfo {author} {\bibfnamefont {M.~J.}\ \bibnamefont {Collins}}, \bibinfo
		{author} {\bibfnamefont {C.}~\bibnamefont {Xiong}}, \bibinfo {author}
		{\bibfnamefont {A.}~\bibnamefont {Schreiber}}, \bibinfo {author}
		{\bibfnamefont {F.}~\bibnamefont {Katzschmann}}, \bibinfo {author}
		{\bibfnamefont {F.}~\bibnamefont {Eilenberger}}, \bibinfo {author}
		{\bibfnamefont {R.}~\bibnamefont {Schiek}}, \bibinfo {author} {\bibfnamefont
			{W.}~\bibnamefont {Sohler}}, \bibinfo {author} {\bibfnamefont
			{A.}~\bibnamefont {Mitchell}}, \bibinfo {author} {\bibfnamefont
			{C.}~\bibnamefont {Silberhorn}}, \bibinfo {author} {\bibfnamefont {B.~J.}\
			\bibnamefont {Eggleton}}, \bibinfo {author} {\bibfnamefont {T.}~\bibnamefont
			{Pertsch}}, \bibinfo {author} {\bibfnamefont {A.~A.}\ \bibnamefont
			{Sukhorukov}}, \bibinfo {author} {\bibfnamefont {D.~N.}\ \bibnamefont
			{Neshev}},\ and\ \bibinfo {author} {\bibfnamefont {Y.~S.}\ \bibnamefont
			{Kivshar}},\ }\bibfield  {title} {\emph {\bibinfo {title}
			{{\color{blue}Generation of nonclassical biphoton states through cascaded
					quantum walks on a nonlinear chip}}},\ }\href@noop {} {\bibfield  {journal}
		{\bibinfo  {journal} {Phys. Rev. X}\ }\textbf {\bibinfo {volume} {4}},\
		\bibinfo {pages} {031007} (\bibinfo {year} {2014})}\BibitemShut {NoStop}%
	\bibitem [{\citenamefont {Blanco-Redondo}\ \emph {et~al.}(2018)\citenamefont
		{Blanco-Redondo}, \citenamefont {Bell}, \citenamefont {Oren}, \citenamefont
		{Eggleton},\ and\ \citenamefont {Segev}}]{BlancoRedondo18}%
	\BibitemOpen
	\bibfield  {author} {\bibinfo {author} {\bibfnamefont {A.}~\bibnamefont
			{Blanco-Redondo}}, \bibinfo {author} {\bibfnamefont {B.}~\bibnamefont
			{Bell}}, \bibinfo {author} {\bibfnamefont {D.}~\bibnamefont {Oren}}, \bibinfo
		{author} {\bibfnamefont {B.~J.}\ \bibnamefont {Eggleton}},\ and\ \bibinfo
		{author} {\bibfnamefont {M.}~\bibnamefont {Segev}},\ }\bibfield  {title}
	{\emph {\bibinfo {title} {{\color{blue}Topological protection of biphoton
					states}}},\ }\href@noop {} {\bibfield  {journal} {\bibinfo  {journal}
			{Science}\ }\textbf {\bibinfo {volume} {362}},\ \bibinfo {pages} {568}
		(\bibinfo {year} {2018})}\BibitemShut {NoStop}%
	\bibitem [{\citenamefont {Elshaari}\ \emph {et~al.}(2020)\citenamefont
		{Elshaari}, \citenamefont {Pernice}, \citenamefont {Srinivasan},
		\citenamefont {Benson},\ and\ \citenamefont {Zwiller}}]{Elshaari20}%
	\BibitemOpen
	\bibfield  {author} {\bibinfo {author} {\bibfnamefont {A.~W.}\ \bibnamefont
			{Elshaari}}, \bibinfo {author} {\bibfnamefont {W.}~\bibnamefont {Pernice}},
		\bibinfo {author} {\bibfnamefont {K.}~\bibnamefont {Srinivasan}}, \bibinfo
		{author} {\bibfnamefont {O.}~\bibnamefont {Benson}},\ and\ \bibinfo {author}
		{\bibfnamefont {V.}~\bibnamefont {Zwiller}},\ }\bibfield  {title} {\emph
		{\bibinfo {title} {{\color{blue}Hybrid integrated quantum photonic
					circuits}}},\ }\href@noop {} {\bibfield  {journal} {\bibinfo  {journal}
			{Nature Photonics}\ }\textbf {\bibinfo {volume} {14}},\ \bibinfo {pages}
		{285} (\bibinfo {year} {2020})}\BibitemShut {NoStop}%
	\bibitem [{\citenamefont {Yue}\ \emph {et~al.}(2018)\citenamefont {Yue},
		\citenamefont {Dou}, \citenamefont {Wang}, \citenamefont {Ma}, \citenamefont
		{Niu},\ and\ \citenamefont {Sun}}]{Yue18}%
	\BibitemOpen
	\bibfield  {author} {\bibinfo {author} {\bibfnamefont {P.}~\bibnamefont
			{Yue}}, \bibinfo {author} {\bibfnamefont {X.}~\bibnamefont {Dou}}, \bibinfo
		{author} {\bibfnamefont {H.}~\bibnamefont {Wang}}, \bibinfo {author}
		{\bibfnamefont {B.}~\bibnamefont {Ma}}, \bibinfo {author} {\bibfnamefont
			{Z.}~\bibnamefont {Niu}},\ and\ \bibinfo {author} {\bibfnamefont
			{B.}~\bibnamefont {Sun}},\ }\bibfield  {title} {\emph {\bibinfo {title}
			{{\color{blue}{Single photon emissions from InAs/GaAs quantum dots embedded
						in GaAs/SiO$_2$ hybrid microdisks}}}},\ }\href@noop {} {\bibfield  {journal}
		{\bibinfo  {journal} {Optics Communications}\ }\textbf {\bibinfo {volume}
			{411}},\ \bibinfo {pages} {114} (\bibinfo {year} {2018})}\BibitemShut
	{NoStop}%
	\bibitem [{\citenamefont {Davanco}\ \emph {et~al.}(2017)\citenamefont
		{Davanco}, \citenamefont {Liu}, \citenamefont {Sapienza}, \citenamefont
		{Zhang}, \citenamefont {Cardoso}, \citenamefont {Verma}, \citenamefont
		{Mirin}, \citenamefont {Nam}, \citenamefont {Liu},\ and\ \citenamefont
		{Srinivasan}}]{Davanco17}%
	\BibitemOpen
	\bibfield  {author} {\bibinfo {author} {\bibfnamefont {M.}~\bibnamefont
			{Davanco}}, \bibinfo {author} {\bibfnamefont {J.}~\bibnamefont {Liu}},
		\bibinfo {author} {\bibfnamefont {L.}~\bibnamefont {Sapienza}}, \bibinfo
		{author} {\bibfnamefont {C.-Z.}\ \bibnamefont {Zhang}}, \bibinfo {author}
		{\bibfnamefont {J.~V. D.~M.}\ \bibnamefont {Cardoso}}, \bibinfo {author}
		{\bibfnamefont {V.}~\bibnamefont {Verma}}, \bibinfo {author} {\bibfnamefont
			{R.}~\bibnamefont {Mirin}}, \bibinfo {author} {\bibfnamefont {S.~W.}\
			\bibnamefont {Nam}}, \bibinfo {author} {\bibfnamefont {L.}~\bibnamefont
			{Liu}},\ and\ \bibinfo {author} {\bibfnamefont {K.}~\bibnamefont
			{Srinivasan}},\ }\bibfield  {title} {\emph {\bibinfo {title}
			{{\color{blue}Heterogeneous integration for on-chip quantum photonic circuits
					with single quantum dot devices}}},\ }\href@noop {} {\bibfield  {journal}
		{\bibinfo  {journal} {Nature Communications}\ }\textbf {\bibinfo {volume}
			{8}},\ \bibinfo {pages} {1} (\bibinfo {year} {2017})}\BibitemShut {NoStop}%
	\bibitem [{\citenamefont {Osada}\ \emph {et~al.}(2019)\citenamefont {Osada},
		\citenamefont {Ota}, \citenamefont {Katsumi}, \citenamefont {Kakuda},
		\citenamefont {Iwamoto},\ and\ \citenamefont {Arakawa}}]{Osada19}%
	\BibitemOpen
	\bibfield  {author} {\bibinfo {author} {\bibfnamefont {A.}~\bibnamefont
			{Osada}}, \bibinfo {author} {\bibfnamefont {Y.}~\bibnamefont {Ota}}, \bibinfo
		{author} {\bibfnamefont {R.}~\bibnamefont {Katsumi}}, \bibinfo {author}
		{\bibfnamefont {M.}~\bibnamefont {Kakuda}}, \bibinfo {author} {\bibfnamefont
			{S.}~\bibnamefont {Iwamoto}},\ and\ \bibinfo {author} {\bibfnamefont
			{Y.}~\bibnamefont {Arakawa}},\ }\bibfield  {title} {\emph {\bibinfo {title}
			{{\color{blue}{Strongly coupled single-quantum-dot--cavity system integrated
						on a CMOS-processed silicon photonic chip}}}},\ }\href@noop {} {\bibfield
		{journal} {\bibinfo  {journal} {Phys. Rev. Applied}\ }\textbf {\bibinfo
			{volume} {11}},\ \bibinfo {pages} {024071} (\bibinfo {year}
		{2019})}\BibitemShut {NoStop}%
	\bibitem [{\citenamefont {Schuhmann}\ \emph {et~al.}(2022)\citenamefont
		{Schuhmann}, \citenamefont {Lazzari}, \citenamefont {Lemaître},
		\citenamefont {Sagnes}, \citenamefont {Amanti}, \citenamefont {Boeuf},
		\citenamefont {Raineri}, \citenamefont {Baboux},\ and\ \citenamefont
		{Ducci}}]{HybridConf2022}%
	\BibitemOpen
	\bibfield  {author} {\bibinfo {author} {\bibfnamefont {J.}~\bibnamefont
			{Schuhmann}}, \bibinfo {author} {\bibfnamefont {L.}~\bibnamefont {Lazzari}},
		\bibinfo {author} {\bibfnamefont {A.}~\bibnamefont {Lemaître}}, \bibinfo
		{author} {\bibfnamefont {I.}~\bibnamefont {Sagnes}}, \bibinfo {author}
		{\bibfnamefont {M.}~\bibnamefont {Amanti}}, \bibinfo {author} {\bibfnamefont
			{F.}~\bibnamefont {Boeuf}}, \bibinfo {author} {\bibfnamefont
			{F.}~\bibnamefont {Raineri}}, \bibinfo {author} {\bibfnamefont
			{F.}~\bibnamefont {Baboux}},\ and\ \bibinfo {author} {\bibfnamefont
			{S.}~\bibnamefont {Ducci}},\ }\bibfield  {title} {\emph {\bibinfo {title}
			{{\color{blue}{AlGaAs Bragg} reflection waveguides for hybrid quantum
					photonic devices}}},\ }\href@noop {} {\bibfield  {journal} {\bibinfo
			{journal} {ICIQP Conference}\ } (\bibinfo {year} {2022})}\BibitemShut
	{NoStop}%
	\bibitem [{\citenamefont {Yu}\ \emph {et~al.}(2018)\citenamefont {Yu},
		\citenamefont {Sun}, \citenamefont {Li},\ and\ \citenamefont
		{Beling}}]{Yu18}%
	\BibitemOpen
	\bibfield  {author} {\bibinfo {author} {\bibfnamefont {Q.}~\bibnamefont
			{Yu}}, \bibinfo {author} {\bibfnamefont {K.}~\bibnamefont {Sun}}, \bibinfo
		{author} {\bibfnamefont {Q.}~\bibnamefont {Li}},\ and\ \bibinfo {author}
		{\bibfnamefont {A.}~\bibnamefont {Beling}},\ }\bibfield  {title} {\emph
		{\bibinfo {title} {{\color{blue}Segmented waveguide photodetector with 90\%
					quantum efficiency}}},\ }\href@noop {} {\bibfield  {journal} {\bibinfo
			{journal} {Optics Express}\ }\textbf {\bibinfo {volume} {26}},\ \bibinfo
		{pages} {12499} (\bibinfo {year} {2018})}\BibitemShut {NoStop}%
	\bibitem [{\citenamefont {Nehra}\ \emph {et~al.}(2020)\citenamefont {Nehra},
		\citenamefont {Chang}, \citenamefont {Yu}, \citenamefont {Beling},\ and\
		\citenamefont {Pfister}}]{Nehra20}%
	\BibitemOpen
	\bibfield  {author} {\bibinfo {author} {\bibfnamefont {R.}~\bibnamefont
			{Nehra}}, \bibinfo {author} {\bibfnamefont {C.-H.}\ \bibnamefont {Chang}},
		\bibinfo {author} {\bibfnamefont {Q.}~\bibnamefont {Yu}}, \bibinfo {author}
		{\bibfnamefont {A.}~\bibnamefont {Beling}},\ and\ \bibinfo {author}
		{\bibfnamefont {O.}~\bibnamefont {Pfister}},\ }\bibfield  {title} {\emph
		{\bibinfo {title} {{\color{blue}Photon-number-resolving segmented detectors
					based on single-photon avalanche-photodiodes}}},\ }\href@noop {} {\bibfield
		{journal} {\bibinfo  {journal} {Optics Express}\ }\textbf {\bibinfo {volume}
			{28}},\ \bibinfo {pages} {3660} (\bibinfo {year} {2020})}\BibitemShut
	{NoStop}%
	\bibitem [{\citenamefont {Dutt}\ \emph {et~al.}(2015)\citenamefont {Dutt},
		\citenamefont {Luke}, \citenamefont {Manipatruni}, \citenamefont {Gaeta},
		\citenamefont {Nussenzveig},\ and\ \citenamefont {Lipson}}]{Dutt15}%
	\BibitemOpen
	\bibfield  {author} {\bibinfo {author} {\bibfnamefont {A.}~\bibnamefont
			{Dutt}}, \bibinfo {author} {\bibfnamefont {K.}~\bibnamefont {Luke}}, \bibinfo
		{author} {\bibfnamefont {S.}~\bibnamefont {Manipatruni}}, \bibinfo {author}
		{\bibfnamefont {A.~L.}\ \bibnamefont {Gaeta}}, \bibinfo {author}
		{\bibfnamefont {P.}~\bibnamefont {Nussenzveig}},\ and\ \bibinfo {author}
		{\bibfnamefont {M.}~\bibnamefont {Lipson}},\ }\bibfield  {title} {\emph
		{\bibinfo {title} {{\color{blue}On-chip optical squeezing}}},\ }\href@noop {}
	{\bibfield  {journal} {\bibinfo  {journal} {Phys. Rev. Applied}\ }\textbf
		{\bibinfo {volume} {3}},\ \bibinfo {pages} {044005} (\bibinfo {year}
		{2015})}\BibitemShut {NoStop}%
	\bibitem [{\citenamefont {Mondain}\ \emph {et~al.}(2019)\citenamefont
		{Mondain}, \citenamefont {Lunghi}, \citenamefont {Zavatta}, \citenamefont
		{Gouzien}, \citenamefont {Doutre}, \citenamefont {De~Micheli}, \citenamefont
		{Tanzilli},\ and\ \citenamefont {D’Auria}}]{Mondain19}%
	\BibitemOpen
	\bibfield  {author} {\bibinfo {author} {\bibfnamefont {F.}~\bibnamefont
			{Mondain}}, \bibinfo {author} {\bibfnamefont {T.}~\bibnamefont {Lunghi}},
		\bibinfo {author} {\bibfnamefont {A.}~\bibnamefont {Zavatta}}, \bibinfo
		{author} {\bibfnamefont {E.}~\bibnamefont {Gouzien}}, \bibinfo {author}
		{\bibfnamefont {F.}~\bibnamefont {Doutre}}, \bibinfo {author} {\bibfnamefont
			{M.}~\bibnamefont {De~Micheli}}, \bibinfo {author} {\bibfnamefont
			{S.}~\bibnamefont {Tanzilli}},\ and\ \bibinfo {author} {\bibfnamefont
			{V.}~\bibnamefont {D’Auria}},\ }\bibfield  {title} {\emph {\bibinfo {title}
			{{\color{blue}Chip-based squeezing at a telecom wavelength}}},\ }\href@noop
	{} {\bibfield  {journal} {\bibinfo  {journal} {Photonics Research}\ }\textbf
		{\bibinfo {volume} {7}},\ \bibinfo {pages} {A36} (\bibinfo {year}
		{2019})}\BibitemShut {NoStop}%
	\bibitem [{\citenamefont {Kashiwazaki}\ \emph {et~al.}(2020)\citenamefont
		{Kashiwazaki}, \citenamefont {Takanashi}, \citenamefont {Yamashima},
		\citenamefont {Kazama}, \citenamefont {Enbutsu}, \citenamefont {Kasahara},
		\citenamefont {Umeki},\ and\ \citenamefont {Furusawa}}]{Kashiwazaki20}%
	\BibitemOpen
	\bibfield  {author} {\bibinfo {author} {\bibfnamefont {T.}~\bibnamefont
			{Kashiwazaki}}, \bibinfo {author} {\bibfnamefont {N.}~\bibnamefont
			{Takanashi}}, \bibinfo {author} {\bibfnamefont {T.}~\bibnamefont
			{Yamashima}}, \bibinfo {author} {\bibfnamefont {T.}~\bibnamefont {Kazama}},
		\bibinfo {author} {\bibfnamefont {K.}~\bibnamefont {Enbutsu}}, \bibinfo
		{author} {\bibfnamefont {R.}~\bibnamefont {Kasahara}}, \bibinfo {author}
		{\bibfnamefont {T.}~\bibnamefont {Umeki}},\ and\ \bibinfo {author}
		{\bibfnamefont {A.}~\bibnamefont {Furusawa}},\ }\bibfield  {title} {\emph
		{\bibinfo {title} {{\color{blue}Continuous-wave {6-dB-squeezed light with
						2.5-THz-bandwidth from single-mode PPLN waveguide}}}},\ }\href@noop {}
	{\bibfield  {journal} {\bibinfo  {journal} {APL Photonics}\ }\textbf
		{\bibinfo {volume} {5}},\ \bibinfo {pages} {036104} (\bibinfo {year}
		{2020})}\BibitemShut {NoStop}%
	\bibitem [{\citenamefont {Masada}\ \emph {et~al.}(2015)\citenamefont {Masada},
		\citenamefont {Miyata}, \citenamefont {Politi}, \citenamefont {Hashimoto},
		\citenamefont {{O'Brien}},\ and\ \citenamefont {Furusawa}}]{Masada15}%
	\BibitemOpen
	\bibfield  {author} {\bibinfo {author} {\bibfnamefont {G.}~\bibnamefont
			{Masada}}, \bibinfo {author} {\bibfnamefont {K.}~\bibnamefont {Miyata}},
		\bibinfo {author} {\bibfnamefont {A.}~\bibnamefont {Politi}}, \bibinfo
		{author} {\bibfnamefont {T.}~\bibnamefont {Hashimoto}}, \bibinfo {author}
		{\bibfnamefont {J.~L.}\ \bibnamefont {{O'Brien}}},\ and\ \bibinfo {author}
		{\bibfnamefont {A.}~\bibnamefont {Furusawa}},\ }\bibfield  {title} {\emph
		{\bibinfo {title} {{\color{blue}Continuous-variable entanglement on a
					chip}}},\ }\href@noop {} {\bibfield  {journal} {\bibinfo  {journal} {Nature
				Photonics}\ }\textbf {\bibinfo {volume} {9}},\ \bibinfo {pages} {316}
		(\bibinfo {year} {2015})}\BibitemShut {NoStop}%
	\bibitem [{\citenamefont {Tasker}\ \emph {et~al.}(2021)\citenamefont {Tasker},
		\citenamefont {Frazer}, \citenamefont {Ferranti}, \citenamefont {Allen},
		\citenamefont {Brunel}, \citenamefont {Tanzilli}, \citenamefont {D’Auria},\
		and\ \citenamefont {Matthews}}]{Tasker21}%
	\BibitemOpen
	\bibfield  {author} {\bibinfo {author} {\bibfnamefont {J.~F.}\ \bibnamefont
			{Tasker}}, \bibinfo {author} {\bibfnamefont {J.}~\bibnamefont {Frazer}},
		\bibinfo {author} {\bibfnamefont {G.}~\bibnamefont {Ferranti}}, \bibinfo
		{author} {\bibfnamefont {E.~J.}\ \bibnamefont {Allen}}, \bibinfo {author}
		{\bibfnamefont {L.~F.}\ \bibnamefont {Brunel}}, \bibinfo {author}
		{\bibfnamefont {S.}~\bibnamefont {Tanzilli}}, \bibinfo {author}
		{\bibfnamefont {V.}~\bibnamefont {D’Auria}},\ and\ \bibinfo {author}
		{\bibfnamefont {J.~C.}\ \bibnamefont {Matthews}},\ }\bibfield  {title} {\emph
		{\bibinfo {title} {{\color{blue}Silicon photonics interfaced with integrated
					electronics for 9 {GHz} measurement of squeezed light}}},\ }\href@noop {}
	{\bibfield  {journal} {\bibinfo  {journal} {Nature Photonics}\ }\textbf
		{\bibinfo {volume} {15}},\ \bibinfo {pages} {11} (\bibinfo {year}
		{2021})}\BibitemShut {NoStop}%
	\bibitem [{\citenamefont {Yan}\ \emph {et~al.}(2022)\citenamefont {Yan},
		\citenamefont {He}, \citenamefont {Liu}, \citenamefont {Iu}, \citenamefont
		{Ahmed}, \citenamefont {Chen}, \citenamefont {Blakey}, \citenamefont
		{Akasaka}, \citenamefont {Ikeuchi},\ and\ \citenamefont {Helmy}}]{Yan22}%
	\BibitemOpen
	\bibfield  {author} {\bibinfo {author} {\bibfnamefont {Z.}~\bibnamefont
			{Yan}}, \bibinfo {author} {\bibfnamefont {H.}~\bibnamefont {He}}, \bibinfo
		{author} {\bibfnamefont {H.}~\bibnamefont {Liu}}, \bibinfo {author}
		{\bibfnamefont {M.}~\bibnamefont {Iu}}, \bibinfo {author} {\bibfnamefont
			{O.}~\bibnamefont {Ahmed}}, \bibinfo {author} {\bibfnamefont
			{E.}~\bibnamefont {Chen}}, \bibinfo {author} {\bibfnamefont {P.}~\bibnamefont
			{Blakey}}, \bibinfo {author} {\bibfnamefont {Y.}~\bibnamefont {Akasaka}},
		\bibinfo {author} {\bibfnamefont {T.}~\bibnamefont {Ikeuchi}},\ and\ \bibinfo
		{author} {\bibfnamefont {A.~S.}\ \bibnamefont {Helmy}},\ }\bibfield  {title}
	{\emph {\bibinfo {title} {{\color{blue}{$\chi$ 2-based {AlGaAs} phase
						sensitive amplifier with record gain, noise, and sensitivity}}}},\
	}\href@noop {} {\bibfield  {journal} {\bibinfo  {journal} {Optica}\ }\textbf
		{\bibinfo {volume} {9}},\ \bibinfo {pages} {56} (\bibinfo {year}
		{2022})}\BibitemShut {NoStop}%
	\bibitem [{\citenamefont {Brodutch}\ \emph {et~al.}(2018)\citenamefont
		{Brodutch}, \citenamefont {Marchildon},\ and\ \citenamefont
		{Helmy}}]{Brodutch18}%
	\BibitemOpen
	\bibfield  {author} {\bibinfo {author} {\bibfnamefont {A.}~\bibnamefont
			{Brodutch}}, \bibinfo {author} {\bibfnamefont {R.}~\bibnamefont
			{Marchildon}},\ and\ \bibinfo {author} {\bibfnamefont {A.~S.}\ \bibnamefont
			{Helmy}},\ }\bibfield  {title} {\emph {\bibinfo {title}
			{{\color{blue}Dynamically reconfigurable sources for arbitrary {Gaussian}
					states in integrated photonics circuits}}},\ }\href@noop {} {\bibfield
		{journal} {\bibinfo  {journal} {Optics Express}\ }\textbf {\bibinfo {volume}
			{26}},\ \bibinfo {pages} {17635} (\bibinfo {year} {2018})}\BibitemShut
	{NoStop}%
	\bibitem [{\citenamefont {Abouraddy}\ \emph {et~al.}(2007)\citenamefont
		{Abouraddy}, \citenamefont {Yarnall}, \citenamefont {Saleh},\ and\
		\citenamefont {Teich}}]{Abouraddy07}%
	\BibitemOpen
	\bibfield  {author} {\bibinfo {author} {\bibfnamefont {A.~F.}\ \bibnamefont
			{Abouraddy}}, \bibinfo {author} {\bibfnamefont {T.}~\bibnamefont {Yarnall}},
		\bibinfo {author} {\bibfnamefont {B.~E.}\ \bibnamefont {Saleh}},\ and\
		\bibinfo {author} {\bibfnamefont {M.~C.}\ \bibnamefont {Teich}},\ }\bibfield
	{title} {\emph {\bibinfo {title} {{\color{blue}Violation of {Bell’s}
					inequality with continuous spatial variables}}},\ }\href@noop {} {\bibfield
		{journal} {\bibinfo  {journal} {Phys. Rev. A}\ }\textbf {\bibinfo {volume}
			{75}},\ \bibinfo {pages} {052114} (\bibinfo {year} {2007})}\BibitemShut
	{NoStop}%
	\bibitem [{\citenamefont {Fabre}\ \emph
		{et~al.}(2022{\natexlab{b}})\citenamefont {Fabre}, \citenamefont {Keller},\
		and\ \citenamefont {Milman}}]{Fabre22}%
	\BibitemOpen
	\bibfield  {author} {\bibinfo {author} {\bibfnamefont {N.}~\bibnamefont
			{Fabre}}, \bibinfo {author} {\bibfnamefont {A.}~\bibnamefont {Keller}},\ and\
		\bibinfo {author} {\bibfnamefont {P.}~\bibnamefont {Milman}},\ }\bibfield
	{title} {\emph {\bibinfo {title} {{\color{blue}Time and frequency as quantum
					continuous variables}}},\ }\href@noop {} {\bibfield  {journal} {\bibinfo
			{journal} {Phys. Rev. A}\ }\textbf {\bibinfo {volume} {105}},\ \bibinfo
		{pages} {052429} (\bibinfo {year} {2022}{\natexlab{b}})}\BibitemShut
	{NoStop}%
	\bibitem [{\citenamefont {Tasca}\ \emph {et~al.}(2011)\citenamefont {Tasca},
		\citenamefont {Gomes}, \citenamefont {Toscano}, \citenamefont {Ribeiro},\
		and\ \citenamefont {Walborn}}]{Tasca11}%
	\BibitemOpen
	\bibfield  {author} {\bibinfo {author} {\bibfnamefont {D.}~\bibnamefont
			{Tasca}}, \bibinfo {author} {\bibfnamefont {R.}~\bibnamefont {Gomes}},
		\bibinfo {author} {\bibfnamefont {F.}~\bibnamefont {Toscano}}, \bibinfo
		{author} {\bibfnamefont {P.~S.}\ \bibnamefont {Ribeiro}},\ and\ \bibinfo
		{author} {\bibfnamefont {S.}~\bibnamefont {Walborn}},\ }\bibfield  {title}
	{\emph {\bibinfo {title} {{\color{blue}Continuous-variable quantum
					computation with spatial degrees of freedom of photons}}},\ }\href@noop {}
	{\bibfield  {journal} {\bibinfo  {journal} {Phys. Rev. A}\ }\textbf {\bibinfo
			{volume} {83}},\ \bibinfo {pages} {052325} (\bibinfo {year}
		{2011})}\BibitemShut {NoStop}%
	\bibitem [{\citenamefont {Gottesman}\ \emph {et~al.}(2001)\citenamefont
		{Gottesman}, \citenamefont {Kitaev},\ and\ \citenamefont
		{Preskill}}]{Gottesman01}%
	\BibitemOpen
	\bibfield  {author} {\bibinfo {author} {\bibfnamefont {D.}~\bibnamefont
			{Gottesman}}, \bibinfo {author} {\bibfnamefont {A.}~\bibnamefont {Kitaev}},\
		and\ \bibinfo {author} {\bibfnamefont {J.}~\bibnamefont {Preskill}},\
	}\bibfield  {title} {\emph {\bibinfo {title} {{\color{blue}Encoding a qubit
					in an oscillator}}},\ }\href@noop {} {\bibfield  {journal} {\bibinfo
			{journal} {Physical Review A}\ }\textbf {\bibinfo {volume} {64}},\ \bibinfo
		{pages} {012310} (\bibinfo {year} {2001})}\BibitemShut {NoStop}%
	\bibitem [{\citenamefont {Kwiat}(1997)}]{Kwiat97}%
	\BibitemOpen
	\bibfield  {author} {\bibinfo {author} {\bibfnamefont {P.~G.}\ \bibnamefont
			{Kwiat}},\ }\bibfield  {title} {\emph {\bibinfo {title}
			{{\color{blue}Hyper-entangled states}}},\ }\href@noop {} {\bibfield
		{journal} {\bibinfo  {journal} {Journal of Modern Optics}\ }\textbf {\bibinfo
			{volume} {44}},\ \bibinfo {pages} {2173} (\bibinfo {year}
		{1997})}\BibitemShut {NoStop}%
	\bibitem [{\citenamefont {Wang}\ \emph
		{et~al.}(2018{\natexlab{c}})\citenamefont {Wang}, \citenamefont {Luo},
		\citenamefont {Huang}, \citenamefont {Chen}, \citenamefont {Su},
		\citenamefont {Liu}, \citenamefont {Chen}, \citenamefont {Li}, \citenamefont
		{Fang}, \citenamefont {Jiang} \emph {et~al.}}]{Wang18_18qubit}%
	\BibitemOpen
	\bibfield  {author} {\bibinfo {author} {\bibfnamefont {X.-L.}\ \bibnamefont
			{Wang}}, \bibinfo {author} {\bibfnamefont {Y.-H.}\ \bibnamefont {Luo}},
		\bibinfo {author} {\bibfnamefont {H.-L.}\ \bibnamefont {Huang}}, \bibinfo
		{author} {\bibfnamefont {M.-C.}\ \bibnamefont {Chen}}, \bibinfo {author}
		{\bibfnamefont {Z.-E.}\ \bibnamefont {Su}}, \bibinfo {author} {\bibfnamefont
			{C.}~\bibnamefont {Liu}}, \bibinfo {author} {\bibfnamefont {C.}~\bibnamefont
			{Chen}}, \bibinfo {author} {\bibfnamefont {W.}~\bibnamefont {Li}}, \bibinfo
		{author} {\bibfnamefont {Y.-Q.}\ \bibnamefont {Fang}}, \bibinfo {author}
		{\bibfnamefont {X.}~\bibnamefont {Jiang}}, \emph {et~al.},\ }\bibfield
	{title} {\emph {\bibinfo {title} {{\color{blue}18-qubit entanglement with six
					photons’ three degrees of freedom}}},\ }\href@noop {} {\bibfield  {journal}
		{\bibinfo  {journal} {Physical review letters}\ }\textbf {\bibinfo {volume}
			{120}},\ \bibinfo {pages} {260502} (\bibinfo {year}
		{2018}{\natexlab{c}})}\BibitemShut {NoStop}%
	\bibitem [{\citenamefont {Steinlechner}\ \emph {et~al.}(2017)\citenamefont
		{Steinlechner}, \citenamefont {Ecker}, \citenamefont {Fink}, \citenamefont
		{Liu}, \citenamefont {Bavaresco}, \citenamefont {Huber}, \citenamefont
		{Scheidl},\ and\ \citenamefont {Ursin}}]{Steinlechner17}%
	\BibitemOpen
	\bibfield  {author} {\bibinfo {author} {\bibfnamefont {F.}~\bibnamefont
			{Steinlechner}}, \bibinfo {author} {\bibfnamefont {S.}~\bibnamefont {Ecker}},
		\bibinfo {author} {\bibfnamefont {M.}~\bibnamefont {Fink}}, \bibinfo {author}
		{\bibfnamefont {B.}~\bibnamefont {Liu}}, \bibinfo {author} {\bibfnamefont
			{J.}~\bibnamefont {Bavaresco}}, \bibinfo {author} {\bibfnamefont
			{M.}~\bibnamefont {Huber}}, \bibinfo {author} {\bibfnamefont
			{T.}~\bibnamefont {Scheidl}},\ and\ \bibinfo {author} {\bibfnamefont
			{R.}~\bibnamefont {Ursin}},\ }\bibfield  {title} {\emph {\bibinfo {title}
			{{\color{blue}Distribution of high-dimensional entanglement via an intra-city
					free-space link}}},\ }\href@noop {} {\bibfield  {journal} {\bibinfo
			{journal} {Nature Communications}\ }\textbf {\bibinfo {volume} {8}},\
		\bibinfo {pages} {1} (\bibinfo {year} {2017})}\BibitemShut {NoStop}%
	\bibitem [{\citenamefont {Vergyris}\ \emph {et~al.}(2019)\citenamefont
		{Vergyris}, \citenamefont {Mazeas}, \citenamefont {Gouzien}, \citenamefont
		{Labont{\'e}}, \citenamefont {Alibart}, \citenamefont {Tanzilli},\ and\
		\citenamefont {Kaiser}}]{Vergyris19}%
	\BibitemOpen
	\bibfield  {author} {\bibinfo {author} {\bibfnamefont {P.}~\bibnamefont
			{Vergyris}}, \bibinfo {author} {\bibfnamefont {F.}~\bibnamefont {Mazeas}},
		\bibinfo {author} {\bibfnamefont {E.}~\bibnamefont {Gouzien}}, \bibinfo
		{author} {\bibfnamefont {L.}~\bibnamefont {Labont{\'e}}}, \bibinfo {author}
		{\bibfnamefont {O.}~\bibnamefont {Alibart}}, \bibinfo {author} {\bibfnamefont
			{S.}~\bibnamefont {Tanzilli}},\ and\ \bibinfo {author} {\bibfnamefont
			{F.}~\bibnamefont {Kaiser}},\ }\bibfield  {title} {\emph {\bibinfo {title}
			{{\color{blue}Fibre based hyperentanglement generation for dense wavelength
					division multiplexing}}},\ }\href@noop {} {\bibfield  {journal} {\bibinfo
			{journal} {Quantum Science and Technology}\ }\textbf {\bibinfo {volume}
			{4}},\ \bibinfo {pages} {045007} (\bibinfo {year} {2019})}\BibitemShut
	{NoStop}%
	\bibitem [{\citenamefont {Kim}\ \emph {et~al.}(2021)\citenamefont {Kim},
		\citenamefont {Kim}, \citenamefont {Im}, \citenamefont {Lee}, \citenamefont
		{Chae}, \citenamefont {Scarcelli},\ and\ \citenamefont {Kim}}]{Kim21}%
	\BibitemOpen
	\bibfield  {author} {\bibinfo {author} {\bibfnamefont {J.-H.}\ \bibnamefont
			{Kim}}, \bibinfo {author} {\bibfnamefont {Y.}~\bibnamefont {Kim}}, \bibinfo
		{author} {\bibfnamefont {D.-G.}\ \bibnamefont {Im}}, \bibinfo {author}
		{\bibfnamefont {C.-H.}\ \bibnamefont {Lee}}, \bibinfo {author} {\bibfnamefont
			{J.-W.}\ \bibnamefont {Chae}}, \bibinfo {author} {\bibfnamefont
			{G.}~\bibnamefont {Scarcelli}},\ and\ \bibinfo {author} {\bibfnamefont
			{Y.-H.}\ \bibnamefont {Kim}},\ }\bibfield  {title} {\emph {\bibinfo {title}
			{{\color{blue}Noise-resistant quantum communications using
					hyperentanglement}}},\ }\href@noop {} {\bibfield  {journal} {\bibinfo
			{journal} {Optica}\ }\textbf {\bibinfo {volume} {8}},\ \bibinfo {pages}
		{1524} (\bibinfo {year} {2021})}\BibitemShut {NoStop}%
\end{thebibliography}

%

\end{document}